\documentclass[aps,pof,superscriptaddress]{revtex4}
\usepackage[parfill]{parskip}    			% Activate to begin paragraphs with an empty line rather than an indent
\usepackage{amssymb}
\usepackage{amsmath}
\usepackage{graphicx}					% Include figure files
\usepackage{epstopdf}					%Package ams, je sais pas trop ce qu'ils font, pour les maths je crois
\usepackage[latin1]{inputenc}				%Ces deux lignes permettent à latex
\usepackage[T1]{fontenc}					%de reconnaitre les accents

\usepackage{version}				
\usepackage{cases}						%Pour les systèmes d'équations

\usepackage{pstricks}					%pour dessiner où on veut ou annoter des figures

\usepackage{mathrsfs}
\usepackage[squaren, Gray]{SIunits}			%pour écrire les unités

\begin{document}

%Title of paper
\title{Global vs local energy dissipation: the energy cycle of  the  turbulent von Kármán flow}
\author{Denis Kuzzay}
\email{denis.kuzzay@cea.fr}
\affiliation{
Laboratoire SPHYNX, Service de Physique de l'Etat Condens\'e, DSM,
CEA Saclay, CNRS UMR 3680, 91191 Gif-sur-Yvette, France
}
\author{Davide Faranda}%
\affiliation{
LSCE CEA Saclay, CNRS UMR 8212 , 91191 Gif-sur-Yvette, France
}
\author{B{\'e}reng{\`e}re Dubrulle}%
\affiliation{
Laboratoire SPHYNX, Service de Physique de l'Etat Condens\'e, DSM,
CEA Saclay, CNRS UMR 3680, 91191 Gif-sur-Yvette, France
}

\begin{abstract}

In this paper, we investigate the relations between global and local energy transfers in a turbulent von Kármán flow. The goal is to understand how and where energy is dissipated in such a flow and to reconstruct the energy cycle in an experimental device where local as well as global quantities can be measured. In order to do so, we use PIV measurements and we model the Reynolds stress tensor to take subgrid scales into account. This procedure involves a free parameter that is calibrated using  angular momentum balance. We then estimate the local and global mean injected and dissipated power for several types of impellers, for various Reynolds numbers and for various flow topologies. These PIV-estimates are then compared with direct injected power estimates provided by torque measurements at the impellers. The agreement between PIV-estimates and direct measurements depends on the flow topology. In symmetric situations, we are able to capture up to $90\%$ of the actual global energy dissipation rate. However, our results become increasingly inaccurate as the shear layer responsible for most of the dissipation approaches one of the impellers, and cannot be resolved by our PIV set-up. Finally, we show that a very good agreement between PIV-estimates and direct measurements is obtained using a new method based on the  work of Duchon and Robert [J. Duchon and R. Robert, Nonlinearity, 13, 249 (2000)] which generalizes the Kármán-Howarth equation to nonisotropic, nonhomogeneous flows. This method provides parameter-free estimates of the energy dissipation rate as long as the smallest resolved scale lies in the inertial range. These results are used to evidence a well-defined stationary energy cycle within the flow in which most of the energy is injected at the top and bottom impellers, and  dissipated within the shear layer. The influence of the mean flow geometry and the Reynolds number on this energy cycle is studied for a wide range of parameters.

 \end{abstract}

\maketitle

\section{Introduction}

Understanding how and where energy is dissipated in turbulent flows has been a great challenge for many years, and would have important implications in many areas such as fundamental research, aeronautics or industry.  In the classical three-dimensional turbulence phenomenology, energy is injected at large scales by the forcing mechanism, transferred downscale at a constant rate $\epsilon$  following a self-similar cascade, and then dissipated into heat at the Kolmogorov  length scale, where viscous effects become dominant. In ideal stationary, homogeneous and isotropic turbulence, the measurement of energy dissipation can therefore be achieved via 3 independent and equivalent means: i) by monitoring the injected energy; ii) by monitoring the dissipated heat; iii) by monitoring the cascade energy rate via multi-scale single points measurements of the velocity (via e.g. anemometers or array of hot wires). The first two measurements are global, the last one local, but given the homogeneity in space and time, they all provide the same information. In most realistic situations, however, the turbulence is anisotropic and/or inhomogeneous and/or non-stationary. In such cases, there is not any necessary equivalence between global energy injection, global energy dissipation and local energy dissipation. The study of these three quantities requires detailed knowledge of the forcing, the heat distribution and velocity over the whole domain.\

In that respect, it is interesting to focus on intermediate situations, where the turbulence is generated by a well controlled mechanism, in a simplified geometry, as achieved for example in classical laboratory experiments such as Taylor-Couette or von Kármán set-ups. In such cases, it is easy to monitor the forcing  and implement a cooling mechanism so as to achieve stationarity, where global energy injection and dissipation equilibrate on average. The local energy dissipation can then be computed from stationary energy budgets derived from Navier-Stokes equation, using local measurements of velocity obtained for example using the now classical Particle Image Velocimetry (PIV). This technique provides measurements of the instantaneous velocity field $u_i$ at several points of a plane (or of a volume) at the same time. From this, one may compute the Reynolds stress tensor $S_{ij}=\frac{1}{2}(\partial_i u_j+\partial_j u_i)$, and study the local dissipated power $\epsilon_v=2\nu S_{ij}S_{ij}$. A detailed comparison between this local estimate and the global energy dissipation rate in a Taylor-Couette flow at various Reynolds number has recently been made by Tokgoz et al. \cite{tokgoz2012} using tomographic PIV measurements. They observe that the local and global estimates coincide within $10 \%$ as long as  the laminar flow and Taylor vortex flow regimes are fully resolved. However, as the Reynolds number increases, the dissipative scale decreases and becomes smaller than the finite resolution of the PIV (set by the camera resolution and the velocity reconstruction algorithm). In that case, the estimate of local energy dissipation based on the velocity gradients and on the viscosity $\epsilon_v$ becomes increasingly inaccurate, and underestimate the global energy dissipation. In order to remedy this problem, it has been suggested to use  techniques borrowed from Large Eddy Simulation (LES) \cite{sheng2000} . This allows us to model the subgrid scales (SGS) in terms of the large scale velocity field resolved by the PIV and allows computation of all terms in the energy budget \cite{saa2000,tanaka2007,delafosse2011,baldihann} including the terms responsible for the scale-to-scale energy transfer. By the Richardson-Kolmogorov cascade picture, this allows an estimate of the local energy dissipation  as long as the scale used in the computation lies in the inertial range. \medbreak

In this paper, we test these methods in a turbulent flow generated by two contra-rotating impellers (von Kármán flow) for Reynolds numbers ranging from $10^3$ up to more than $10^6$. At such Reynolds numbers, the dissipative scale ranges from a few millimeters to a few tens of microns. With fixed velocity of the impellers and cooling, the resulting flow is stationary, highly anisotropic and inhomogeneous \cite{cortet2009}, thereby providing a unique laboratory flow to test local energy dissipation procedure. This closed flow geometry permits direct estimates of the global energy injection by torque monitoring at the two impellers. Using a large scale Helium facility with calorimetric measurements \cite{roussetrsi}, we were able to show that in a stationary state, the global injected power and the global dissipated heat coincide within a few percent at large Reynolds number, and for a wide range of operating conditions (including differential rotation of the impellers). Moreover, the simple cylindrical geometry allows for Stereoscopic PIV measurements of the velocity field over a vertical plane spanning the whole experiment, at  a resolution of a few millimeters. Since measurements have only been made in a meridional plane, we do not have  access to orthoradial derivatives. Hence, it is challenging to have an accurate estimate of the local energy dissipation rate despite this lack of information. To do this we first use a LES method using statistical axisymmetry and angular momentum budget to calibrate the model. Then, we test a generalization of the Kármán-Howarth formula derived by Duchon and Robert \cite{duchonrobert} that provides a parameter-free estimate of the local energy dissipation for any nonhomogeneous, nonisotropic flow. Both estimates are then averaged over the whole volume for comparison with the global estimate of the energy dissipation based on torque monitoring. This is done for a wide range of control parameters, varying both the Reynolds number, the mean flow geometry and the flow asymmetry. We then use these measurements to evidence a stationary energy cycle within the flow where energy is injected at the top and bottom impellers, and dissipated within the shear layer.

\bigbreak

This paper is organized as follows: in Section II, we summarize the theoretical tools needed for the implementation of the LES technique and the Duchon-Robert (DR) formula in our analysis of the energy dissipation rate. In Section III, we review the von Kármán geometry and specialize these formulae to the case of cylindrical geometry. The flow diagnostics based on PIV measurements are derived and summarized. In  Section IV, we apply these results to a set of measurements drawn from our data base of von Kármán flow. We first tune our LES model using angular momentum balance. Then, we compare PIV estimates of the global dissipated power with direct torque measurements using both the LES technique and the DR formula. Finally, in section \ref{results}, we evidence the energy cycle of the von Kármán flow, and we discuss its evolution, as well the evolution of our diagnostics, as a function of the flow topology.

\section{Theoretical background and methodology}

In this section, we use a filtering approach to derive energy and angular momentum balance at a given scale $\ell$ from Navier-Stokes equations, and we give the expression of the different terms that we will use in our analysis of the injected and dissipated power. We consider Navier-Stokes equations

\begin{numcases}
\strut
\label{NSeq}
\partial_t u_i + u_j \partial_j u_i = -\partial_i P + \nu \partial_j \partial_j u_i\\
\label{incomp}
\partial_j u_j =0,
\end{numcases}

where we use Einstein summation convention over repeated indices.

\subsection{The filtering approach}

Following the procedure in \cite{germano92}, we define a coarse-grained velocity field at scale $\ell$ as

\begin{equation}
u^\ell_i(\vec x) = \int d\vec r \ G_\ell(\vec r) u_i(\vec x + \vec r),
\end{equation}

where $G$ is a smooth filtering function, nonnegative, spatially localized and such that $\int d\vec r \ G(\vec r)=1$. The function $G_\ell$ is rescaled with $\ell$ as $G_\ell (\vec r) = \ell^{-3}G(\vec r/\ell)$. Coarse-graining the Navier-Stokes equations gives

\begin{numcases}
\strut
\label{NSeqcg}
\partial_t u^\ell_i + u^\ell_j \partial_j u^\ell_i = -\partial_j \tau^{ij} -\partial_i P^\ell+ \nu \partial_j \partial_j u^\ell_i \\
\label{incomp}
\partial_j u^\ell_j =0.
\end{numcases}

In equation (\ref{NSeqcg}) we introduced $\tau^{ij}=(u_iu_j)^\ell-u^\ell_iu^\ell_j$ which is the stress tensor from the SGS. 

In what follows, in order to simplify the formulae and for readability considerations, we will drop the index $\ell$. Unless specified otherwise \emph{$u_i$ now denotes the $i^{th}$ component of the coarse-grained velocity field} (same thing for $P$). 

\subsection{The energy balance equation}

We now take the scalar product of equation (\ref{NSeqcg}) with $u_i$, and after a few lines of algebra we get the local energy balance equation

\begin{equation}
\partial_t \overline{E} + \partial_j (\overline{u_jE}) = -\partial_j (\overline{u_i\tau^{ij}}) + \overline{S_{ij}\tau^{ij}} - \partial_i (\overline{u_iP})+ \nu (\partial_j \partial_j \overline{E} + \partial_i\partial_j (\overline{u_iu_j)}) - 2\nu \overline{S_{ij}S^{ij}},
\label{Ebalance}
\end{equation}

where $E=\frac{u_iu_i}{2}$ is the large scale kinetic energy per unit mass, and $S_{ij}=\frac{1}{2}(\partial_i u_j + \partial_j u_i) $ is the large scale strain rate tensor. The overline denotes the statistical average of a quantity. Let us now define a vector $J^j=u_j(E+P)+u_i\tau^{ij}-\nu(\partial_jE+\partial_i (u_iu_j))$. We can then rewrite (\ref{Ebalance}) as

\begin{equation}
\partial_t \overline{E} + \partial_j \overline{J^j} = \overline{S_{ij}\tau^{ij}} - 2\nu \overline{S_{ij}S^{ij}}.
\label{Econservation}
\end{equation}

This local equation is valid in any geometry, for any type of flow, for any filtering with the properties given above. In section III D, we apply this formula to the specific axisymmetric von Kármán geometry to derive the local energy production and dissipation rate per unit mass.

\subsection{The angular momentum balance equation}

In a very similar fashion, it is also possible to derive an angular momentum balance equation at scale $\ell$. We take the cross product of $r_j$ and equation (\ref{NSeqcg}), and after a few lines of algebra we get

\begin{equation}
\partial_t \overline{L_i} + \partial_j (\overline{u_jL_i}) = -\epsilon_{ijk}\partial_k(\overline{r_jP}) - \epsilon_{ijk}\partial_m(\overline{r_j\tau^{km}}) + \nu (\partial_j \partial_j \overline{L_i} + \epsilon_{ijk}\partial_m\partial_k (\overline{r_ju_m)}) - \nu\epsilon_{ijk}\partial_j\overline{u_k},
\label{Lbalance}
\end{equation}

where $\epsilon_{ijk}$ is the three dimensional, total antisymmetric Levi-Civita symbol so that $L_i = \epsilon_{ijk}r_ju_k$ is the $i^{th}$ component of the angular momentum per unit mass. Of course, it can be checked that it is also possible to derive (\ref{Lbalance}) from (\ref{Ebalance}) using (\ref{NSeqcg}) and the identity $E=\frac{1}{2\vec{r}^2}(\vec{L}^2+(\vec{r}.\vec{u})^2)$. As before, we can write (\ref{Lbalance}) as an equation of conservation for each component of $\vec{L}$

\begin{equation}
\partial_t \overline{L_i} + \partial_j \overline{T^{ij}} = 0,
\label{Lconservation}
\end{equation}

where $T^{ij}=u_jL_i-\epsilon_{ijk}r_kP+\epsilon_{imk}r_m\tau^{jk}-\nu(\partial_jL_i-\epsilon_{ijk}\partial_m(r_ku_m)+\epsilon_{ijk}u_k)$. 

In section IV A, we use this angular momentum balance as a constraint to calibrate the LES model, that we now describe.

\subsection{LES method for balance equations}

The computation of the different terms in (\ref{Econservation}) and (\ref{Lconservation}) requires the knowledge of both the velocity field at scale $\ell$ and the Reynolds stress $\tau^{ij}$. In most practical situations, e.g. when the flow is turbulent and the velocity is measured through a PIV system, only the former is available, since we cannot resolve the dissipative scale. A traditional way to overcome this problem is to use a LES-PIV approach \cite{tokgoz2012,sheng2000} to model $\tau^{ij}$ in terms of the large scale velocity field \cite{lesieurbook,eyink2005}. Several models exist. In the present paper, we choose the gradient model \cite{clark79} where 

\begin{equation}
\tau^{ij}=C\Delta_r^2\partial_ku_i\partial_ku_j,
\label{clarkmodel}
\end{equation}

 where $C$ is a constant to be calibrated and $\Delta_r$ is the width of the filtering. This model ensures forward scatter and backscatter of energy between resolved scales and SGS.

\subsection{Duchon-Robert energy balance equation}
\label{DRsubsec}

An alternative local energy balance equation has been derived by Duchon and Robert \cite{duchonrobert} using Leray's weak solution formalism and Onsager's ideas \cite{onsager49}. The latter amounts to consider a sequence of coarse-grained solutions of the Navier-Stokes equations Eq. (\ref{NSeqcg}) in the limit $\ell\to 0$ and derive the corresponding energy balance, that reads
\begin{equation}
\partial_t E + \partial_j \left(u^j(E+P)-\nu\partial_j E \right)= -\nu\partial_j u_i\partial^j u^i -\mathscr{D}(u^i),
\label{DREbalance}
\end{equation}

where  $\mathscr{D}(u^i)$ is expressed in terms of velocity increments $\delta \vec u (\vec r) = \vec u(\vec x + \vec r) - \vec u(\vec x)$ as

\begin{equation}
\mathscr{D}(\vec u) \equiv \lim_{\ell\rightarrow 0} \mathscr{D}_\ell (\vec u) = \lim_{\ell\rightarrow 0} \frac{1}{4\ell} \int_\mathcal{V} d\vec r \ (\vec\nabla G_\ell)(\vec r) \cdot \delta\vec u(\vec r) \ |\delta\vec u (\vec r)|^2,
\label{DRfieldnotGeneral}
\end{equation}

where the dependence of $\delta\vec u$ and $\mathscr{D}$ in $\vec x$ is implied. As the Reynolds number tends to infinity, the scale $\ell$ can be chosen as small as one wants, and the quantity $\mathscr{D}(u^i)$  can be seen as the contribution to dissipation coming from a generalized cascade process (possibly linked with the formation of small-scale singularities). Since the result cannot depend on the filtering function $G_\ell$, Duchon and Robert specialized the expression to a radially symmetric filter to get the alternative expression, devoid of any free parameter

\begin{equation}
\mathscr{D}(\vec u) = -\frac{3}{16\pi} \lim_{\epsilon\rightarrow 0} \frac{1}{\epsilon} \int_{|\vec\chi|=1} d\Sigma(\vec\chi) \ |\delta\vec u (\epsilon\vec\chi)|^2 \ \delta\vec u (\epsilon\vec\chi) \cdot \vec\chi \ ,
\label{DRfield}
\end{equation}

where $d\Sigma$ denotes the area measure on the sphere. As noticed by Duchon and Robert, this expression coincides with the statistical mean rate of inertial energy dissipated per unit mass derived from the anisotropic version of the Kármán-Howarth equation. They therefore argue that the previous formula provides a local non-random form of the Kármán-Howarth equation, valid even for anisotropic, inhomogeneous flows. In the sequel, we apply this formula to our PIV measurements to test whether it can provide a parameter-free estimate of the global dissipation, as well as local instantaneous maps of the local energy dissipation.

\section{Application to a von Kármán geometry}

The goal of the present paper is to compare global estimates of dissipated power with torque measurements, check their coincidence, and study maps of local energy dissipation rate to get some insight of the detailed processes governing energy transfers in a von Kármán flow.

\subsection{Experimental set-up}

The von Kármán experiment has been extensively studied over the past years \cite{mariephd,raveletphd,monchauxphd,bricephd,cortet2009,cortet2010,brice2014,brice2014shrek,ravelet2005}. We give here a brief review of the main features of the set-up. \smallbreak

\begin{figure}
\centering
\includegraphics[width=8cm]{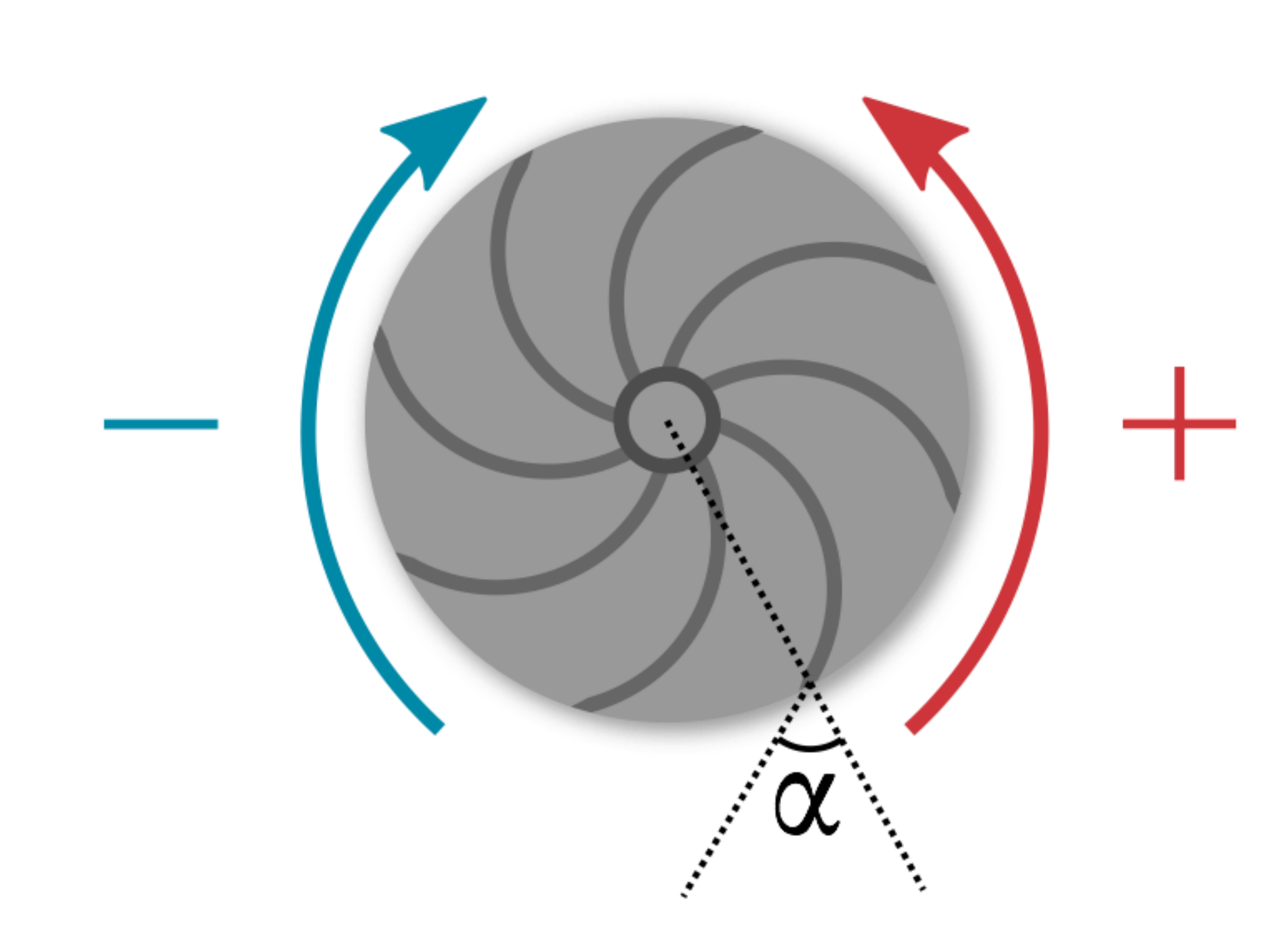}
\caption{Illustration of a TP87 type impeller, with 8 blades. The convention used to name the two different forcing conditions is represented: (+) when the convex face of the blade goes forward, (-) for the other. The angle $\alpha$ characterizes the curvature of the blades and is equal to $\alpha=72\degree$ in the case of TP87 impellers. TM60 impellers look essentially the same, with 16 blades instead of 8.}
\label{impeller}
\end{figure}

Our von Kármán flow is generated in a vertical cylinder by two coaxial, contra-rotating impellers providing energy and momentum flux at the upper and lower end of the cylinder. The inner radius of the cylinder is $R=100$mm and the distance between the inner face of the impellers is $H=180$mm, which gives an aspect ratio of $H/R=1.8$. The turbulence properties (anisotropy, fluctuations, dissipation) are influenced by the geometry of the impellers, i.e their non dimensional radius $R_t$, the oriented angle $\alpha$ between the blades (see Fig. \ref{impeller}), the number $n$ of blades and their heights $h_b$ \cite{raveletphd}. In the present paper, we consider only impellers with $h_b/R=0.2$ and $\alpha=\pm 72\degree$. Those impellers are the so-called ``TM60'' (with 16 blades) and ``TP87'' impellers (with 8 blades), the characteristics of which are summarized in Table  \ref{table:param}. They are essentially similar, except for their number of blades and the material they are made of. A single impeller can be used to propel the fluid in two opposite directions, respectively associated to the concave or convex face of the blades going forward. This can be taken into account by a change of sign of the parameter $\alpha$. In the sequel, we denote $(-)$ ({\it resp.} $(+)$ ) an impeller used with the concave ({\it resp.} convex) face of its blades going forward. The impellers are driven by two independent motors which can rotate at frequencies up to $10\hertz$. The motor frequencies can be either set equal to get exact counter-rotating regime, or set to different values $f_1\ne f_2$. To change the viscosity, we have used either water or glycerol at different dilution rates. 

\subsection{Control parameters}

In the sequel, we choose  $R$ and $\Omega^{-1}=(\pi(f_1+f_2))^{-1}$ as unit of length and time. The von Kármán experiment is then characterized by two control parameters:

\medbreak

\begin{itemize}

\item the Reynolds number:
$$Re=\pi (f_1+f_2) R^2 \nu^{-1},$$ 

where $\nu$ is the fluid kinematic viscosity, ranges from $10^2$ to more than $10^6$ so that we can span a full range of regimes, from the purely laminar to the fully turbulent one.

\item the rotation number:
$$\theta=\frac{f_1-f_2}{f_1+f_2},$$ 

measures the relative influence of global rotation over a typical turbulent shear frequency. Indeed, the exact counter-rotating regime corresponds to $\theta=0$. For a nonzero rotation number, our experimental system is similar, within lateral boundaries, to an exact counter-rotating experiment at frequency $f=(f_1+f_2)/2$, with an overall global rotation at frequency $(f_1-f_2)/2$ \cite{mariephd,ravelet2005}. In our experiments we vary $\theta$ from $-1$ to $+1$, exploring a regime of relatively weak rotation to shear ratio.

\end{itemize}

\medbreak

Table \ref{table:param} summarizes the parameter space explored in this paper.  The Reynolds variation is done at $\theta=0$ while the rotation variation has been explored at $Re\approx10^5$.

\begin{table*}[ht]
\centering
\resizebox{0.8\textwidth}{!}{
\begin{tabular}
{||c||c|c|c|c|c||}%
\hline \hline
    impellers&material &number of blades&$\alpha$  (in degrees)       &$Re$&$\theta$\\
\hline \hline
%\hline
TM60(+)&stainless steel&16&$72$&$[10^3,10^6]$&$ 0$\\
%\hline
TP87(+)&polycarbonate&8&$72$&$10^5$&$ [-0.5 , 0.5]$\\
%\hline
TP87(-)&polycarbonate&8&$-72$&$10^5    $&$ [-0.5 , 0.5]$\\
\hline \hline
\end{tabular}
}

{\caption{Parameter space explored in this paper.} \label{table:param}}
\end{table*}

\subsection{Measurements}

The set-up allows for both global and local flow diagnostics. Torque (global) measurements at each impeller are performed with
SCAIME technology and provide values over the $\kilo\hertz$ range of $C_1$ and $C_2$, being respectively the torque applied to the bottom and top shafts. Following the procedure described in \cite{mariephd}, they are calibrated using measurements at different mean frequencies, so as to remove spurious contributions from genuine offsets or mechanical frictions. From this, we compute the nondimensional value $K_{p1}$ and $K_{p2}$ of the torque as: $K_{pi}=C_i/(\rho R^5\Omega^2)$, where $\rho$ is the density of the working fluid. 

Local measurements of the velocity field of the flow have been made by PIV techniques in the stationary regime. The typical size of the particles used is a few tens of micrometers and their density is 1.4. Two cameras take between 600 and a few thousand successive pictures of the flow at a $15\hertz$ frequency. The resolution of our camera frame is  1600x1200 pixels, and the reconstruction is done using peak correlation performed over overlapping windows of size 16 to 32 pixels. As a result, we get measurements of velocity field on a grid of approximate size $60^2$ in a vertical plane containing the axis of symmetry (Oz), in a cylindrical system of coordinate. The maximum spatial resolution we can reach for the velocity field with this set-up is therefore of the order of $200/60\approx 3$mm, about 10 to 100 times larger than the dissipative scale. For more details about the experimental set-up or the measurement techniques see for instance \cite{bricephd}. We can therefore estimate the derivatives of the velocity field only along the r and z directions, but we do not have access to derivatives along $\theta$. As a consequence, we will either set these derivatives to zero in the sequel, or take them into account using hypothesis of statistical axisymmetry. We have checked that the two procedure give essentially the same result. In any case, we use incompressibility to estimate $\partial_\theta u_\theta$.

\subsection{Diagnostics}

\subsubsection{Flow geometry}

\begin{figure}[!h]
\centering
\includegraphics[width=7.5cm]{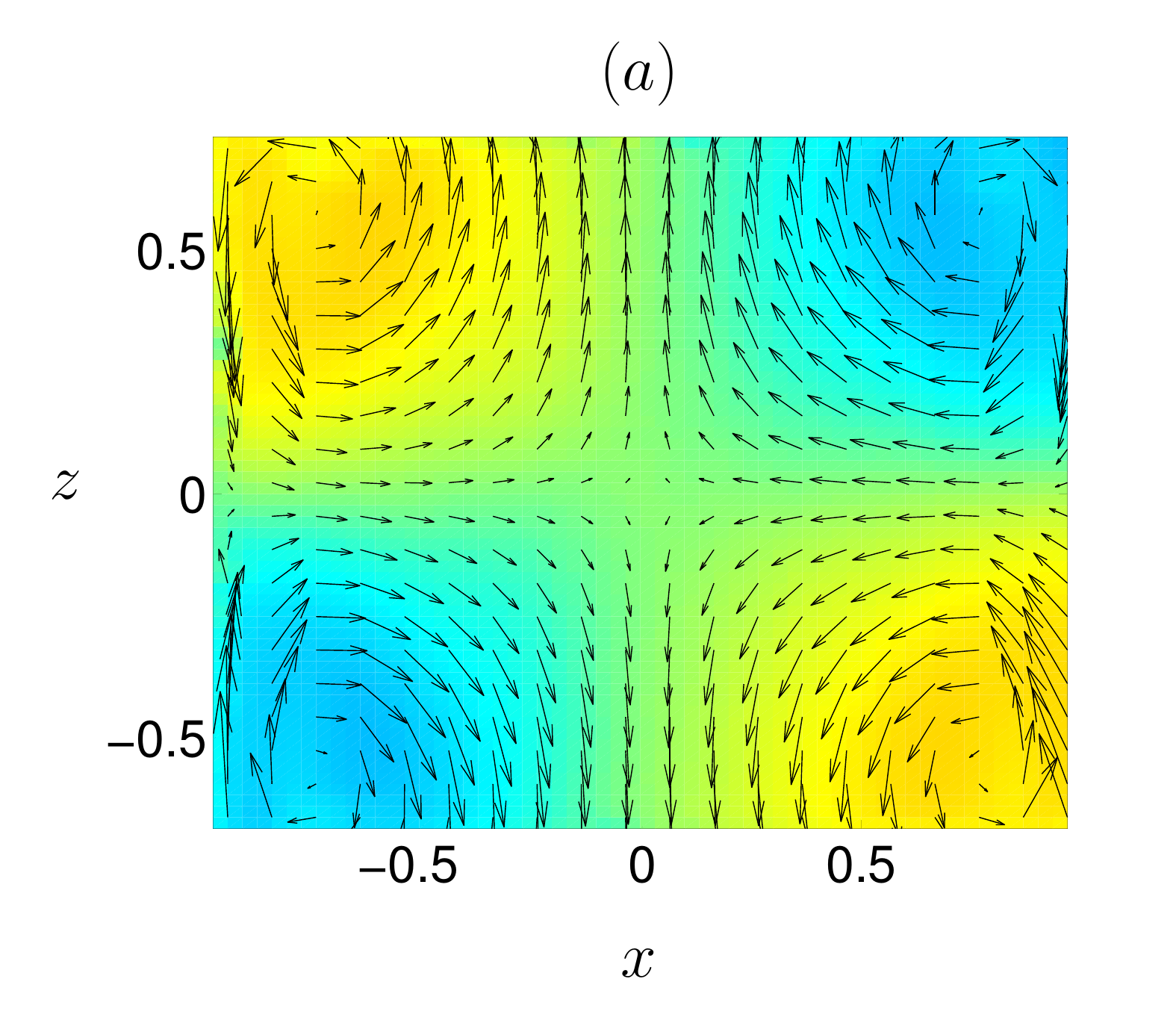}
\includegraphics[width=7.5cm]{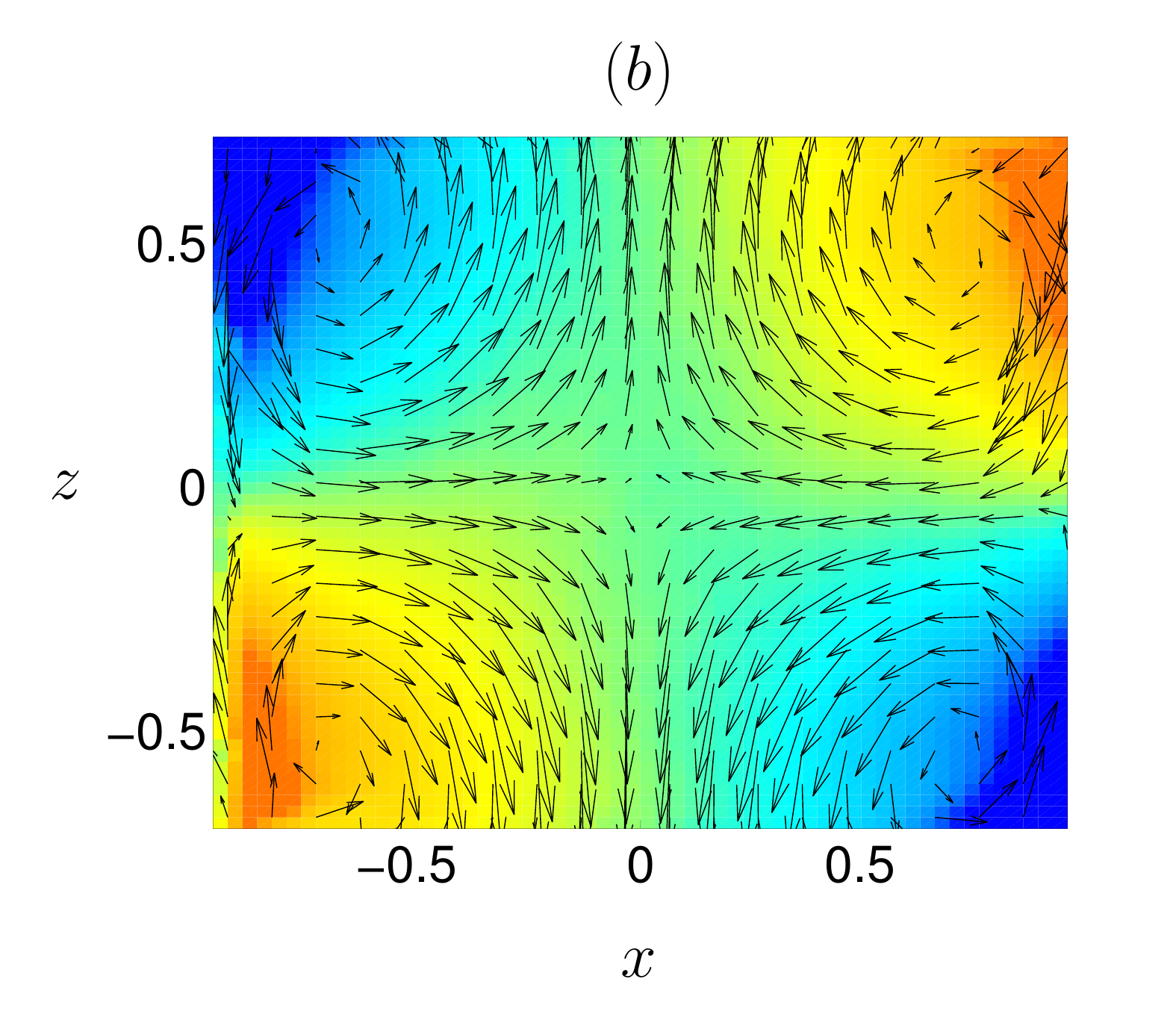}
\includegraphics[width=7.5cm]{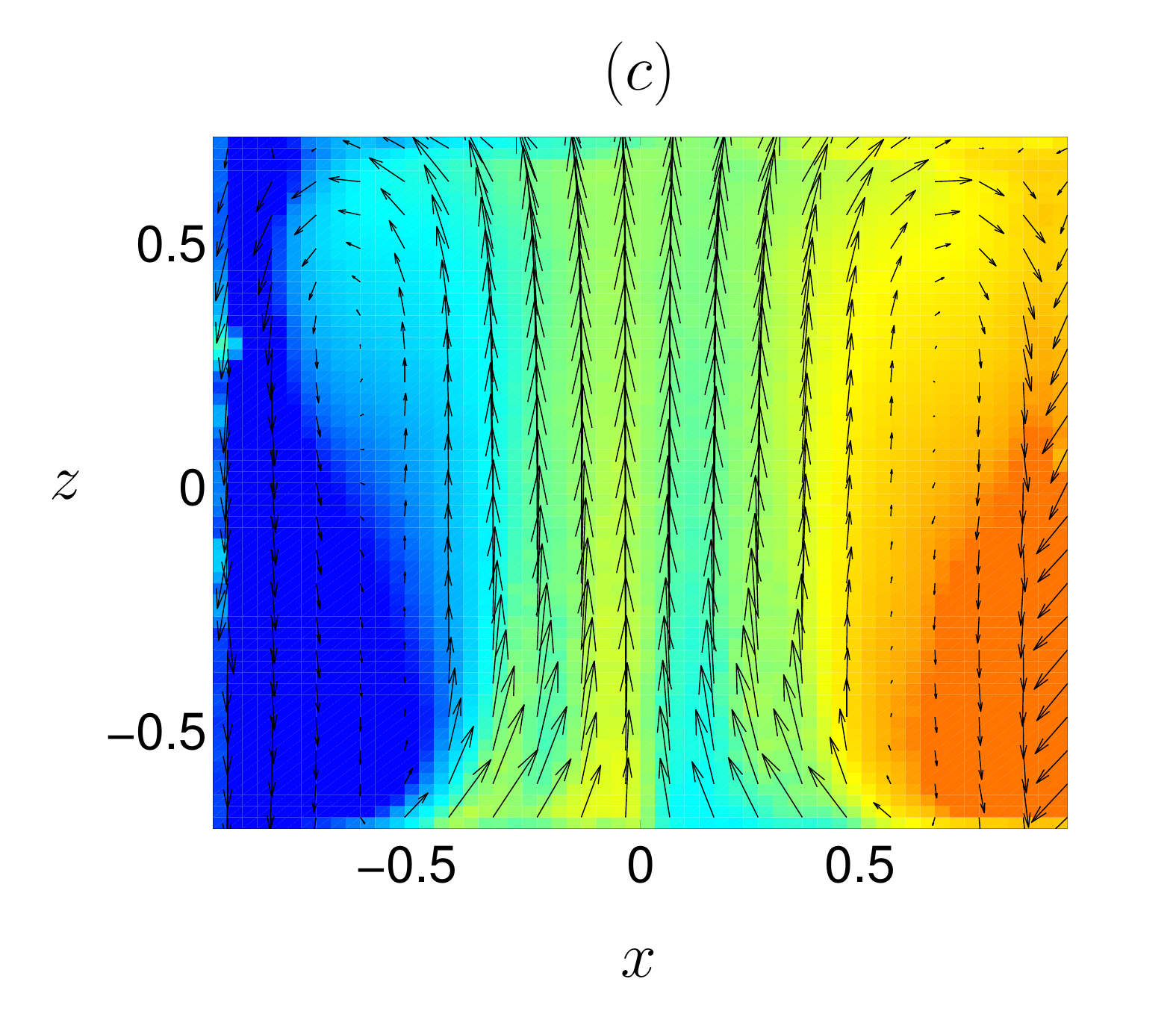}
\includegraphics[width=7.5cm]{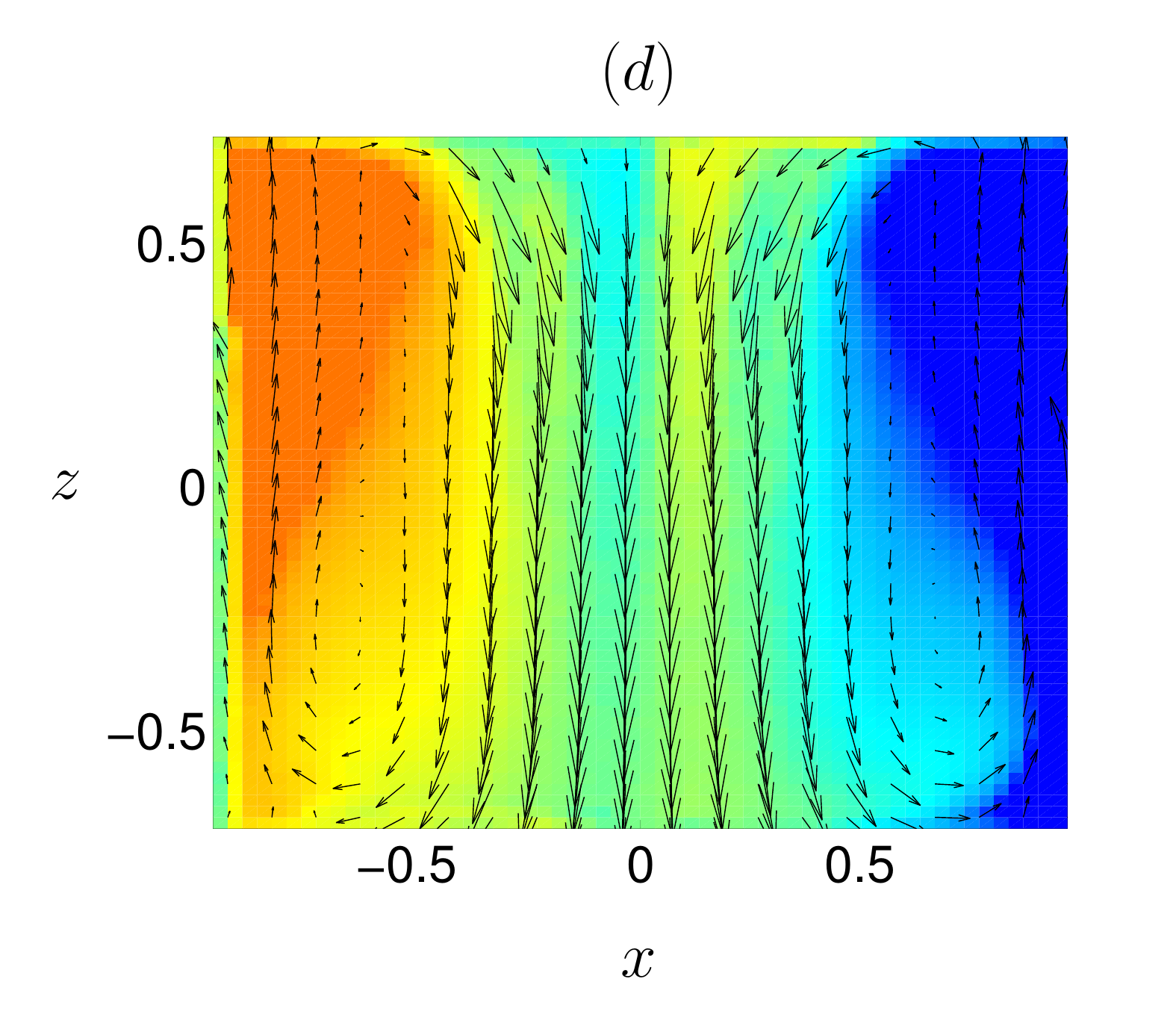}
\includegraphics[width=15cm]{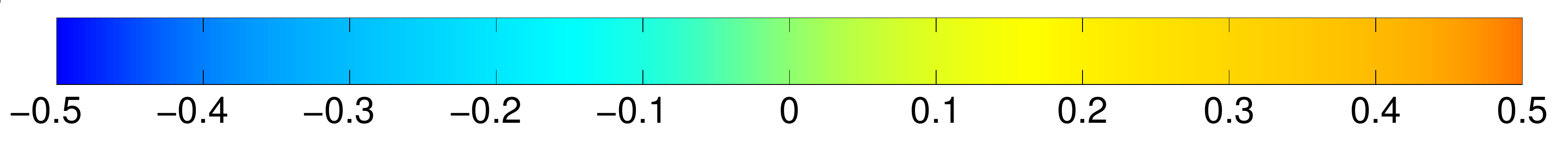}
\caption{The four types of flow characterized by their mean velocity profiles. Arrows represent the velocity field in the radial plane while colours represent the orthogonal component to that plane. a) The flow is forced under the (+) condition, no phase transition can be observed. b), c) and d) show the three different states of a flow forced under the (-) condition. b) no phase transition has occurred, c) (resp. d)) the shear layer has been sent downwards (resp. upwards). In each of these flows $Re\approx3\times10^5$, and the two TP87 impellers rotate in opposite directions at $\theta=0$.}
\label{4types}
\end{figure}

We are going to study different types of von Kármán flows (see Fig. \ref{4types}). These types of flow happen depending on the forcing condition (+) or (-), and whether the system undergoes a spontaneous phase transition (bifurcation) or not \cite{cortet2010}. There is, then, four types of flows: one corresponding to the (+) forcing condition where a phase transition cannot be observed (Fig. \ref{4types}(a)), one corresponding to the (-) forcing condition where a phase transition is not observed (Fig. \ref{4types}(b)), and two more corresponding to the two states of the flow that can be observed once the flow has undergone its phase transition (Fig. \ref{4types}(c)-(d)). The difference between these flows can be characterized through our PIV measurements by their mean velocity profile in the vertical plane of measurements (see Fig. \ref{4types}). In the first two types, the mean flow is symmetric with respect to the equatorial plane $z=0$ and there is a strong shear layer in the middle. In the bifurcated states, the flow is no longer symmetric with respect to the equatorial plane. It consists in a one-cell flow in the vertical direction, with a strong shear layer at the impeller that rotates in the direction opposite to the orthoradial mean flow. The two bifurcated states are symmetric to each other. The interest of considering these different mean flow geometries is that they are characterized by well resolved (resp. badly resolved)  shear layer for the symmetric state (resp. bifurcated state). Since we expect an important fraction of energy dissipation to be localized where there is a strong velocity gradient, this difference may be a large source of error in local estimates based on PIV measurements, as we demonstrate later.

\subsubsection{Global dissipation}

Using torque measurements, we can get an accurate estimate of the global power injected  into the flow. Indeed, since we study a stationary situation, this input must balance the rate of energy dissipation within the flow. This has been checked in a scale 4:1 version of our experiment (see  \cite{roussetrsi}). From these measurements, we get the global mean dissipation rate as $D=2\pi(C_1 f_1+C_2 f_2)$. From this, we compute diagnostics:

\medbreak

\begin{itemize}

\item the dimensionless mean dissipation rate:
\begin{equation}
\mathcal{D}=\frac{D}{\rho R^5 \Omega^3}=K_{p1}(1+\theta)+K_{p2}(1-\theta),
\label{nondimdissipation}
\end{equation}

\item the dimensionless mean dissipated power per unit mass:
\begin{equation}
\epsilon=\mathcal{D}\frac{R^3}{{\mathcal V}}=\frac{\mathcal{D}R}{\pi H},
\label{nondimepsilon}
\end{equation}
which should not be confused with the Levi-Civita symbol. $\mathcal{V}=\pi R^2 H$ is the volume of the experiment.

\end{itemize}

\medbreak

\begin{figure}[!h]
\centering
\includegraphics[width=8.9cm]{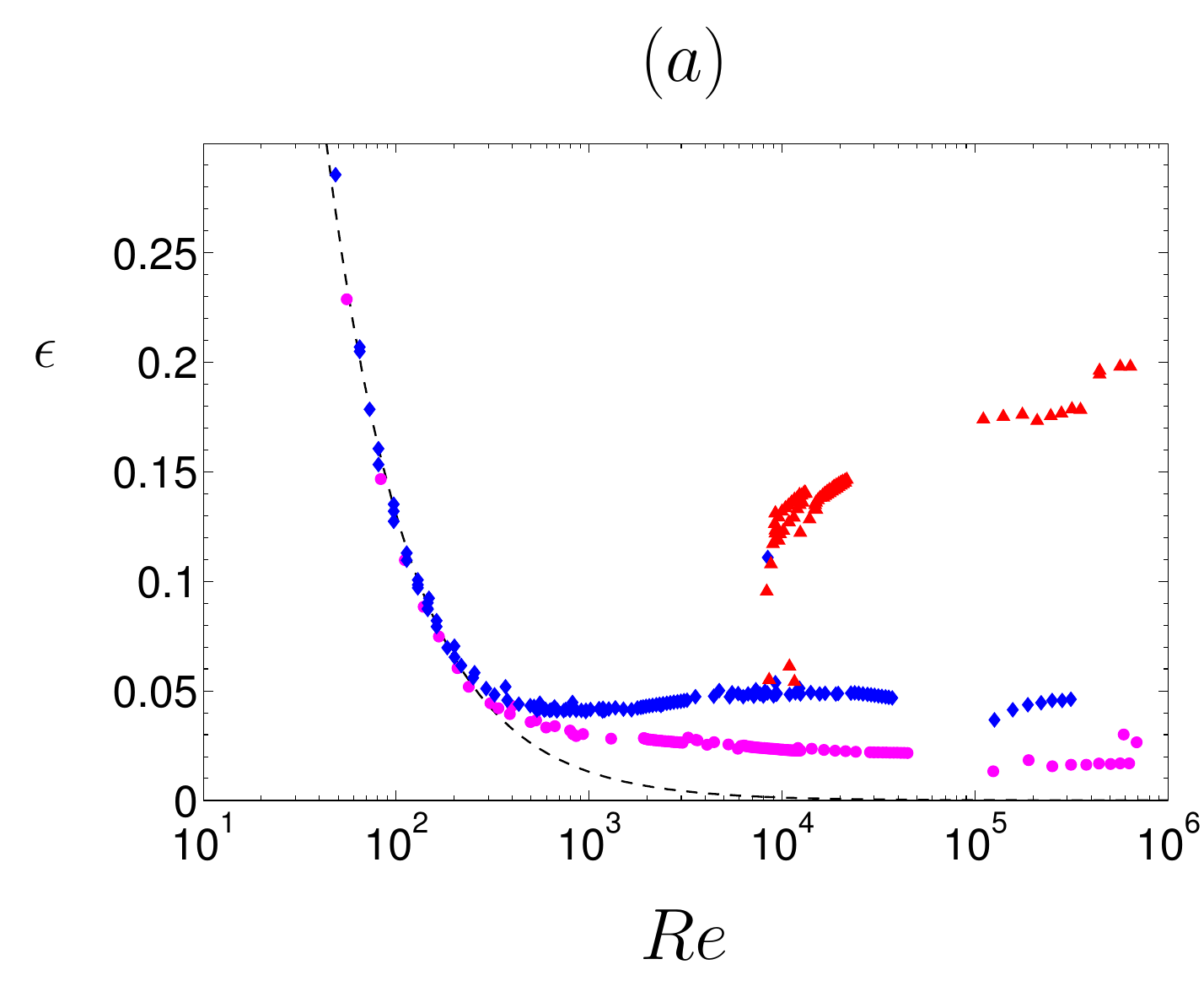}
\includegraphics[width=8.9cm]{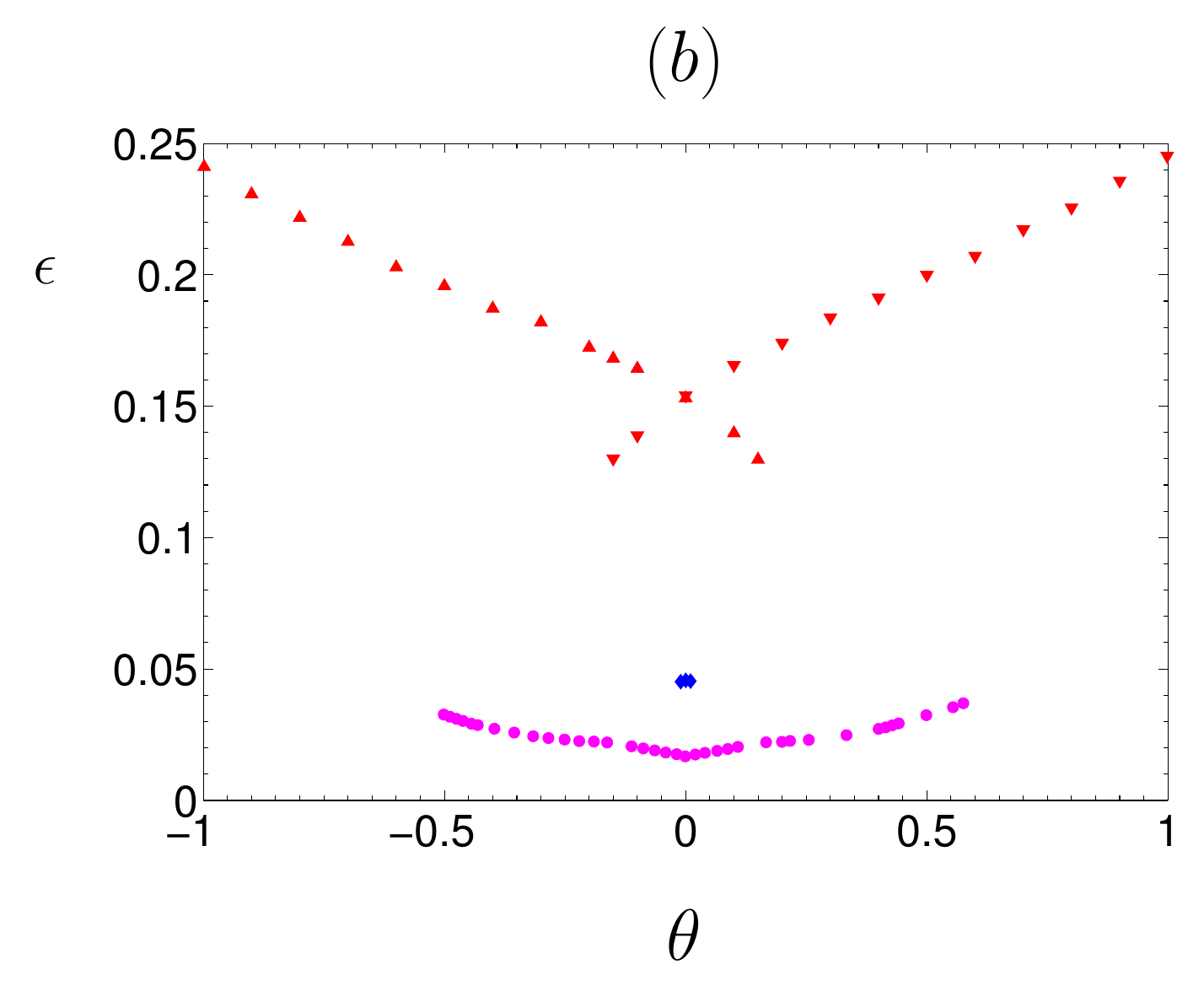}
\caption{Plots of the dimensionless dissipated power per unit mass: a) As a function of Re for TP87 and TM60 impellers for the four geometries depicted in Fig. \ref{4types} at $\theta=0$. Dotted line: Fit $\epsilon\approx37Re^{-1}$. Pink disks: symmetric state (+). Blue rhombi: symmetric state (-). Red triangles: bifurcated states (-). b) As a function of $\theta$ for TP87 impellers, $Re=3\times 10^5$. Pink disks: symmetric state (+). Blue rhombi:  symmetric state (-). Red triangles: bifurcated states (-).}
\label{epsilonReet theta}
\end{figure}

These quantities depend on the Reynolds number, the rotation number, the characteristics of the impellers and the mean flow geometry \cite{ravelet2008}. Examples of variation of $\epsilon$ as a function of Re are provided in Fig. \ref{epsilonReet theta}(a) at $\theta=0$, with TP87 and TM60 impellers, for the different flow geometries illustrated in Fig. \ref{4types}. At low Reynolds numbers, the dissipation rate decays as $Re^{-1}$, until $Re\approx 300$ where the turbulence sets in. The dissipation rate then levels off at a value which depends upon the flow geometry: it is the lowest for the symmetric (+) flow, then increases for the (-) symmetric flow and is the largest for the two bifurcated (-) flows. In Fig. \ref{epsilonReet theta}(b), the mean dissipation rate is shown as a function of $\theta$ for TP87 $(\pm)$  impellers at $Re=3\times 10^5$. In the $(+)$ case (pink disks), it may be seen that the minimum of $\epsilon$ is obtained at $\theta=0$. When $\theta$ is varied from, say, $0$ to $0.5$, $\epsilon$ increases and reaches a value twice as large as its minimum. In the $(-)$ case, there is a discontinuity of the energy dissipation at $\theta=0$ due to the global symmetry breaking. We see that the symmetric state (blue rhombi) dissipates less energy than the two bifurcated states (red triangles). Another striking feature of Fig. \ref{epsilonReet theta}(b), is the coexistence of two branches of dissipation, corresponding to the two bifurcated states, for a certain range of values of $\theta$. This coexistence vanishes around $\theta=\pm0.2$ (this value depends on the Reynolds number, see \cite{brice2014shrek}).

\subsubsection{Local production and dissipation rates}

Since the von Kármán flow is neither isotropic, nor homogeneous, it is interesting to study more locally the production and dissipation of energy. Our set-up being statistically axisymmetric, it is sufficient to perform the study in the vertical plane including the rotation axis, where we perform our velocity measurements. We thus consider the main stationary terms of the local energy balance Eq. (\ref{Econservation}), and divide them by the experimental volume $\mathcal{V}=\pi R^2 H$ to get three local quantities:

\medbreak

\begin{itemize}

\item the local production rate per unit mass:
\begin{equation}
\Gamma=-\frac{1}{\mathcal{V}}\partial_j \overline{J^j} .
\label{prod}
\end{equation}

Integrating over the whole volume, one may then get from this the corresponding total production rate per unit mass $\Gamma_T$:

\begin{equation}
\Gamma_T=-\frac{1}{\mathcal{V}} \int_{\mathcal{V}} \partial_i \overline{J^i} d\mathcal{V}=-\frac{1}{\mathcal{V}} \oint_{\mathcal{S}} \overline{J^i} d\mathcal{S}_i  ,
\label{prodglob}
\end{equation}

where the last equality comes from  Green-Ostrogradsky formula.

\item the positive local dissipation (transfer) rate per unit mass:
\begin{equation}
\Pi=\frac{1}{\mathcal{V}}\left(-\overline{S_{ij}\tau^{ij}} + 2\nu \overline{S_{ij}S^{ij}}\right).
\label{transfer}
\end{equation}

The first term  represents the energy dissipation due to energy transfers towards SGS while the second one is the laminar energy dissipation. 

\item the positive local singularity dissipation (transfer) rate per unit mass:
\begin{equation}
\Pi_{DR}=\frac{1}{\mathcal{V}} \overline{\mathscr{D}(\vec u)}.
\label{transferSingu}
\end{equation}
As before, we denote by $\Pi_T$ and $\Pi_{DRT}$ the total dissipation rate per unit mass, obtained by volume integration of $\Pi$ and $\Pi_{DR}$.

\end{itemize}

\medbreak

With our stereoscopic PIV measurements, we have access to all terms to compute these quantities, except for the pressure $P$ in $J^i$, and the terms involving derivatives with respect to the azimuthal angle. In the sequel, we will present maps of the local quantities without the terms we do not have access to. In other words, we neglect all terms involving pressure and azimuthal derivatives. Moreover, the computation of the transfer terms requires the calibration of the constant $C$ involved in the gradient model (see Eq. (\ref{clarkmodel})). In section IV A, we describe a calibration procedure using the angular momentum balance equation.

\section{Data processing}

\subsection{Calibration of the gradient model using angular momentum balance}

\begin{figure}[htbp]
\centering
\includegraphics[width=0.36\textwidth]{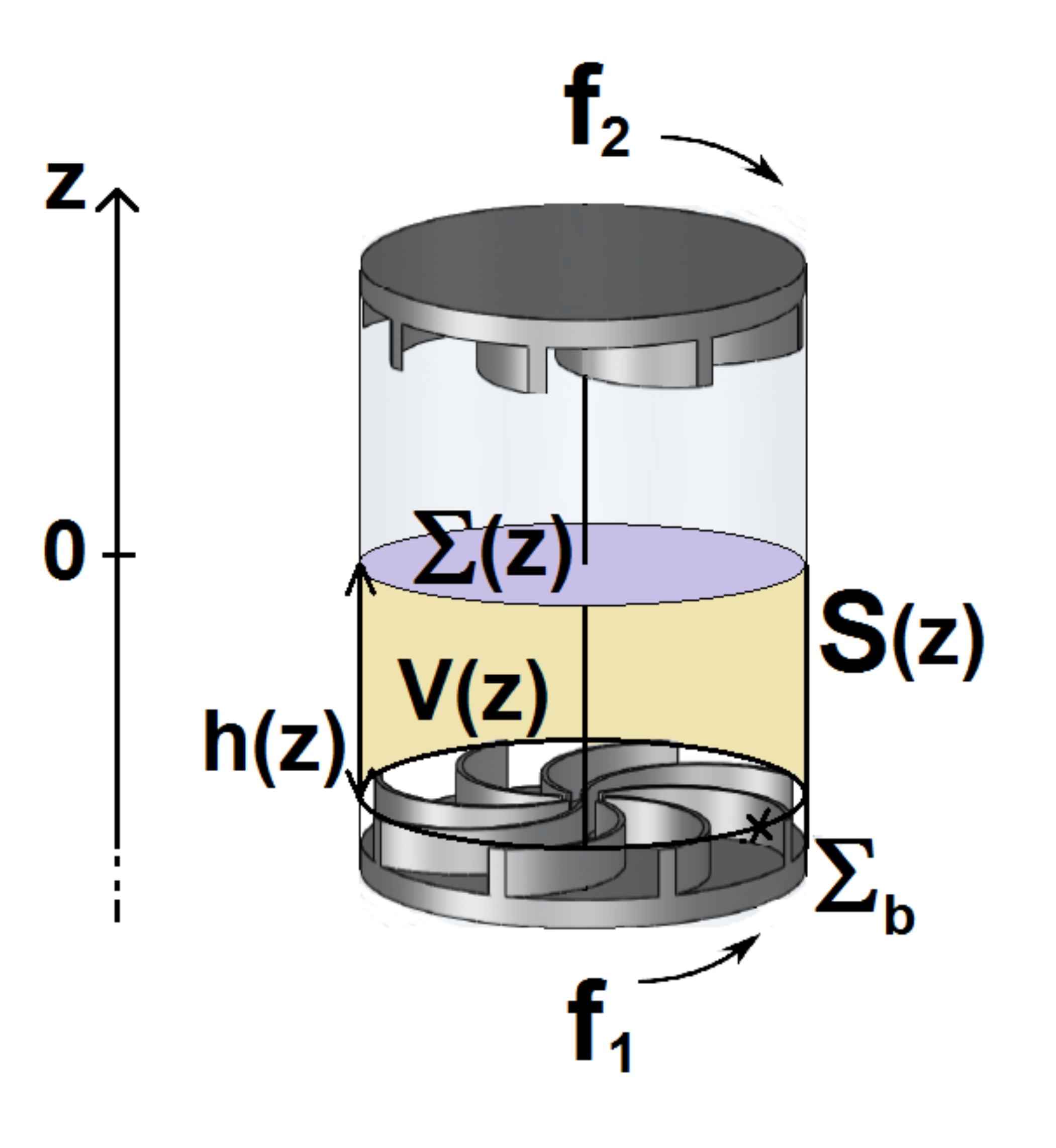}
\includegraphics[width=0.63\textwidth]{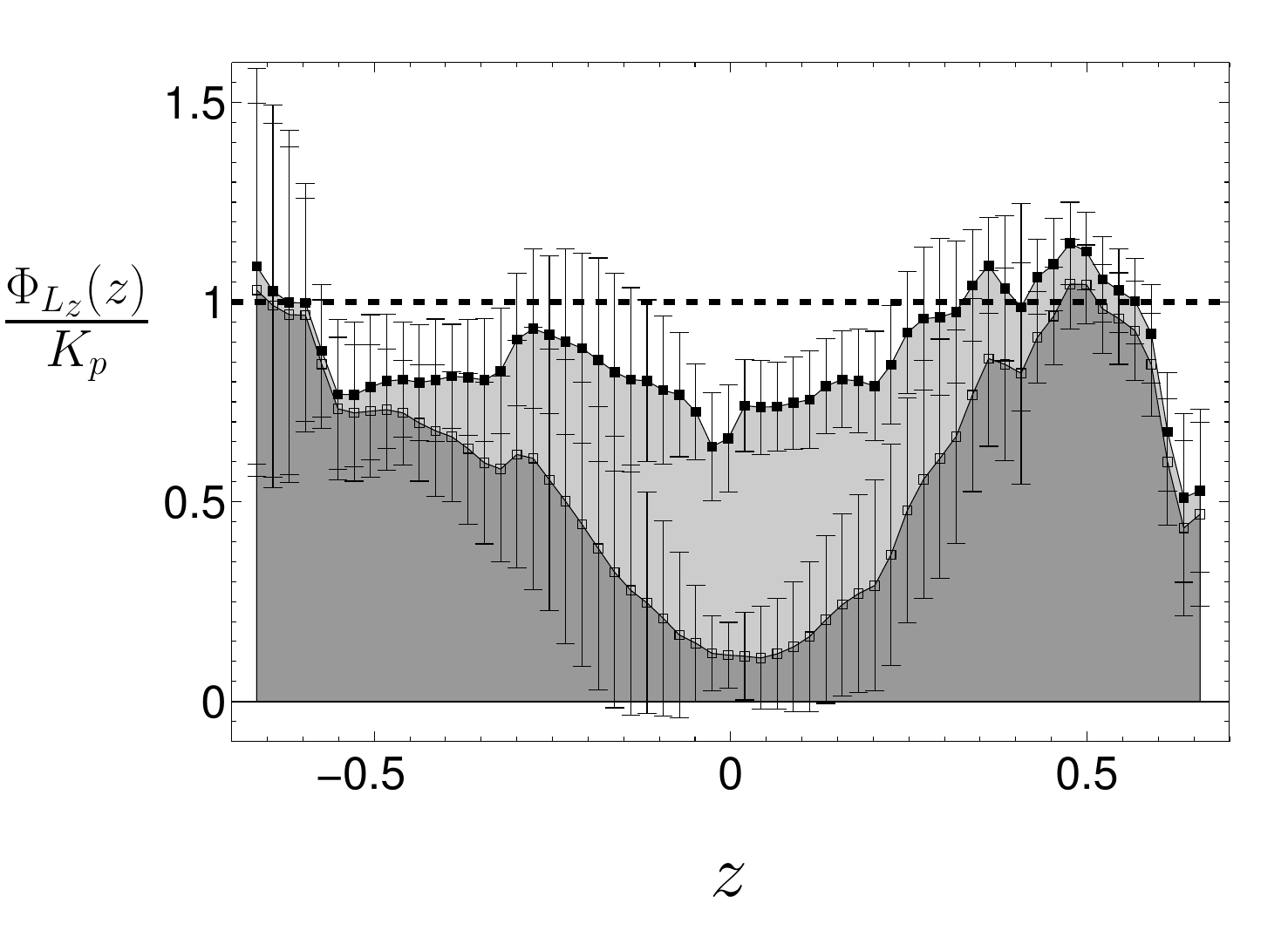}
\caption{Left: Control volume for the angular momentum budget. Right: Plot of the normalized vertical flux of vertical angular momentum as a function of $z$. The dark grey area represents convective transport due to the mean flow while the light grey area represents convective transport due to fluctuations. Filled squares represent the sum of the two contributions so that the total convective transport is almost constant with $z$ and equals to Kp. Each point has been obtained by taking the mean of several computations obtained from several flows at different $Re$. The error bars represent the standard deviation.}
\label{flux_moment}
\end{figure}

The first step is then to find the value of the C constant in order to be able to estimate the contribution of the terms containing $\tau^{ij}$ in (\ref{prod}) and (\ref{transfer}). For this, we follow the work of \cite{marie2004} who has shown, by using high-resolution Laser Doppler Velocimetry (LDV) measurements, that in a symmetric situation ($\theta=0$), the vertical flux of angular momentum is a constant equal to the torque injected by the impellers. This result stems from the  $z$-component of the angular momentum balance Eq. (\ref{Lconservation}) integrated over a volume $V(z)$ that describes a cylinder extending from the bottom impeller, to an altitude $h(z)=H/2+z$ (see Fig. \ref{flux_moment}). In this case, the global $z$-component of the angular momentum balance equation (\ref{Lbalance}) reads

\begin{eqnarray}
K_{p1}&=&-Re^{-1}\left(\int_{\mathcal{S}(z)} (\overline{u_\theta}+\overline{r\partial_r u_\theta}) d\mathcal{S}-\int_{\Sigma(z)} \overline{r\partial_z u_\theta} d\Sigma\right)+\int_{\mathcal{S}(z)} \overline{r(u_ru_\theta + \tau^{r\theta})} d\mathcal{S}+ \int_{\Sigma(z)} \overline{r(u_\theta u_z + \tau^{\theta z})} d\Sigma\nonumber\\
&\equiv& -K_v(z)+\int_{\Sigma(z)} \overline{r(u_\theta u_z + \tau^{\theta z})} d\Sigma,
\label{globalLbalanceVK}
\end{eqnarray}

where $\Sigma(z)$ is a surface at altitude $h(z)$ from the bottom impeller, and $S(z)$ is the lateral boundary (see Fig. \ref{flux_moment}). As discussed in \cite{marie2004}, this equation states that the angular momentum transmitted by the motor to the fluid (measured by the dimensionless torque at the bottom $K_{p1}$) is either evacuated through the lateral boundary, or transported to the upper layers of the fluid to be received by the upper motor as a drag. It was further shown in \cite{marie2004} that in a symmetric situation ($\theta=0$), where the torques at the bottom and at the top are equal, the $K_v(z)$ term is negligible, so that there is a constant flux of angular momentum from the bottom to the top. For each $\theta=0$ case, we thus compute the quantity

\begin{equation}
\Phi_{L_z}(z)=\int_{\Sigma_z} \overline{r(u_\theta u_z + \tau^{\theta z})} d\mathcal{S},
\label{defiPhi}
\end{equation}

 and adjust the constant $C$, so that $\Phi_{L_z}(z)=K_{p1}=K_{p2}\equiv K_p$ for any $z$. In Fig. \ref{flux_moment}, we present the results of our computations for TM60(+) impellers. The adjustment of the constant $C$ is done statistically, by assuming that it does not depend on the Reynolds number (as it is supposed to depend only on the velocity reconstruction algorithm). We then fix  $\Delta_r^2=\Delta_x\Delta_z$, where $\Delta_x=2R/58$ and $\Delta_z=(H-2h_b)/63$ being the spatial resolutions of our PIV and we adjust $C$  using  38 symmetrical flows at different Reynolds numbers so that their statistical mean provides a constant value of $\Phi(z)/K_p=1$ within the (statistical) standard deviation.

The optimal value of $C$ has to be taken equal to $C\approx 4$. This is  around 50 times more than the conventional choice made in \cite{leonard74,eyink2005} for a Gaussian filter. We have identified several factors that may explain this difference: our data are not filtered in a Gaussian way; since we ignored azimuthal derivatives, the constant must be higher to compensate; the turbulence is neither isotropic nor homogenous.

Dividing the flux into the contribution due to the time-average flow and the contribution due to the fluctuations, we also recover that near the impellers the flux is only due to the mean flow while at the center it is almost only due to fluctuations \cite{marie2004}. We also checked that the diffusive terms in (\ref{globalLbalanceVK}), going as $Re^{-1}$, are negligible at high Reynolds number. Likewise, the flux through $S(z)$ is also small, meaning that almost all of the flux goes in the vertical direction. \smallbreak

\subsection{Implementation of the Duchon-Robert formula}

\begin{figure}[htbp]
\centering
\includegraphics[width=0.49\textwidth]{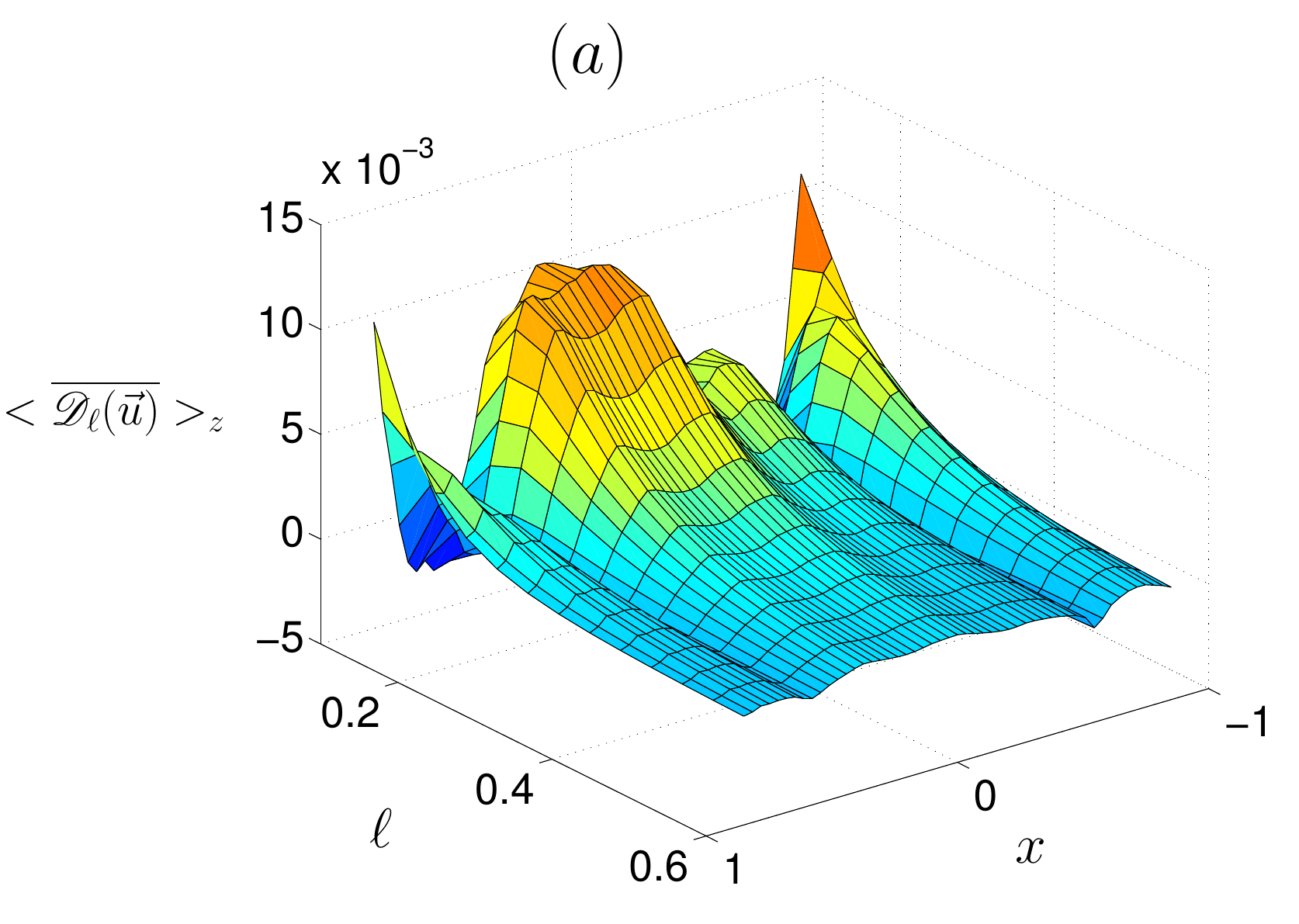}
\includegraphics[width=0.49\textwidth]{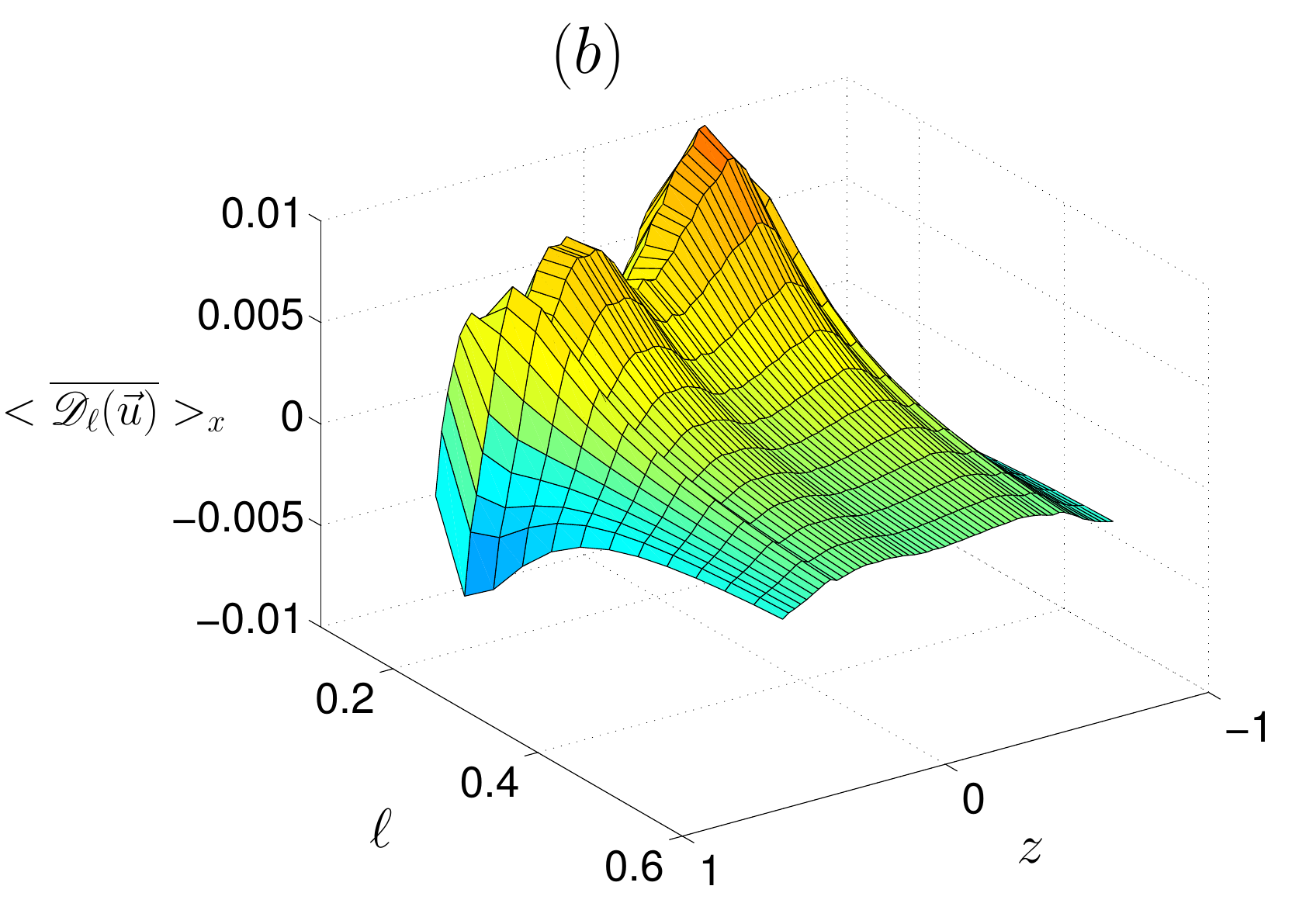}
\caption{Vertical (a) and radial (b) average of the local dissipation $\overline{\mathscr{D}_\ell(\vec u)}$ for TM60(+) at $Re=8\times 10^5$ as a function of $x$, $z$ and $\ell$.}
\label{checkDRaverage}
\end{figure}

\begin{figure}[!h]
\centering
\includegraphics[width=0.49\textwidth]{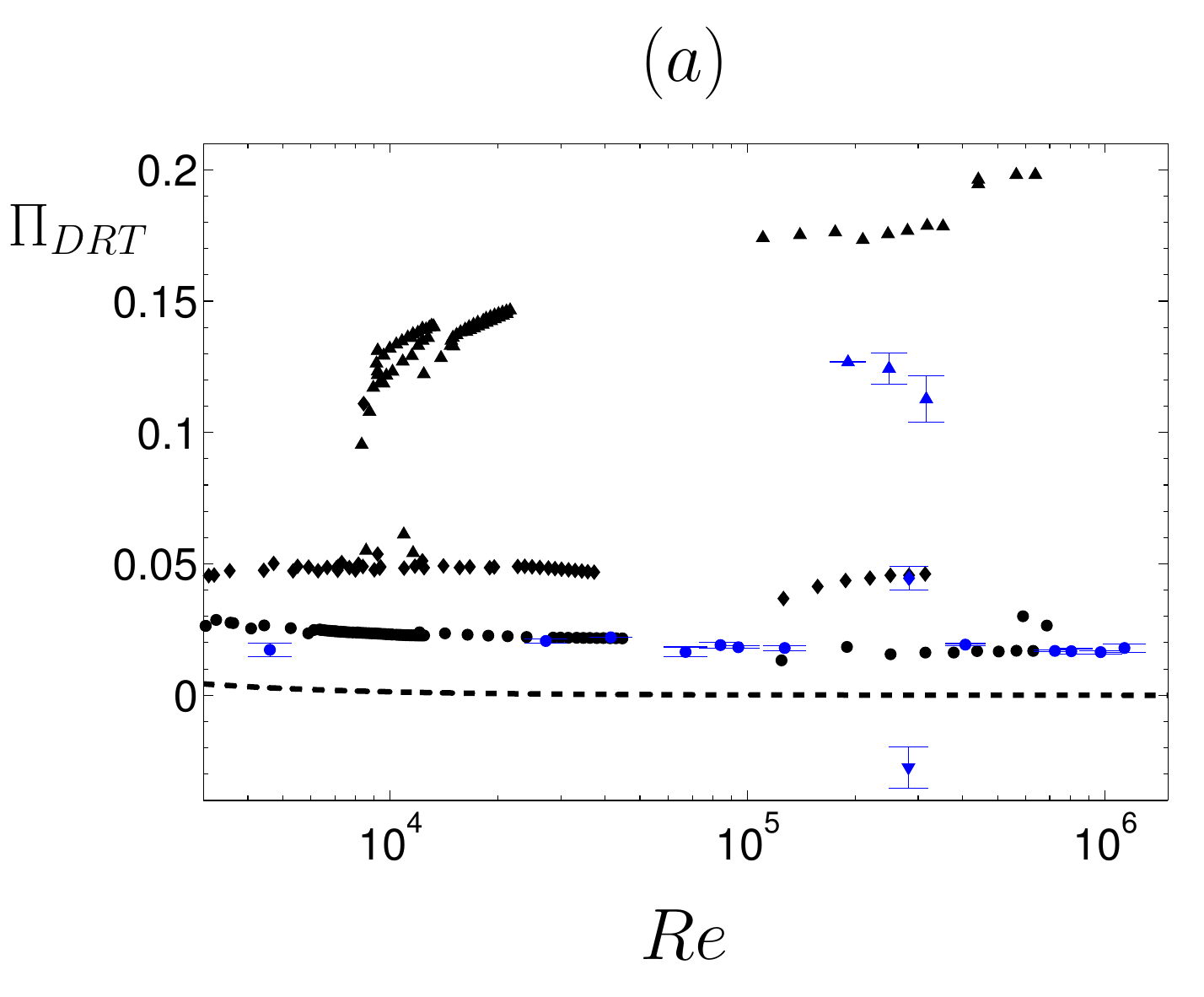}
\includegraphics[width=0.49\textwidth]{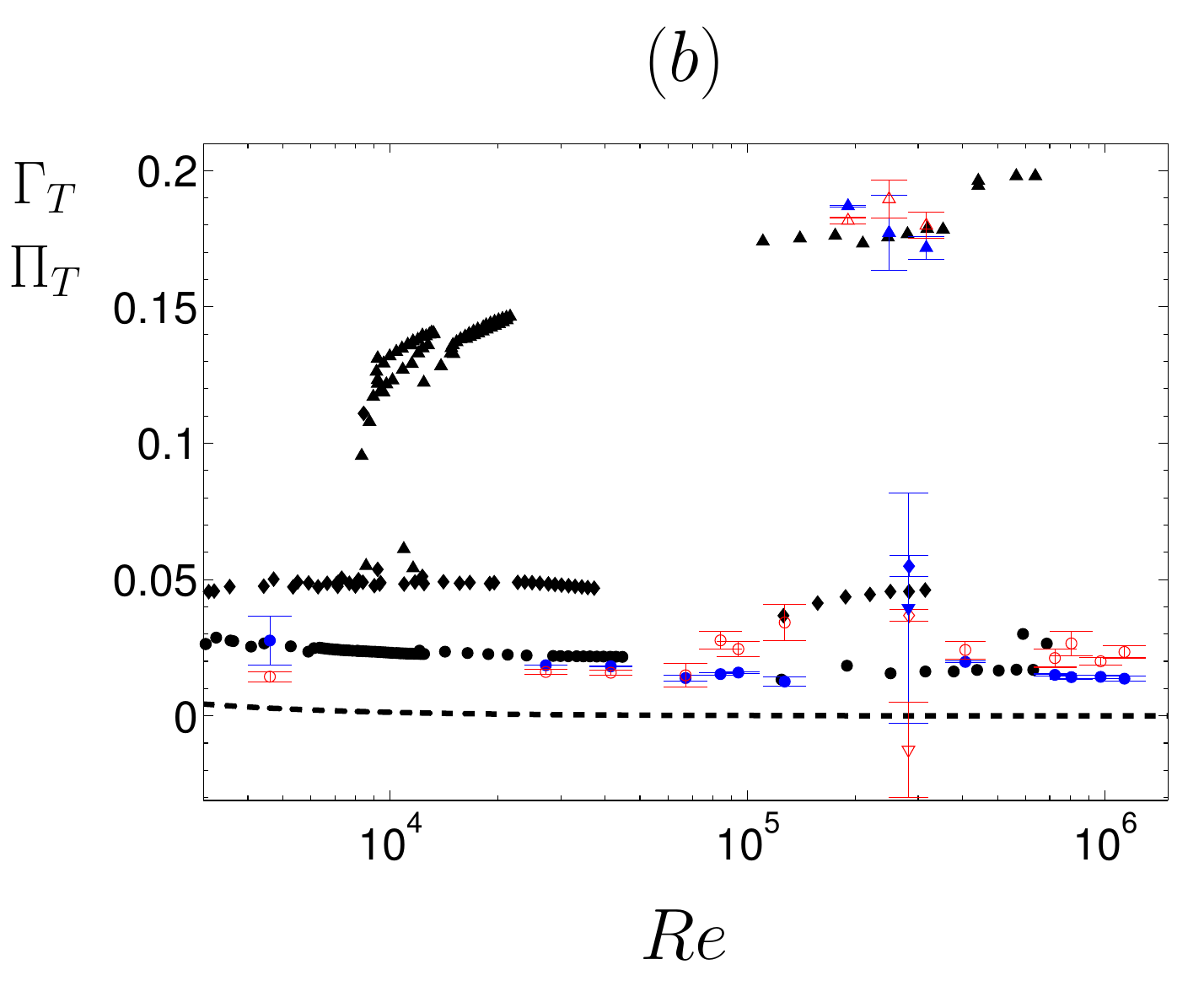}
\caption{Comparison between direct measurements of energy injection obtained using torque measurements (black symbols) and PIV estimates at various Reynolds numbers, for TM60  and TP87 impellers for the four different mean state geometries displayed in Fig. \ref{4types}: disks: (+) symmetric. Rhombi: (-) symmetric. Up triangles: (-) shear layer sent downwards. Down triangle: (-) shear layer sent upwards. a) energy dissipation $\Pi_{DRT}$ using the DR method (blue symbol). b) energy injection $\Gamma_T$ (red symbols) and dissipation $\Pi_T$ (blue symbols) using the LES-PIV method. The dotted line represents the "laminar fit"  $\epsilon=37Re^{-1}$.  The estimates are computed based on 2 to 15 realizations of the experiment where at least 600 instantaneous velocity snapshots have been taken for each of them. The symbols represent the mean of our computations while the error bars represent the standard deviation.}
\label{Dissi_log(Re)}
\end{figure}

 As we have seen in section \ref{DRsubsec}, $\mathscr{D}_\ell (\vec u)$ provides the local dissipation for a given velocity field $\vec u$, at a given length scale $\ell$. Since this formula is a generalization of the Kármán-Howarth equation to any kind of flow, it provides an estimate of the subgrid energy transfer as long as the considered scale $\ell$ is in the inertial range. Moreover, the considered scale must be sufficiently large with respect to the PIV smallest resolved scale, so as to guarantee statistical convergence through sufficient average in the scale space (on the sphere of radius $\ell$). To check these two points, we present on Fig. \ref{checkDRaverage}  $\mathscr{D}_\ell (\vec u)$ in two plots, averaged in the radial (resp. vertical)  direction, as a function of $z$ (resp. $x$) and $\ell$, for an experiment using TM60(+) impellers, at very large Reynolds number ($Re\approx8\times 10^5$).  We see that this quantity is close to zero at large scales ($\ell>0.4$), but that it increases at small scales in the domain $\vert x\vert <0.4, \vert z\vert <0.7$, i.e. at the location of the median shear layer. For $\ell$ between $0.1$ and $0.15$ (i.e approximately 4 to 5 times the smallest scale resolved by our PIV set-up), there is the start of a saturation, indicating the beginning of the inertial range. While the extent of the inertial range is likely to vary (increase) with the Reynolds number, its largest scale is likely to be independent of the Reynolds number, as long as the flow is turbulent. Indeed, as discussed in \cite{simon2015}, the geometry of the largest scales of the von Kármán flow appear fairly independent of the Reynolds number, except around $Re=10^5$ where they may experiment abrupt changes due to the equivalent of a phase transition \cite{cortet2010}. Since in the Kolomogorov picture energy cascades from large to small scales, it is reasonable to assume that the beginning of the inertial range is solely determined by the large scale topology, thereby becoming independent of the Reynolds number (except maybe around $Re=10^5$). To check this, we report on Fig. \ref{Dissi_log(Re)}(a)  the comparison between the total dissipated power $\Pi_{DRT}$ at $\ell=0.1$ (blue symbols) and direct measurements of the injected power (black symbols).  We observe that our estimates are in good  agreement with direct measurements, especially for symmetric flows. In Fig. \ref{Dissi_log(Re)}(b), the results of our computations to estimate the global dissipated power using the LES method are displayed. These results will be discussed in more detail later, but we can already observe that we obtain a good agreement with direct measurements. We can conclude from all our computations that the method using the DR formula may be seen as an interesting alternative to the widespread LES-PIV method, since it relies on very few arbitrary hypotheses. We explore its performances with other flow geometries in the sequel.

\section{Results}
\label{results}

\subsection{Symmetric case $\theta=0$}

\paragraph{Energy production and transport}

To understand the flux of energy in the von Kármán flow, it is interesting to focus first on the symmetric case ($\theta=0$), specifically  on the case when both impellers rotate in the (+) sense with the same frequency, in a stationary regime. In such regime, the dissipated power equals the injected power. Through Green-Ostrogradski theorem, the total energy production $\Gamma_T$ is equal to the entering flux of $J^i$ at the boundaries. In this symmetric case, most of the flux is provided by the component $J^z$, and appears to provide a fairly good estimate of the injected power at large enough Reynolds numbers ($Re>=10^5$). Indeed, we show on Fig. \ref{Dissi_log(Re)}(b) the results of our computations for $\Gamma_T$ (empty red symbols). We see that for $Re>=10^5$, these estimates coincide within $20\%$. However, at lower Reynolds numbers ($Re\approx 4\times10^3$), the PIV estimates only capture $55\%$ of the actual injected power. This may be due to the fact that as the Reynolds number is decreased, an increasing part of the injected power is either through pressure effects, azimuthal variations or viscous boundary layers that are not resolved by our measurements.\

\begin{figure}[!h]
\centering
\includegraphics[width=0.30\textwidth]{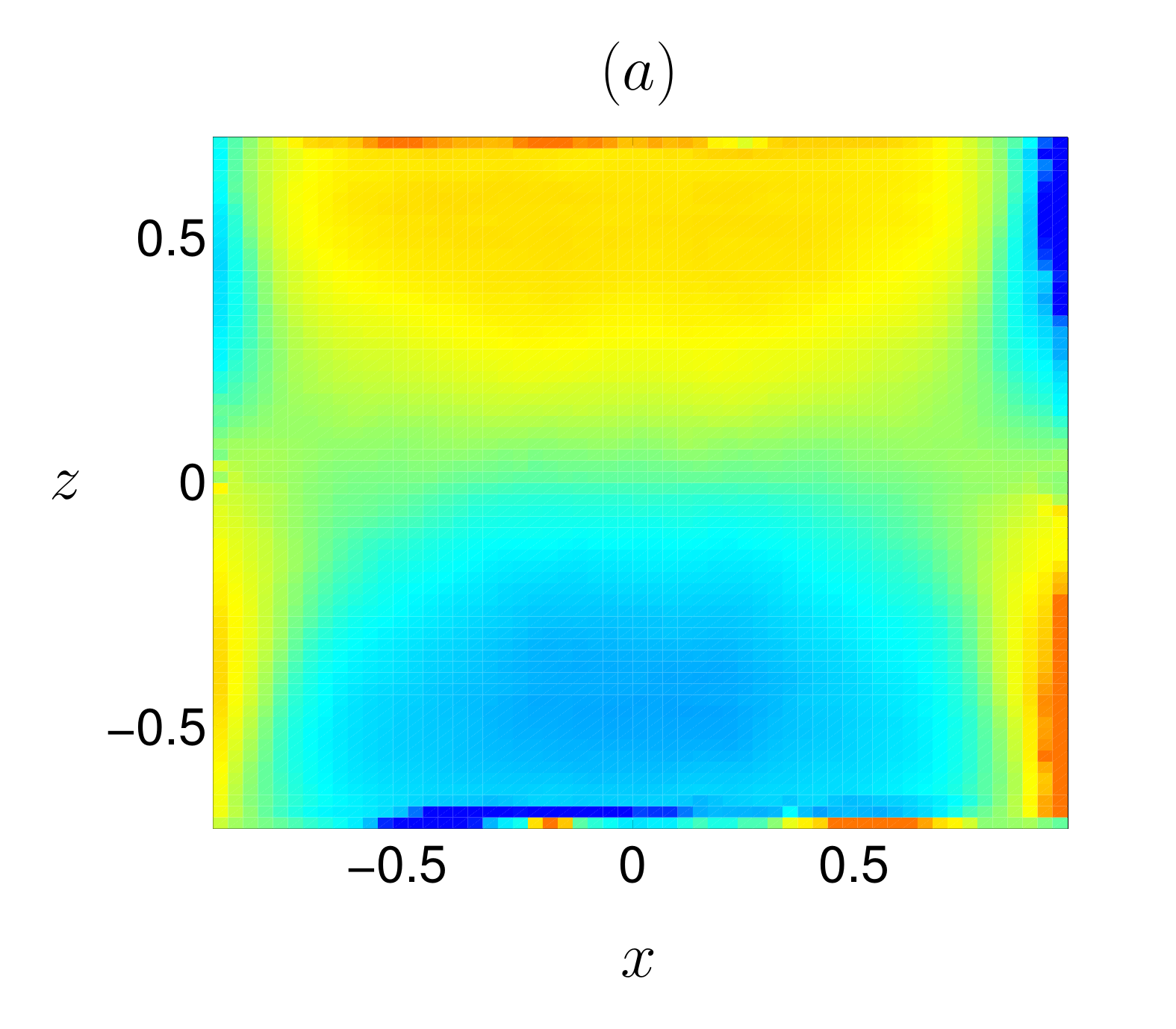}
\includegraphics[width=0.30\textwidth]{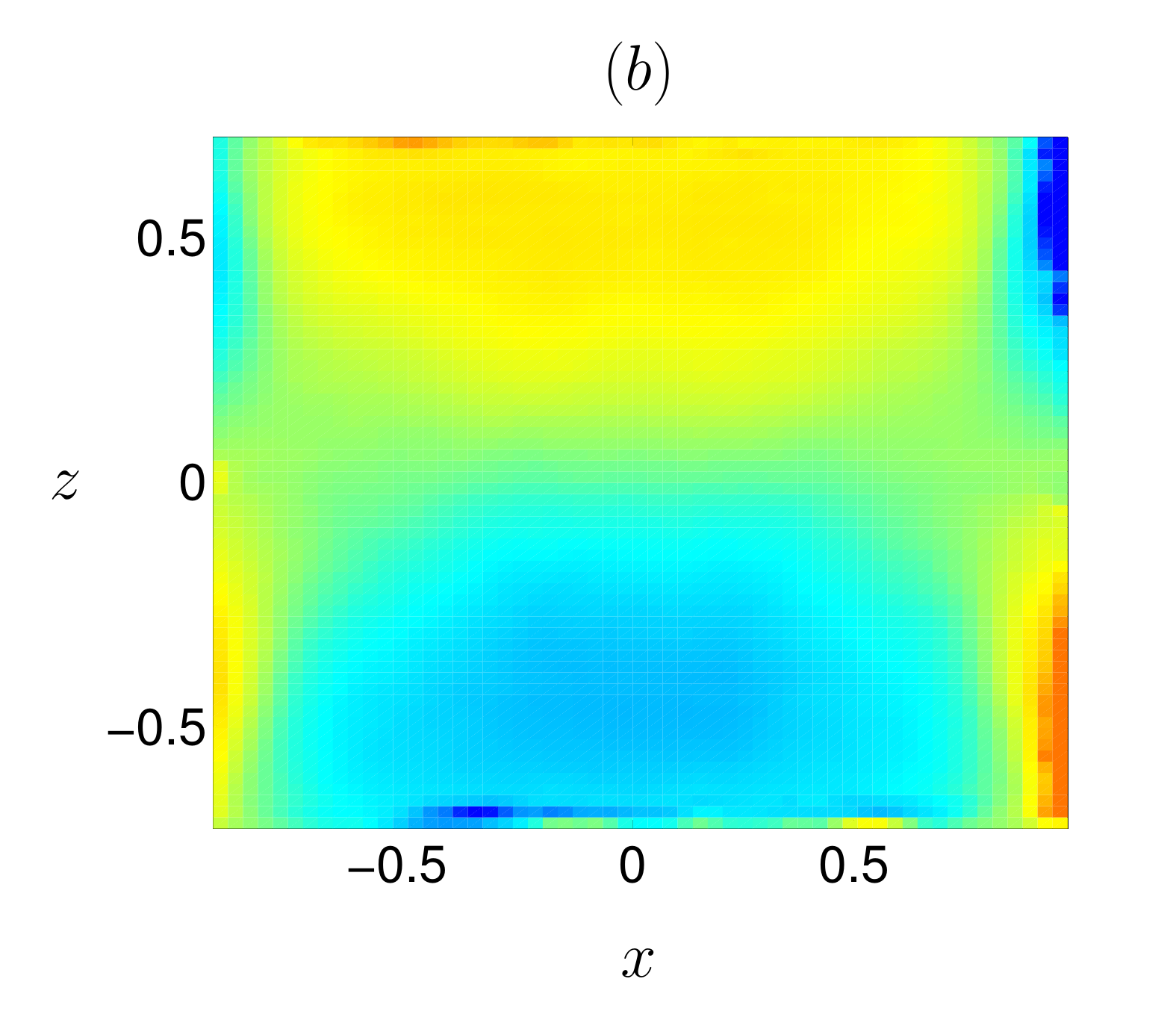}
\includegraphics[width=0.30\textwidth]{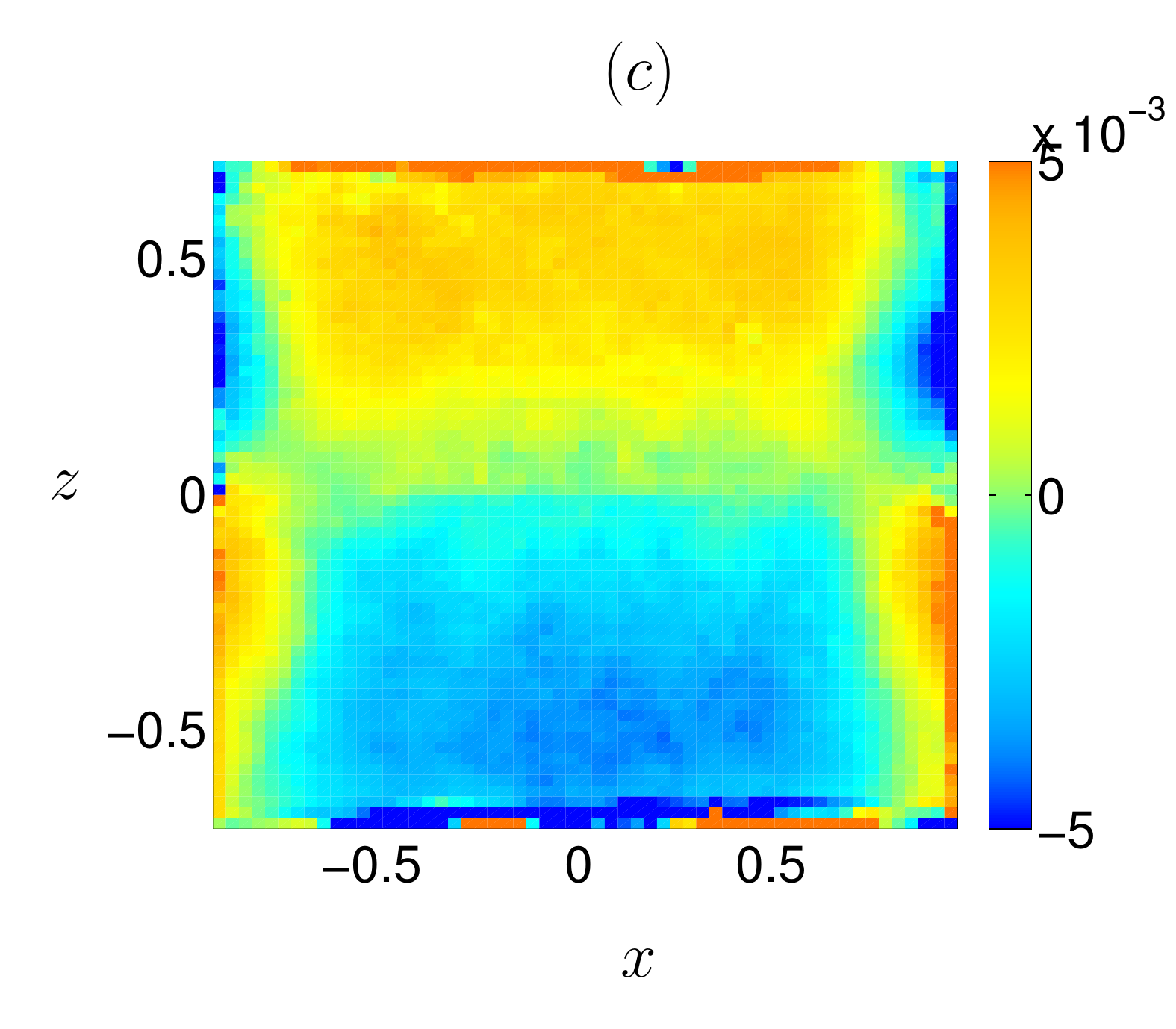}
\includegraphics[width=0.99\textwidth]{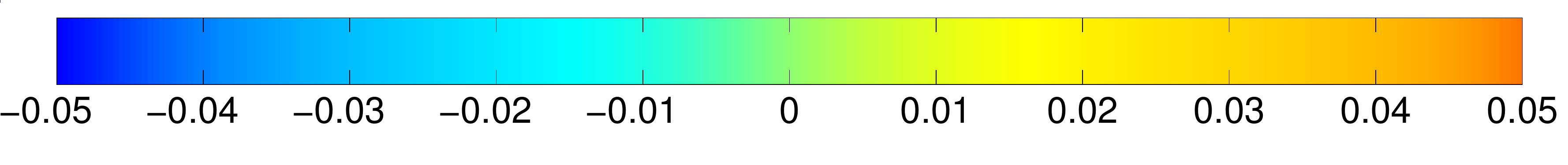}
\caption{Typical maps of $\overline{J^z}$: a) mean energy transfer through the system, b) contribution of the mean flow, c) contribution of the fluctuations}
\label{Jz}
\end{figure}

Given that the $J^z$ contribution dominates the total energy production, it is interesting to focus on the spatial variation of this quantity. Fig. \ref{Jz}(a) shows such a local map of $\overline{J^z}$ for a symmetric flow at $Re\approx 3\times 10^4$.

\begin{figure}[!h]
\centering
\includegraphics[width=15cm]{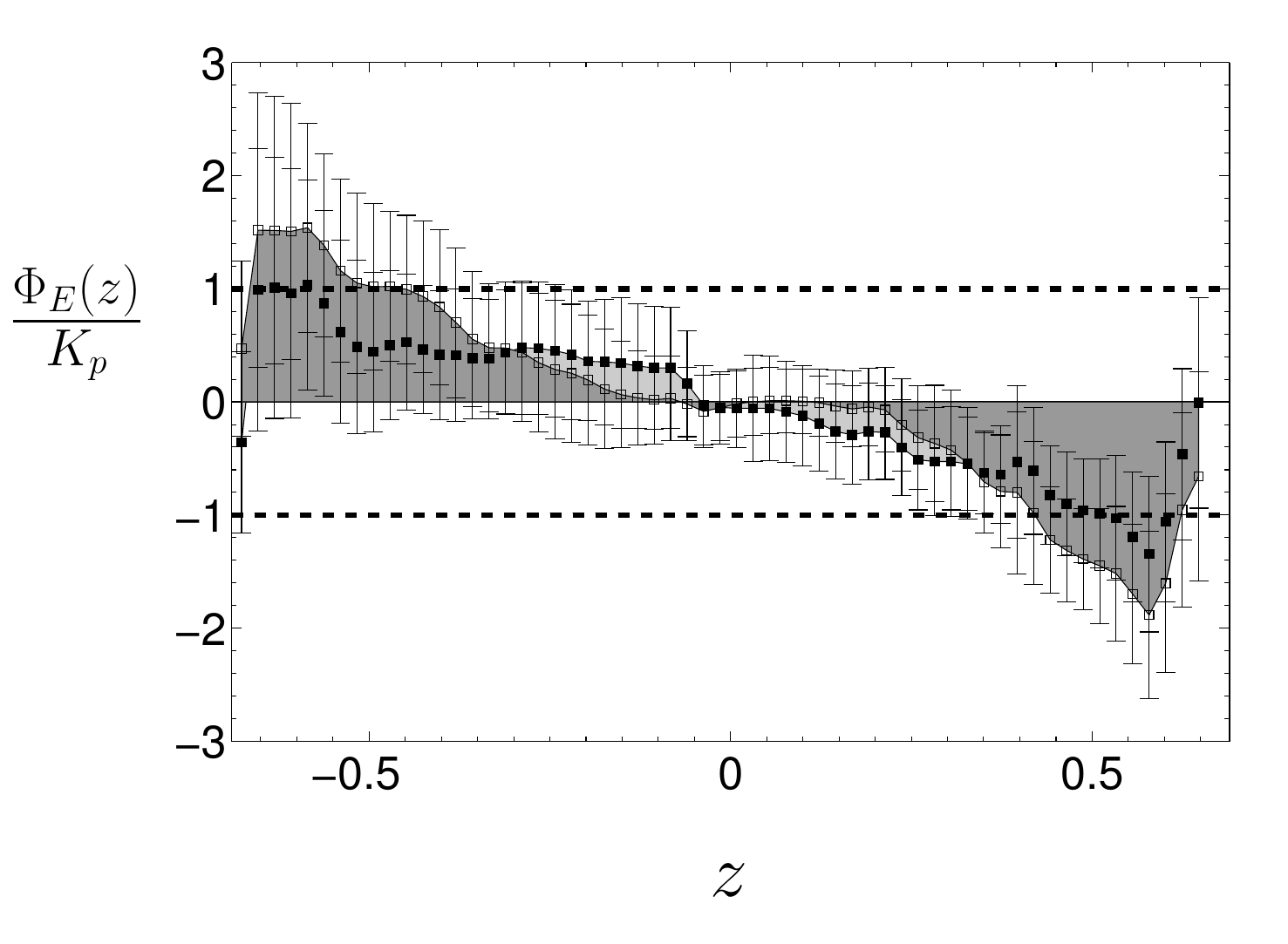}
\caption{Plot of the normalized vertical flux of energy as a function of $z$. The dark grey area represents convective transport due to the mean flow while the light grey area represents convective transport due to fluctuations. Filled squares represent the sum of the two contributions so that the total convective transport is a decreasing function of $z$. Each point has been obtained by taking the mean of several computations obtained from several flows, and the error bars represent the statistical standard deviation.}
\label{flux_energie}
\end{figure}

As can be seen, at the center of the cylinder, i.e. $x/R\in[-0.8 \ 0.8]$, big structures reflect the advection of energy towards the impellers, through the mean Ekman pumping by the impellers. In contrast, at the walls, smaller structures of opposite sign are observed, reflecting the injection of energy within the flow. These two kinds of structures are mirrored by the recirculation cells that we observed on Fig. \ref{4types}. The local structure  of $\overline{J^z}$ can be further used to get information about energy transport in the flow. Indeed, performing an integration of equation (\ref{Econservation}) over the height-varying volume $V(z)$ (see Fig. \ref{flux_moment}) and setting $\partial_t \overline{E}=0$, we can get an equation for the energy transport in the flow as

 \begin{equation}
\oint_{\mathcal{S}} \overline{J^i} d\mathcal{S}_i =-K_{p1}(\theta+1)-K'_v(z)+\int_{\mathcal{S}_z} \overline{J^z} d\mathcal{S},
\label{globalEbalanceVK}
\end{equation}

where $K'_v$ is the contribution due to the lateral boundaries at height $z$. Ignoring this contribution, we can write

 \begin{equation}
\oint_{\mathcal{S}} \overline{J^i} d\mathcal{S}_i =-K_{p1}+\Phi_{E}(z),
\label{enerflux}
\end{equation}

where $\Phi_{E}(z)=\int_{\mathcal{S}_z} \overline{J^z} d\mathcal{S}$. This quantity is displayed on Fig. \ref{flux_energie} .

We observe that near the impellers, the total flux of energy (filled black squares) equals the energy injected by the impellers. The change of sign comes from the fact that the impeller at the bottom injects energy in the $+z$ direction whereas the upper impeller injects energy in the $-z$ direction. At the center of the experiment the flux is zero, meaning that on average there is not any energy going from one half of the cylinder to the other through the shear layer. Finally, it is interesting to see that, as before, the mean flow plays an important role in the transport of energy near the impellers, whereas near the shear layer energy is carried by fluctuations. A slight difference is that near the impellers, as the mean flow sends energy towards the center of the cylinder, fluctuations create a flux which goes in the opposite direction and tries to send energy back to the impellers.\

\begin{figure}[!h]
\begin{minipage}{0.99\textwidth}
\centering
\includegraphics[width=0.49\textwidth]{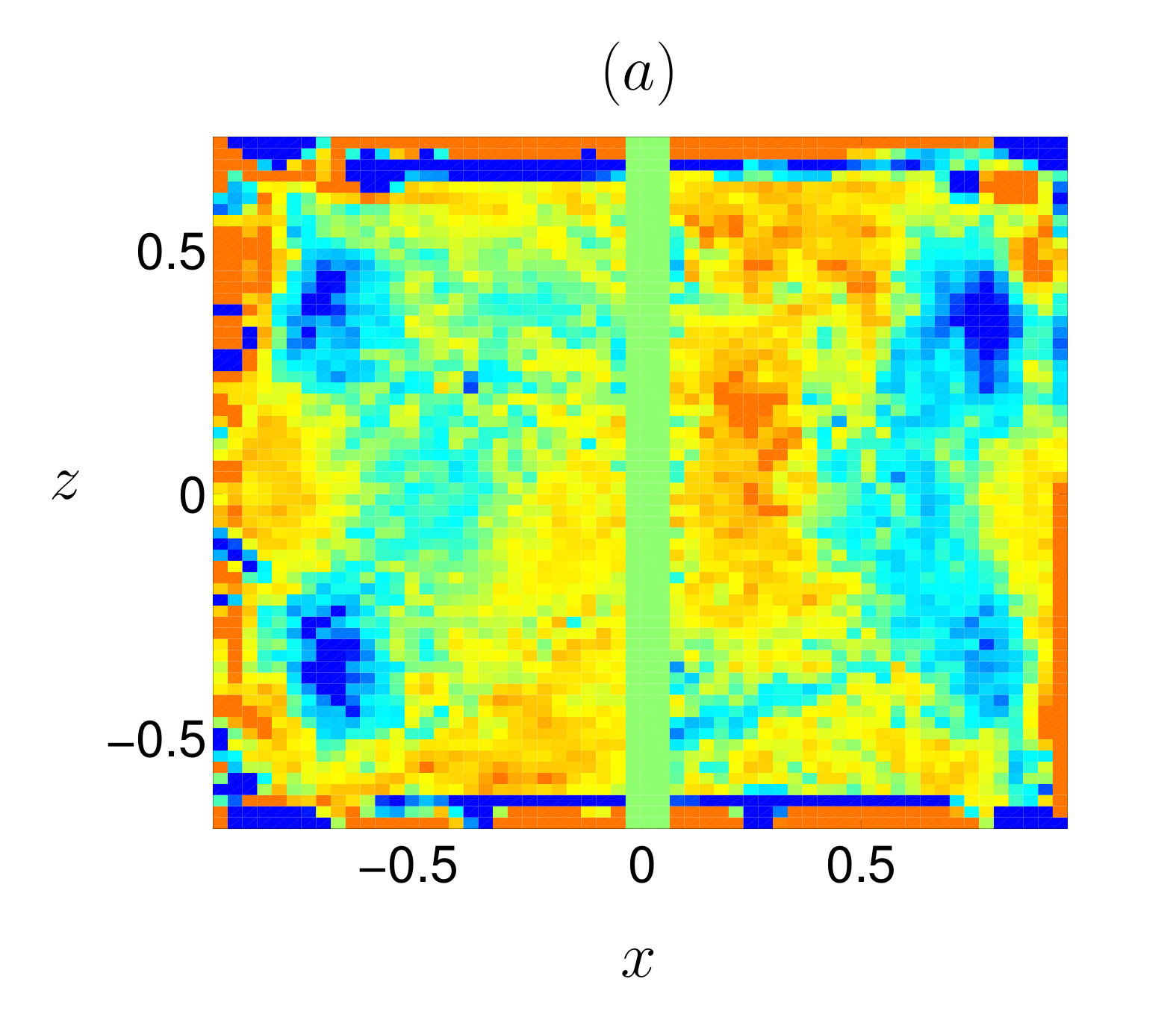}
\includegraphics[width=0.49\textwidth]{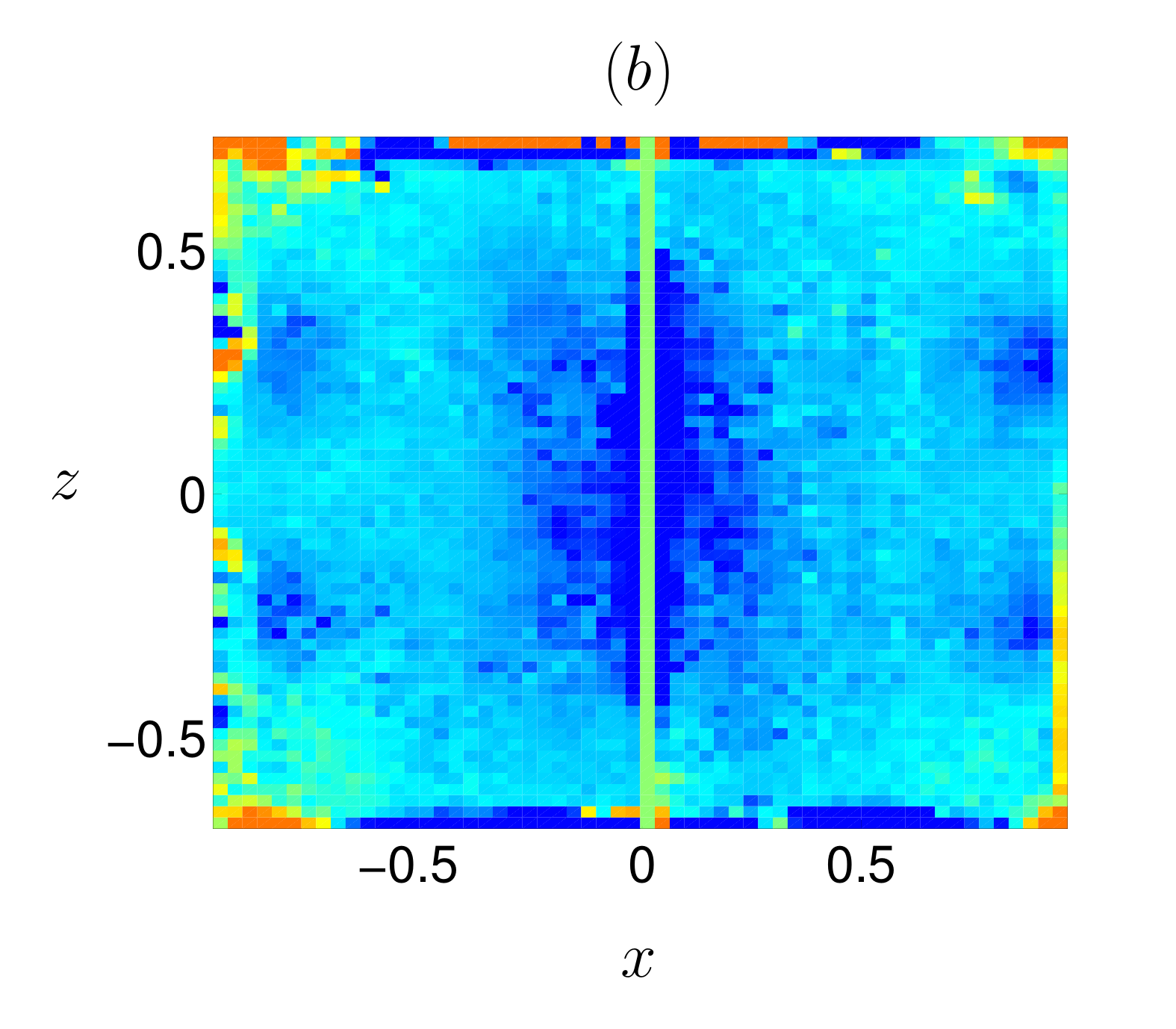}
\includegraphics[width=0.99\textwidth]{colorbar2}
\includegraphics[width=0.49\textwidth]{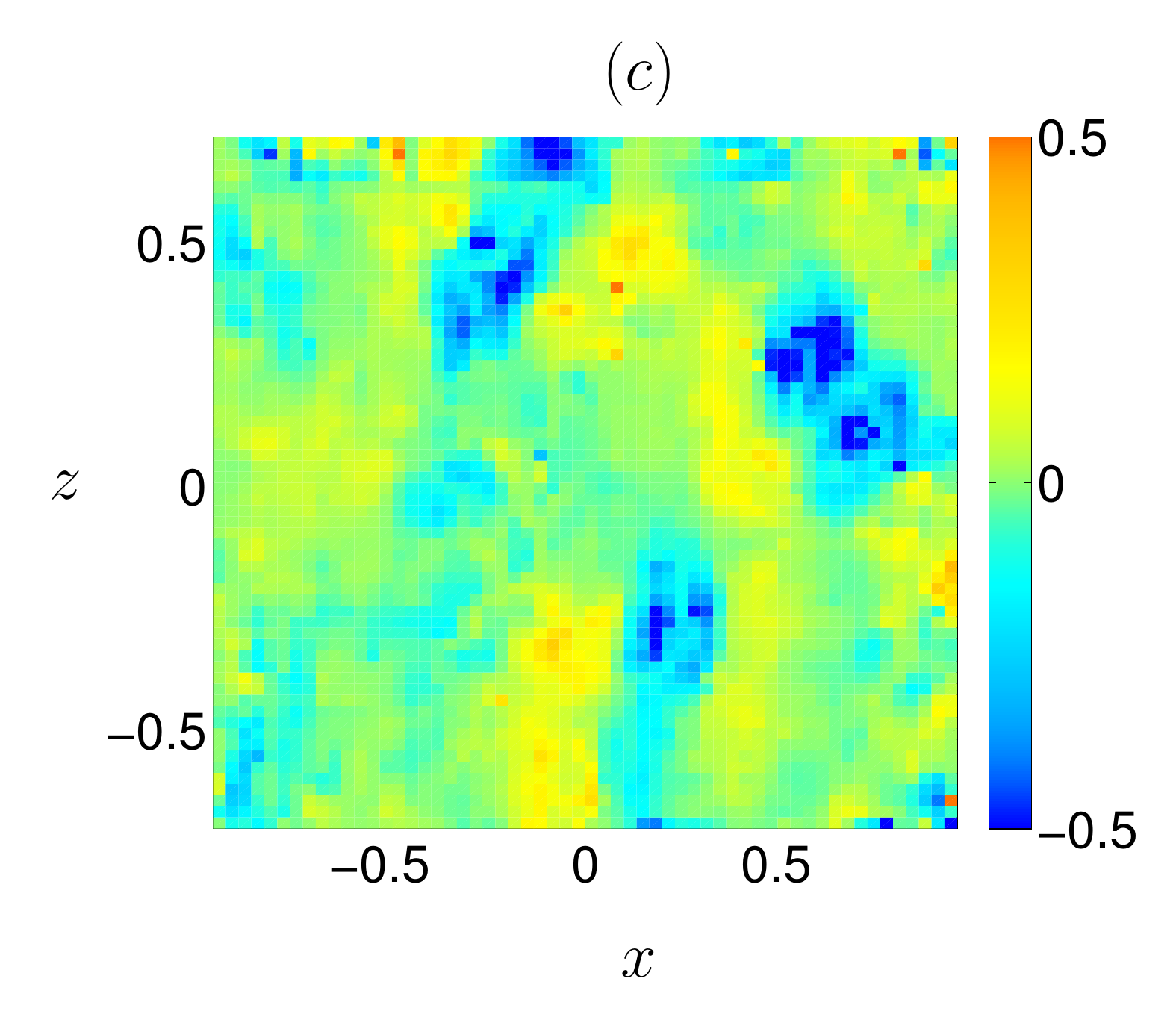}
\includegraphics[width=0.49\textwidth]{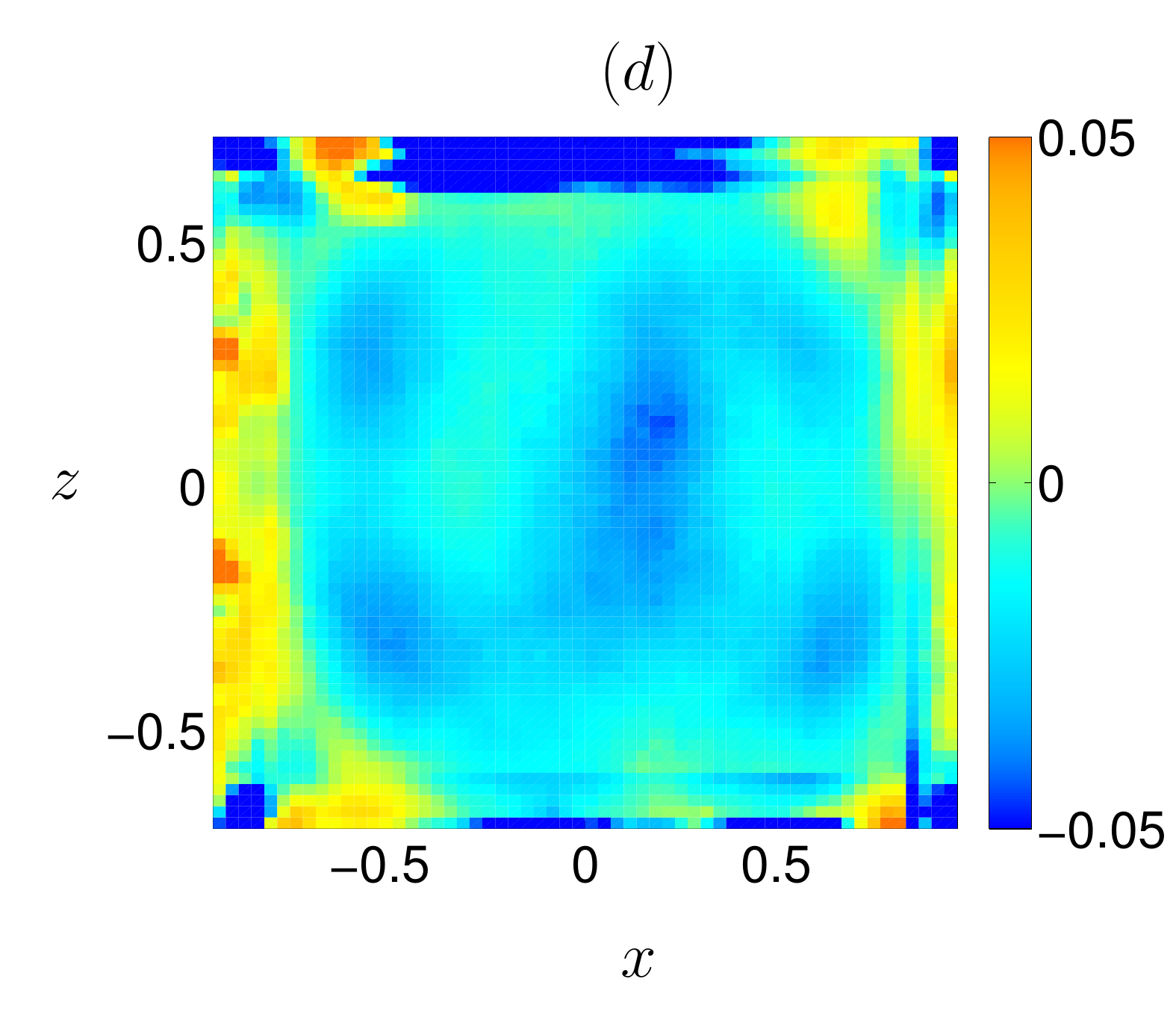}
\end{minipage}
\caption{Top: Maps of the energy production rate (a) and energy transfer rate (b) for the same flow as in Fig. \ref{4types}(a), using LES-PIV estimates. Bottom: Instantaneous (c)  and time averaged (d) maps of the local dissipation $\mathscr{D}_\ell(\vec u)$ at $\ell=0.1$ for the same flow. Areas where energy accumulates are represented in red, while those where energy leaves are represented in blue.}
\label{Mapscon}
\end{figure}

Altogether, the local map of injected power is shown on  Fig. \ref{Mapscon}(a).
We see that near the impellers, the divergence term brings energy into the system as is expected. We also see that energy leaves the center of the recirculation cells and tends to be advected towards the center of the impeller through Ekman pumping.

\paragraph{Energy dissipation}

We now turn to  the total energy dissipation estimated through the LES method $\Pi_T$. 

Its values at various Reynolds numbers in the symmetric case are reported in Fig.  \ref{Dissi_log(Re)}(b). We observe that it is in very good agreement with direct measurements at $Re\approx 4\times 10^3$. At such Reynolds number, the dissipative scale is of the order of $1mm$, close to the PIV resolution. This result is therefore in agreement with the observation of Tokgoz et al. \cite{tokgoz2012} obtained for a Taylor-Couette flow. At larger Reynolds numbers, the estimates using the LES method are in good agreement with respect to direct measurements since we are able to capture up to $90\%$ of the actual energy dissipation. The map of the local dissipation with this method is provided in Fig. \ref{Mapscon}(b), and appears fairly uniform across the vessel, with larger intensity along the vertical axis around $r=0$ and near the lateral boundaries around $z=\pm 0.3$.
 
This map can be compared with the one obtained using the DR formula $\overline{\mathscr{D}_\ell (\vec u)}$ at $\ell=0.1$ Fig. \ref{Mapscon}(d). One observes the same localized structures of energy dissipation near the walls at $x=\pm 0.6$, which are symmetrically distributed with respect to the mid-plane. These structures are statistically significant, since they are not observed on plots of the instantaneous local dissipation $\mathscr{D}_\ell (\vec u)$ (see Fig.  \ref{Mapscon}(c)). They may, therefore, trace the intense vortices of the shear layer. In addition, one observes a clear localization of energy injection (red areas) at the tip of the impellers, with an energy dissipation in the middle part of the cells. Regarding instantaneous maps, it is interesting to note that it is characterized by intense, localized events, that can reach 10 to 20 times the maximum average local dissipation. Whether these intense events are connected to localized quasi singular structures is an interesting open question that we leave for future work.\
 
Overall, these dissipation maps provide clear evidence that the maximum energy dissipation lies within the shear layer and that the DR formula provides a better estimate of the energy dissipation than the LES method in the symmetric case.

\paragraph{Description of the energy cycle}

\begin{figure}[!h]
\centering
\includegraphics[width=0.49\textwidth]{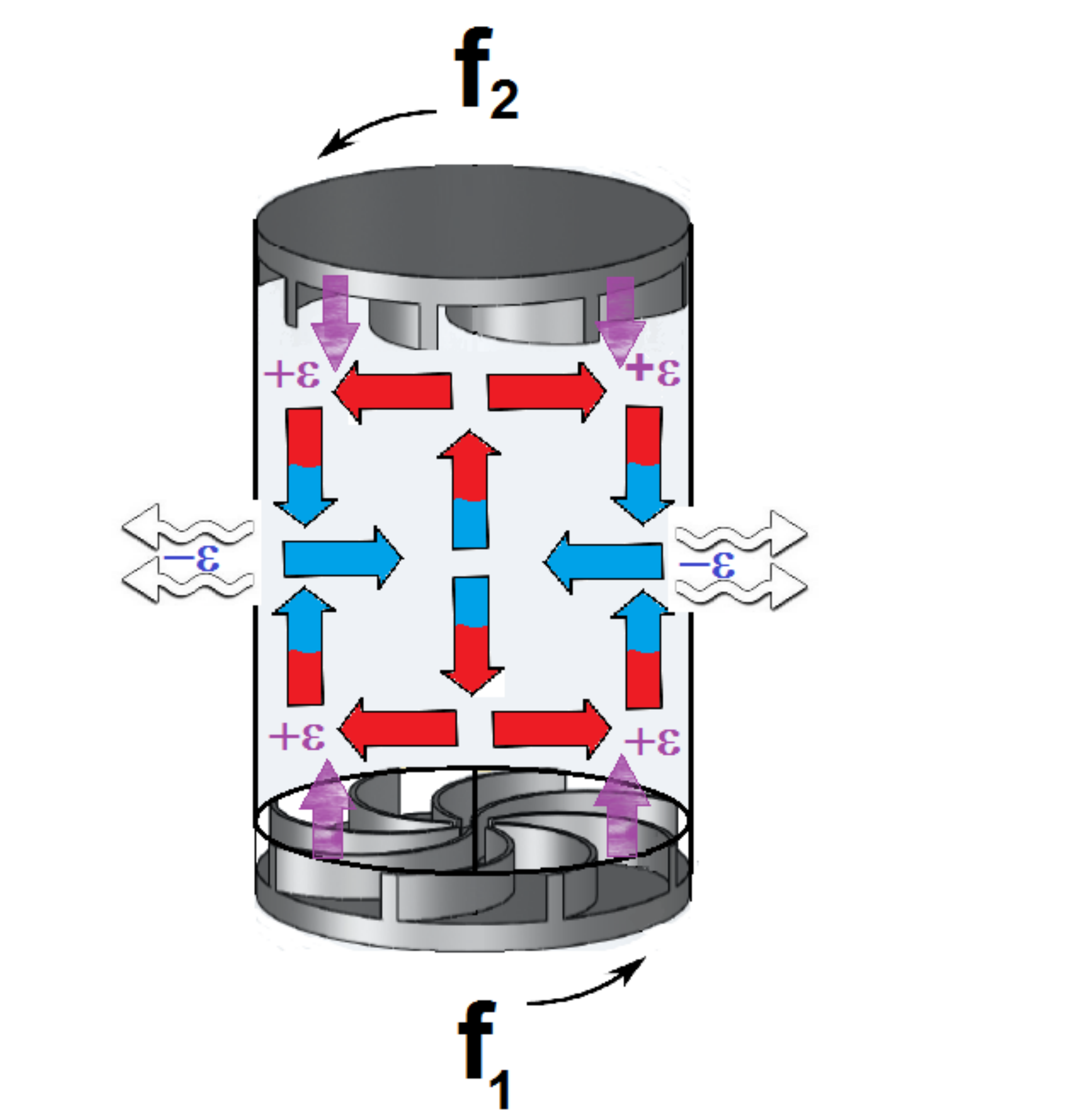}
\caption{Energy cycle in the von Kármán experiment: the energy is advected to the impellers via the Ekman pumping, the flow is then accelerated inside the impellers and expelled at the walls, providing an energy injection $\epsilon$ towards the mixing layer. In that mixing layer, turbulent fluctuations dissipate an energy equal to $\epsilon$. The fluid is then pumped again into the impellers for further reinjection, closing the energy cycle and providing a stationary situation with energy dissipation $\epsilon$.}
\label{EnergyCycle}
\end{figure}

Altogether, our results regarding energy production, transport and dissipation can be summarized into a simple picture of the "energy cycle" for the von Kármán flow, which is sketched in Fig. \ref{EnergyCycle}: the energy is advected to the impellers via the Ekman pumping, the flow is then accelerated inside the impellers and expelled at the walls, providing an energy injection towards the mixing layer. In that mixing layer, turbulent fluctuations dissipate an energy equal to $\epsilon$. The fluid is then pumped again into the impellers for further reinjection, closing the energy cycle and providing a stationary situation with energy injection and dissipation equal to $\epsilon$. 

In the sequel, we study the influence the flow topology and the  forcing conditions onto this energy cycle.  Since in this picture most of the energy is dissipated within the middle shear layer, there is good hope that we can capture its main contribution by the PIV measurements, \emph{provided that the shear layer is not too close to the impeller} since in that case, the PIV measurements cannot resolve the flow. We check in section V B-C that this is indeed the case, and provide detailed informations about the localization of energy dissipation to complete the picture of the energy cycle.

\subsection{Influence of the flow topology}

Let us now consider the case when both impellers rotate in the (-) sense at $\theta=0$.  For this type of forcing, at sufficiently high Reynolds numbers, there is coexistence of three different flow geometries for TP87 type impellers.

\begin{figure}[!h]
\begin{minipage}{0.99\textwidth}
\centering
\includegraphics[width=0.43\textwidth]{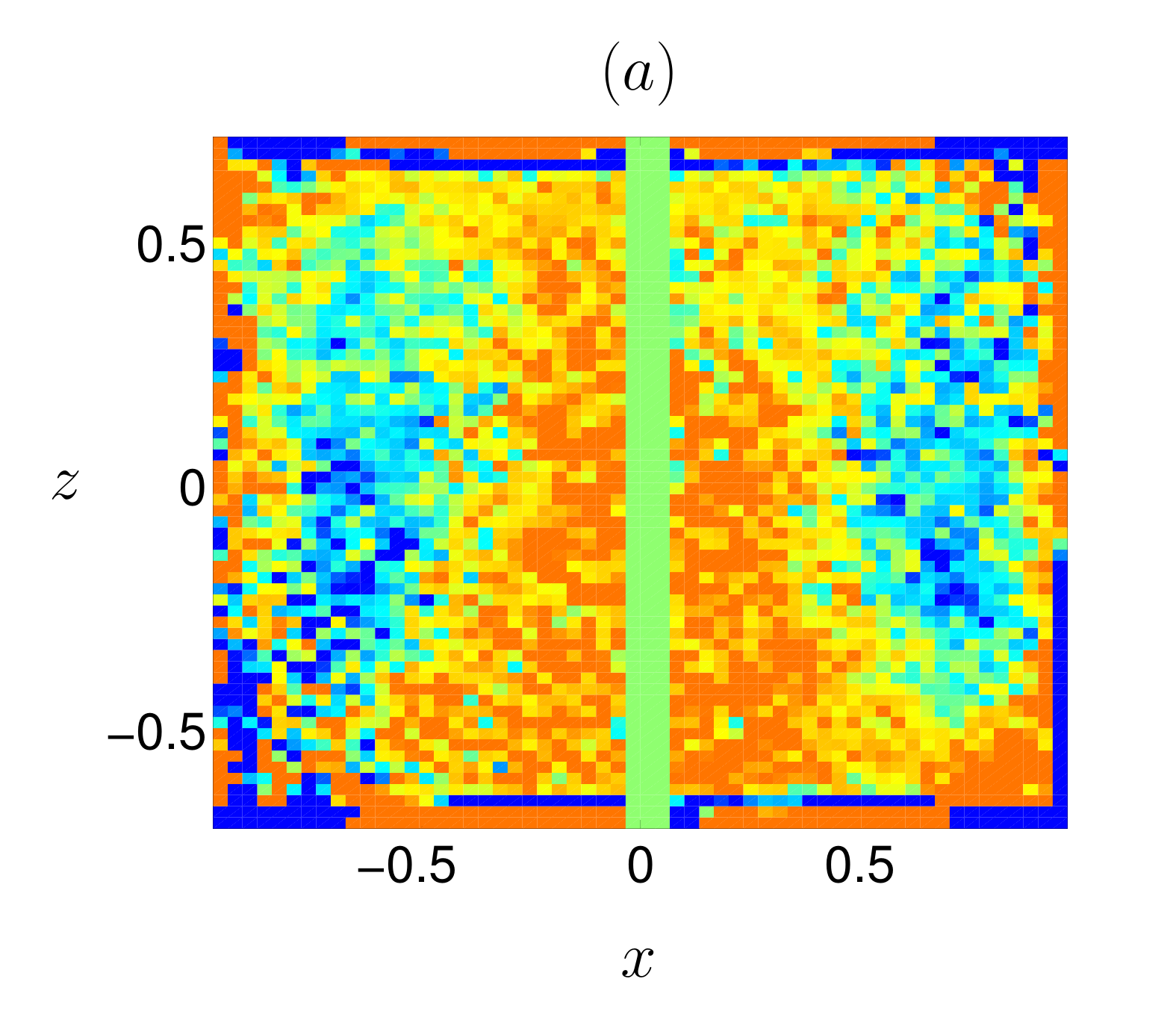}
\includegraphics[width=0.43\textwidth]{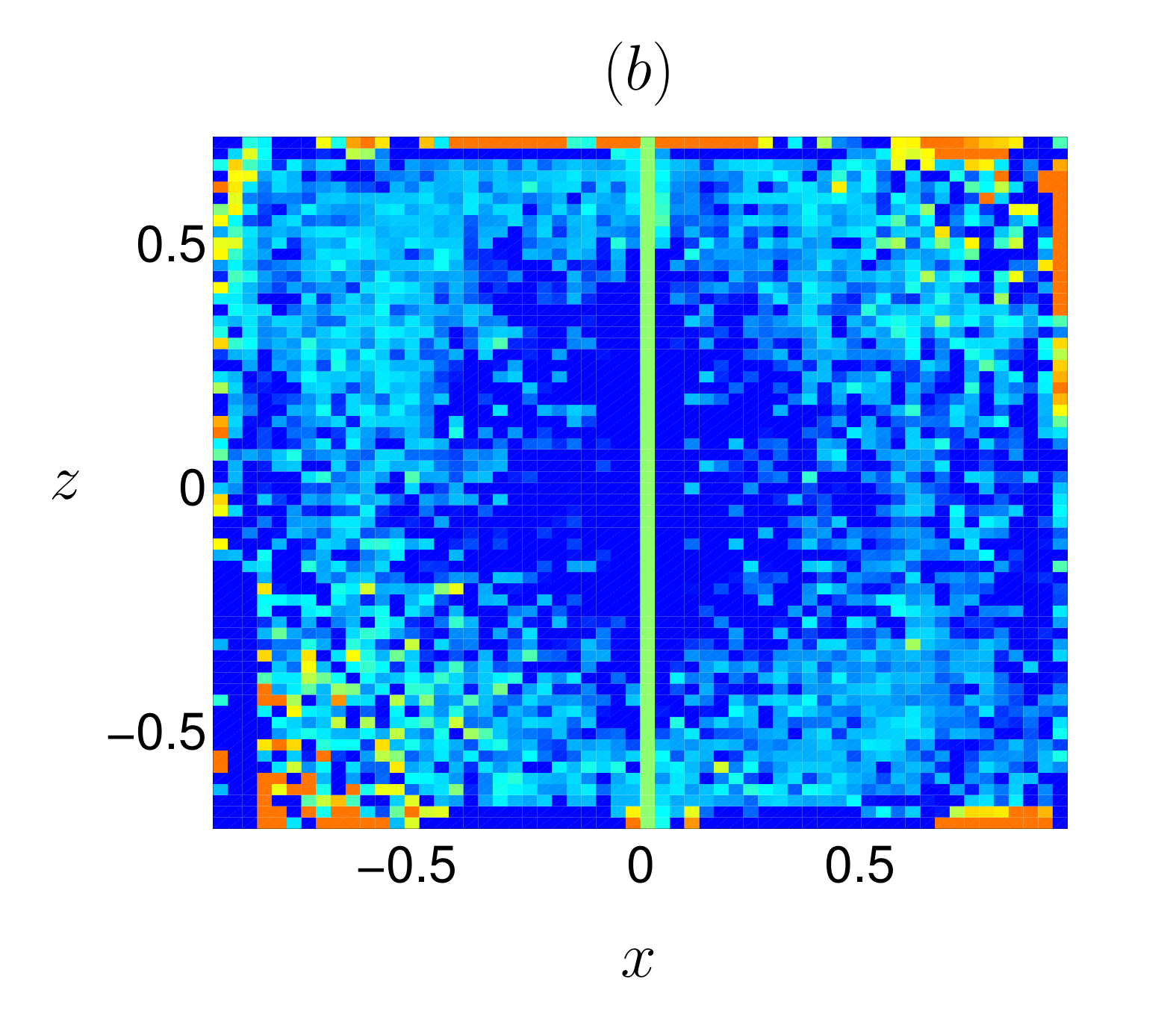}
\includegraphics[width=0.91\textwidth]{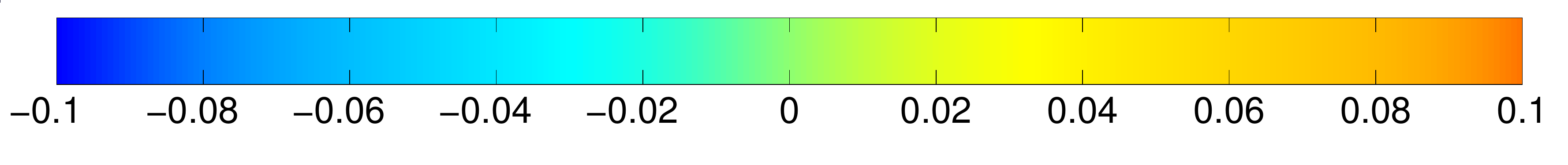}
\includegraphics[width=0.43\textwidth]{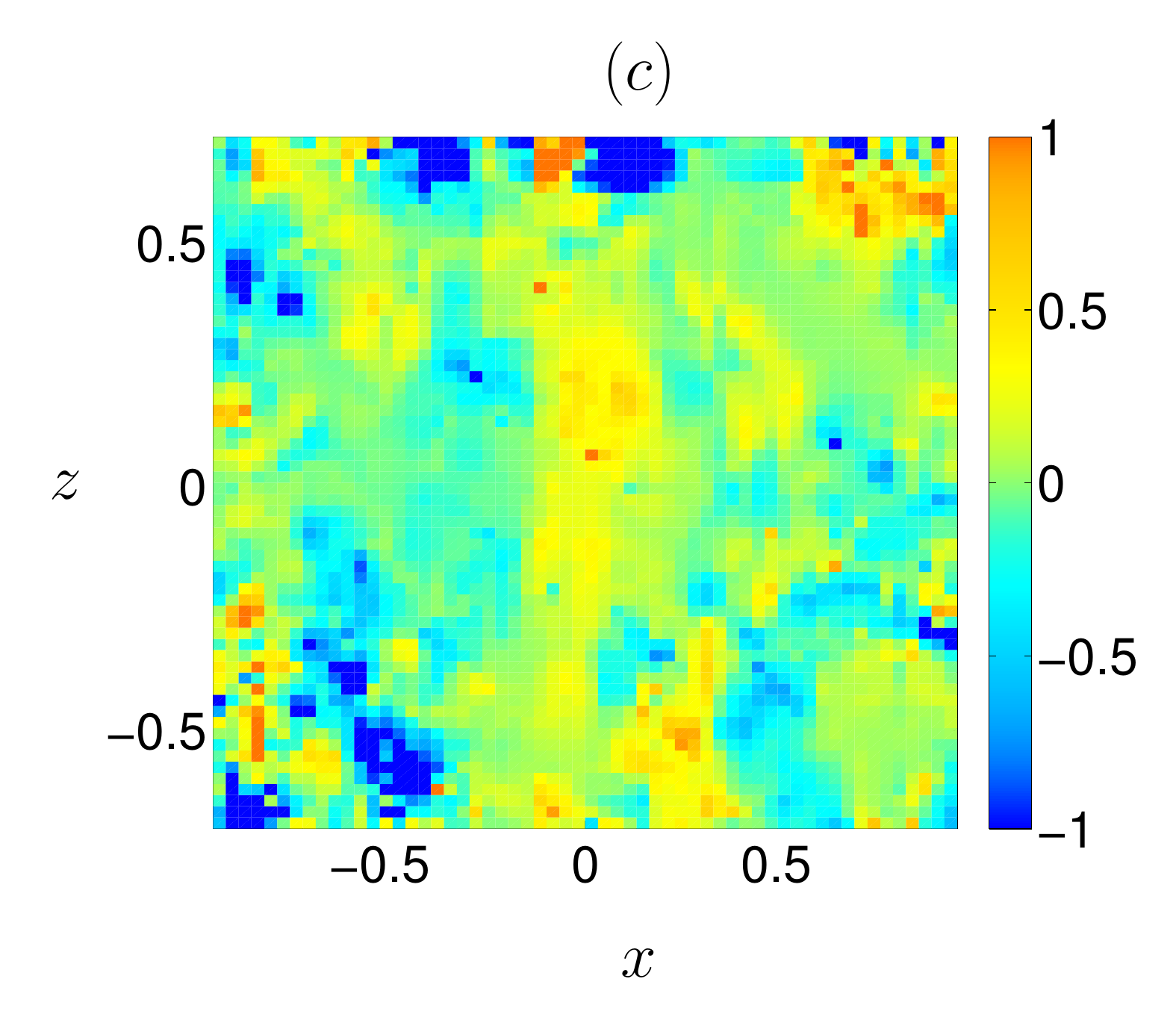}
\includegraphics[width=0.43\textwidth]{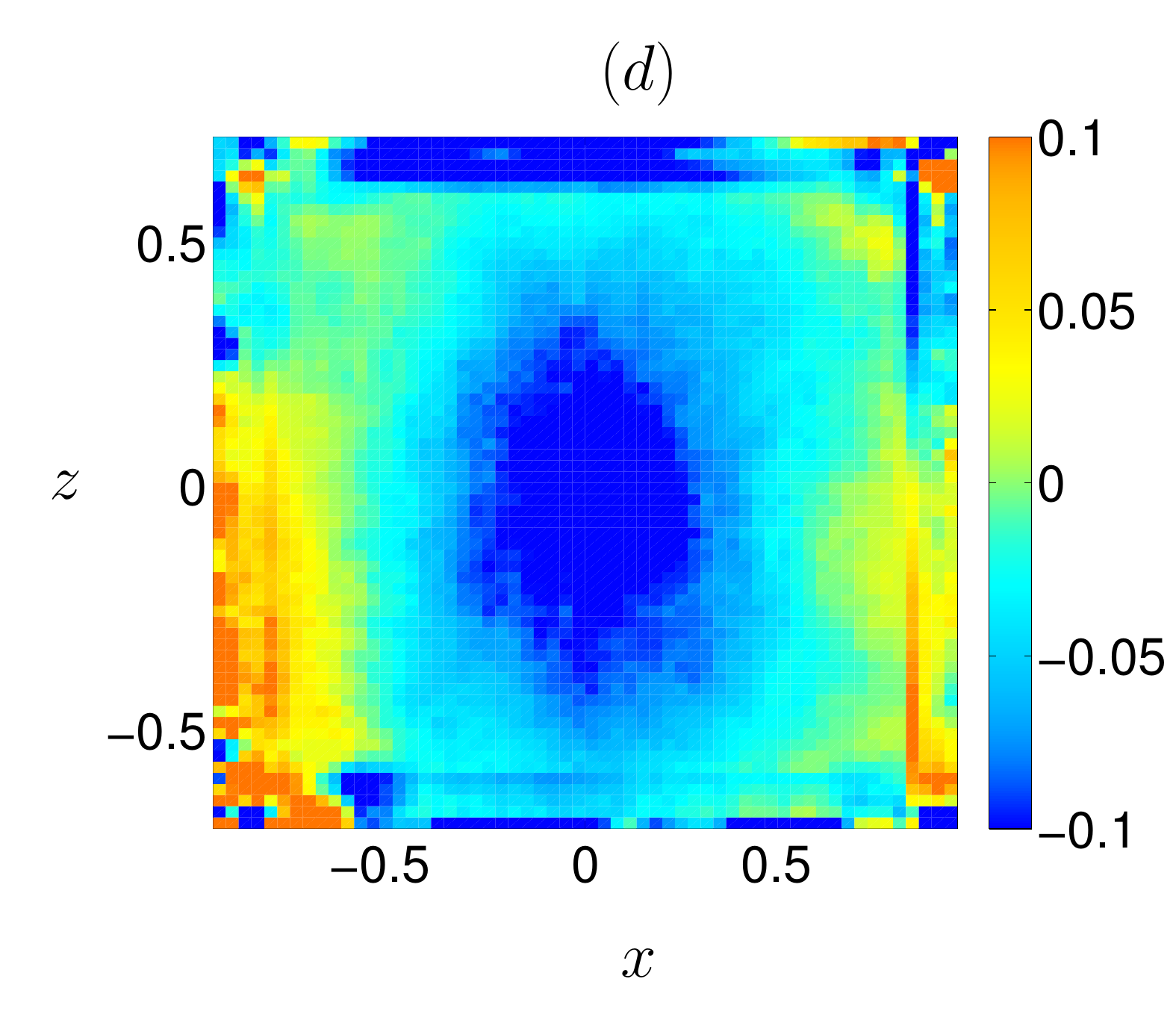}
\end{minipage}
\caption{Top: Maps of the energy production rate (a) and energy transfer rate (b) for symmetric (-) geometry corresponding to the flow in Fig. \ref{4types}(b), using LES-PIV estimates. Bottom: Instantaneous (c)  and time averaged (d) maps of the local dissipation $\mathscr{D}_\ell(\vec u)$ at $\ell=0.1$ for the same flow. Areas where energy accumulates are represented in red, while those where energy leaves are represented in blue.}
\label{Mapanti}
\end{figure}

Considering the case where the flow has not undergone any phase transition, Fig. \ref{Dissi_log(Re)} shows that for this type of forcing we measure a dissipation almost three times bigger than what was observed with the previous forcing condition. In this symmetric case, our estimates of the injected and dissipated power are within $20\%$ of the measured value using the LES-PIV method, whereas we reach $98\%$ of the actual dissipation rate of energy with the DR formula (see Fig. \ref{Dissi_log(Re)}(a)). Local maps of injected and dissipated power are plotted on Fig. \ref{Mapanti}. They correspond to the flows (b) displayed on Fig. \ref{4types} where the forcing is in the (-) sense. Here Fig. \ref{Mapanti}, (a) and (b) represent LES-PIV estimates while  Fig. \ref{Mapanti}, (c) and (d) represents instantaneous and time averaged maps using the DR method. Here, the remarks are essentially the same as in Fig. \ref{Mapscon}: near the impellers, the divergence term brings energy into the system while energy leaves the center of the recirculation cells to accumulate at the center of the cylinder. The dissipation term does not change much either and is approximately constant throughout the plane of measurements. A noticeable difference however, is that the color scale is bigger for this configuration. We thus recover the fact that the flow dissipates more energy than for a forcing in the (+) sense and, as a consequence, the torque of the impellers has to be higher.

The situation changes for the bifurcated (-) states. The shear layer is now very close to the upper or lower impeller, depending on the state "chosen" by the system. The mean velocity field for these two states is essentially the same, differing  only by the transformation $z\rightarrow -z$. The measured injected power is also the same in these two states. However, in terms of PIV-estimated injected or dissipated power, a clear asymmetry occurs. Indeed, when the shear layer is sent downwards (up-pointing triangles), the LES-PIV method provides good estimates of both the injected and the dissipated power (see Fig. \ref{Dissi_log(Re)}(b)), while the DR-method provides about $65\%$  of the dissipated power. In the case where the shear layer is sent upwards (down pointing  triangles), however, both the LES-PIV method and the DR method totally fail to reproduce the measured dissipated energy (giving the wrong sign). Since there is not any asymmetry observed in the mean flow, this difference must be attributed only to fluctuations and tiny asymmetries of the experimental set-up (laser sheet location, focalization of cameras, ..). This points out the  the importance of resolving the shear layer in the PIV-estimates of the injection or dissipation energy rate. We thus conclude that neither the LES-PIV, nor the DR method are appropriate in cases when the shear layer is at the location of the impellers, where it cannot be resolved by our PIV set-up.

\begin{figure}[!h]
\begin{minipage}{0.99\textwidth}
\includegraphics[width=0.4\textwidth]{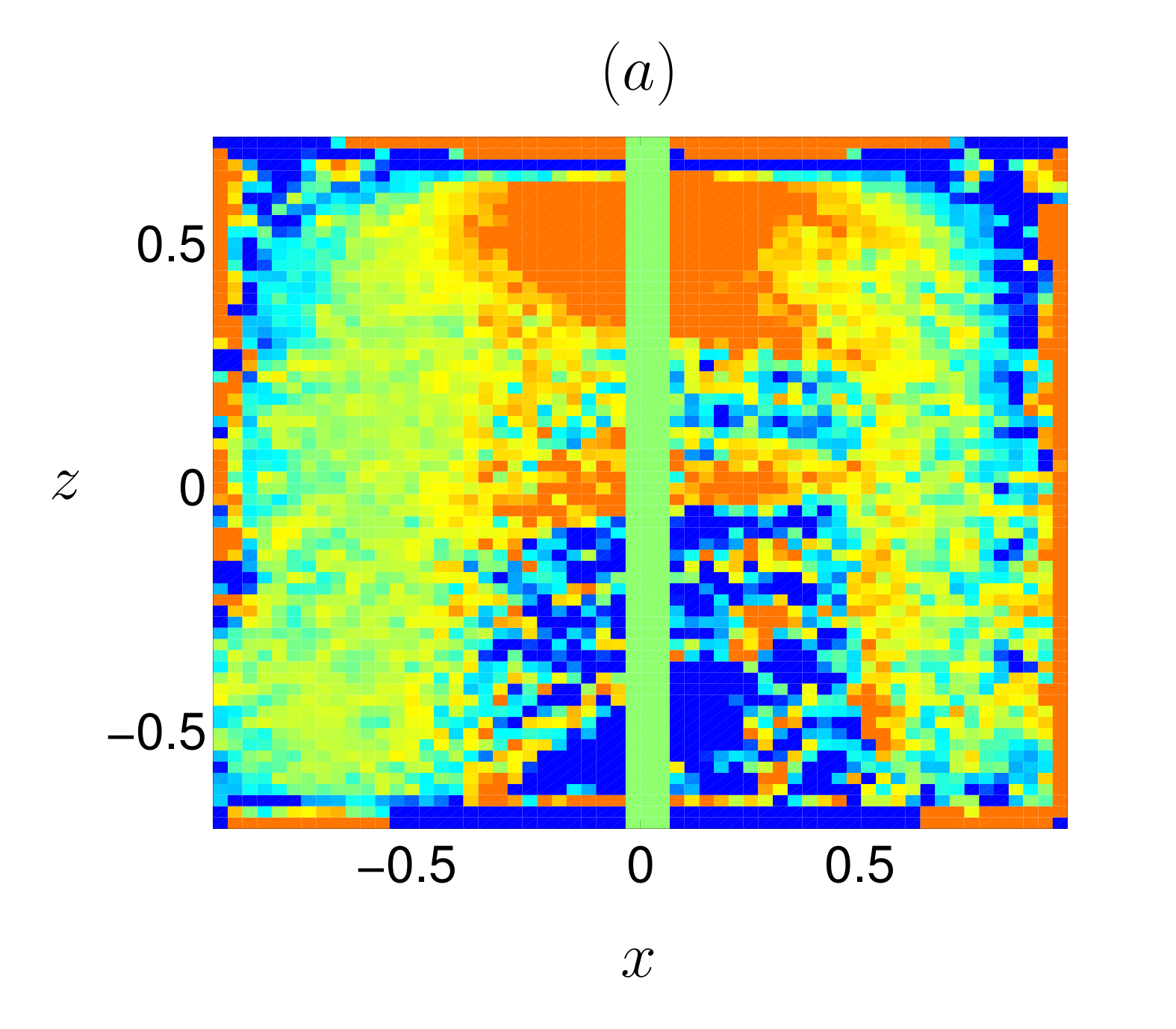}
\includegraphics[width=0.4\textwidth]{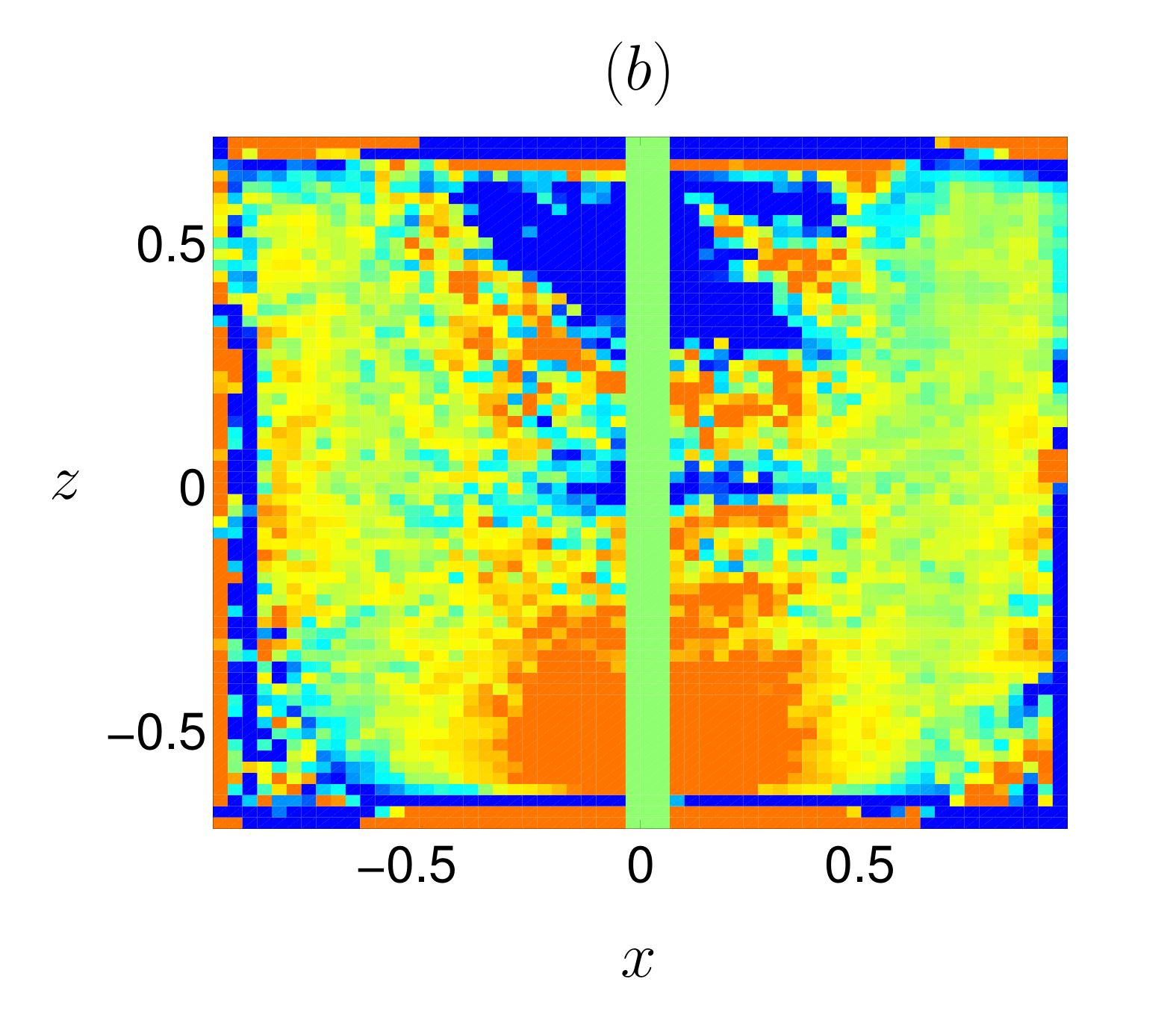}
\includegraphics[width=0.4\textwidth]{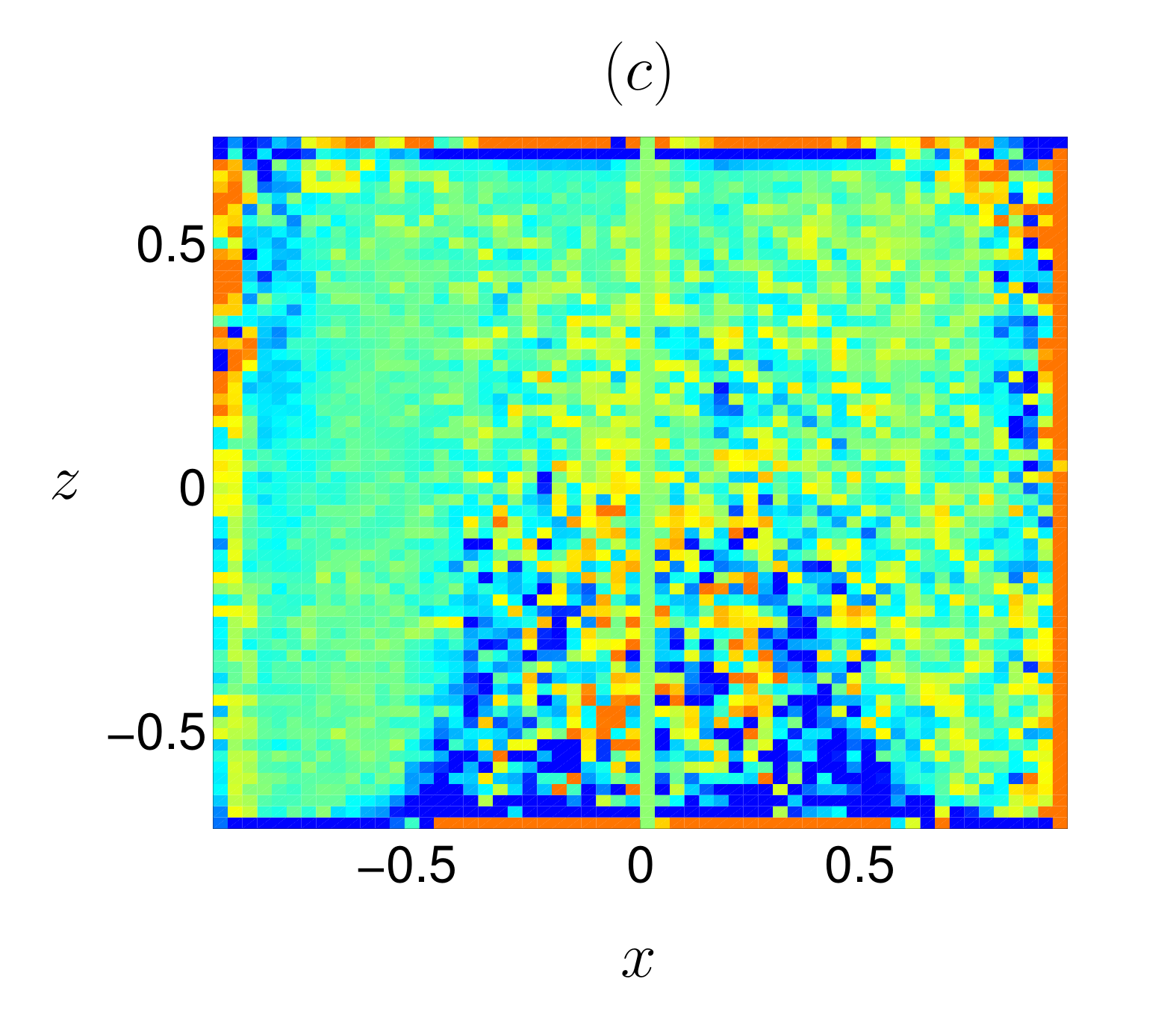}
\includegraphics[width=0.4\textwidth]{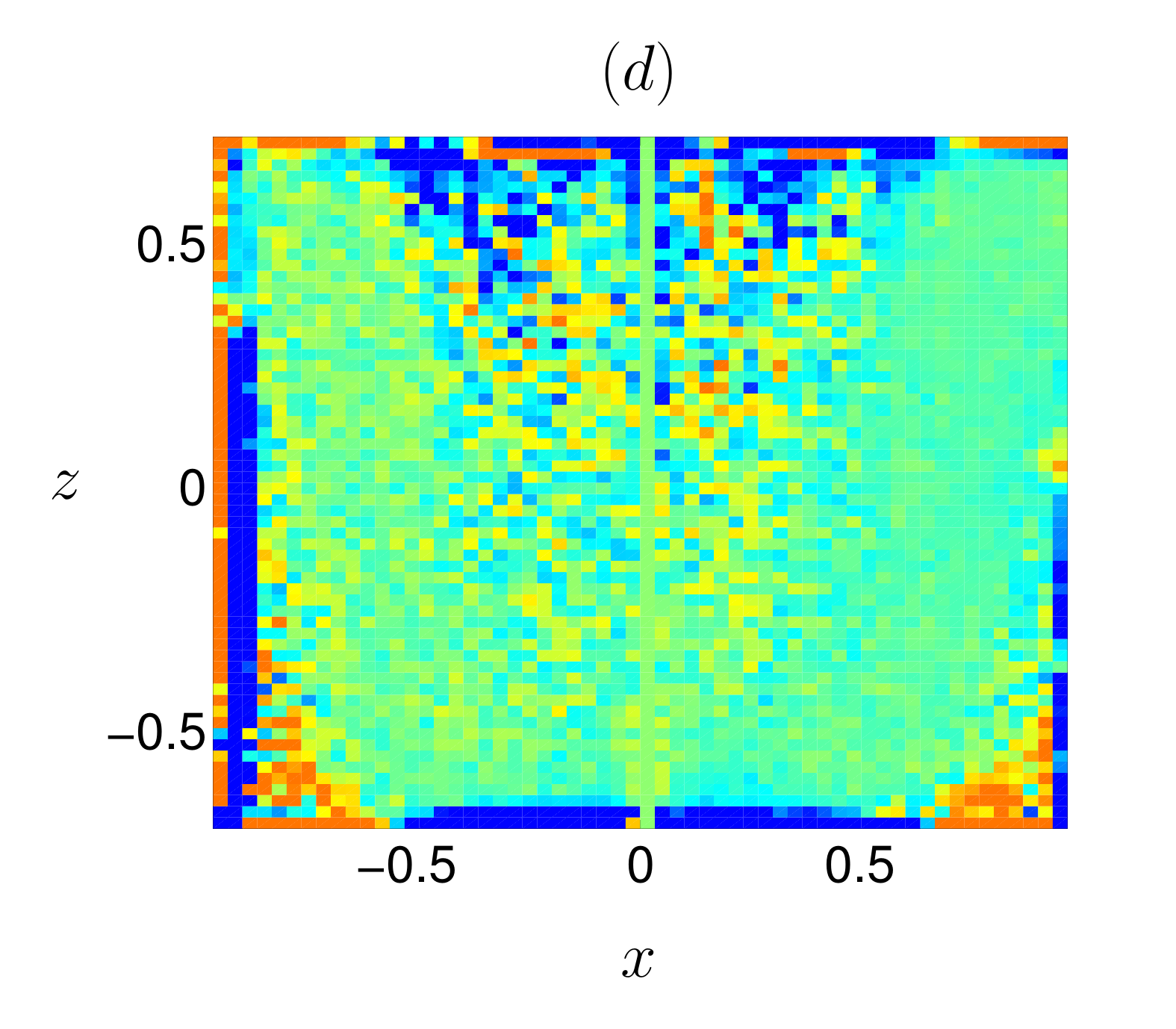}
\includegraphics[width=0.4\textwidth]{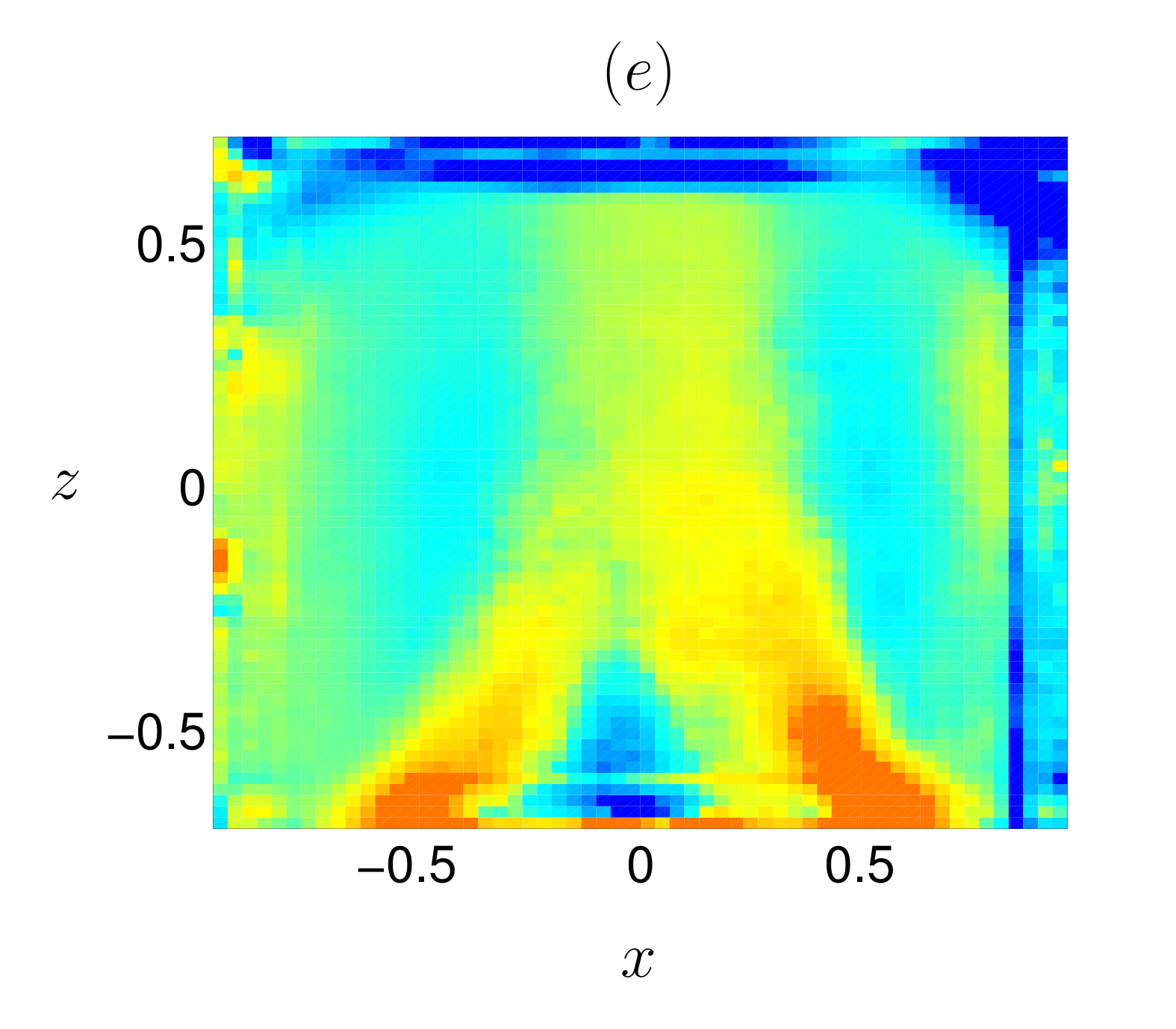}
\includegraphics[width=0.4\textwidth]{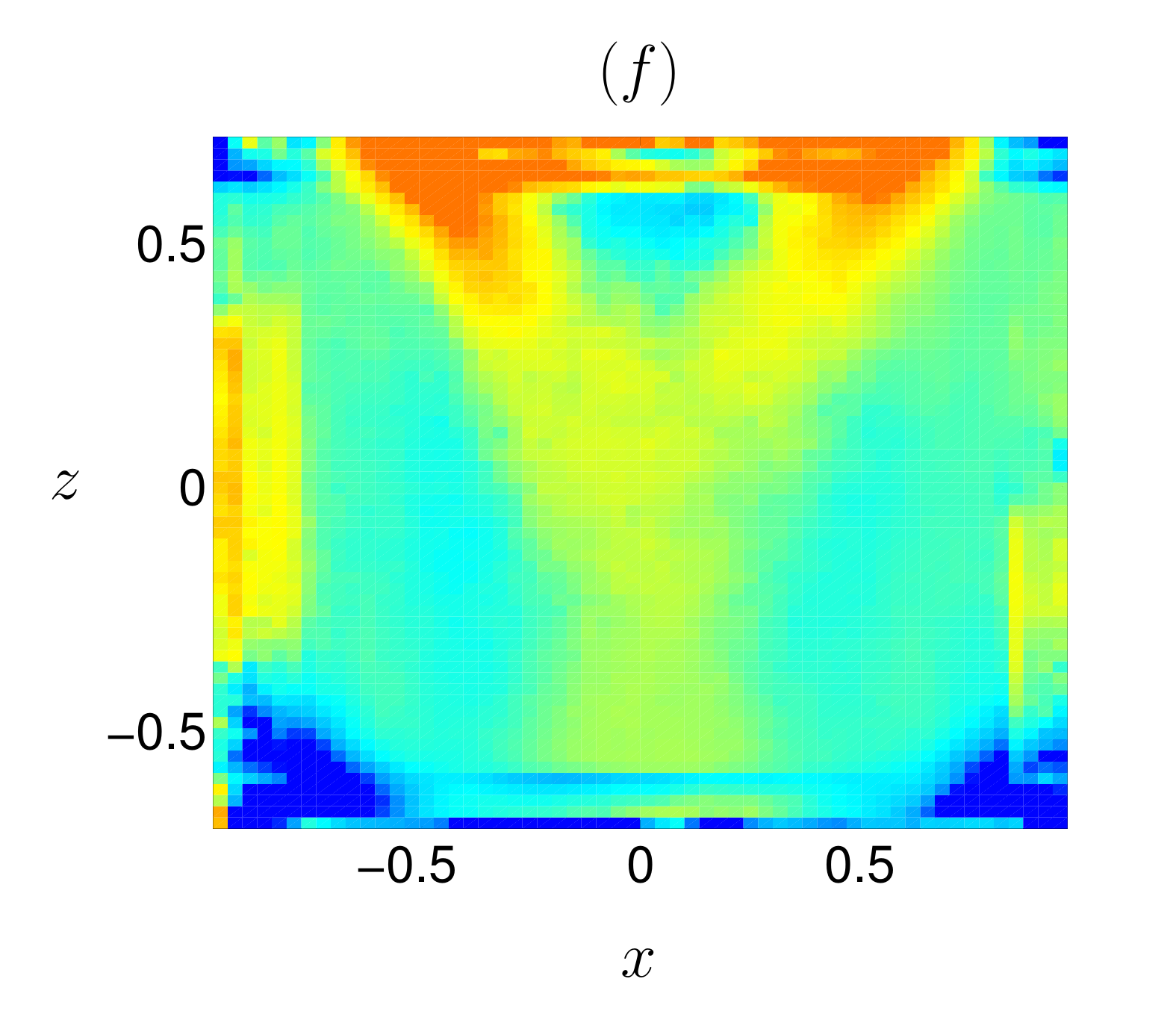}
\includegraphics[width=0.9\textwidth]{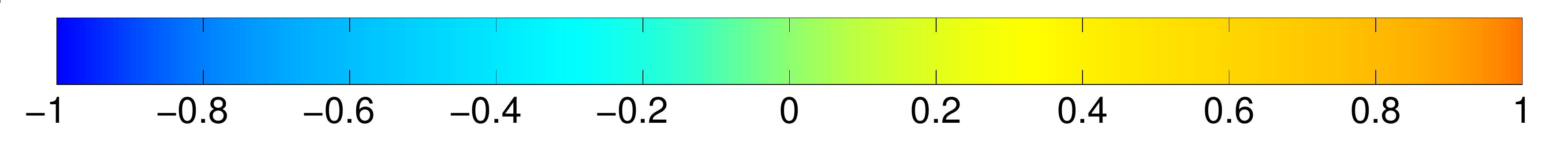}
\end{minipage}
\caption{Maps of the energy production rate and energy transfer rate for the two bifurcated geometries (-): Left: for the flow represented in Fig. \ref{4types}(c). Right: for the flow represented in Fig. \ref{4types}(d). Top line represents the estimated injected energy, using LES-PIV method. Middle line represents the estimated dissipated energy using the LES-PIV method. Bottom line represents the estimated dissipated energy using the DR method. Areas where energy accumulates are represented in red, and those where energy leaves are represented in blue.}
\label{Mapantibif}
\end{figure}

Local maps of injected and dissipated power in the bifurcated states are plotted on Fig. \ref{Mapantibif}. They correspond to the flows (c) and (d) displayed on Fig. \ref{4types} where the forcing is in the (-) sense.  Maps (a)  (c) and (e) correspond to a flow where the shear layer has been sent near the lower impeller. As a consequence, we observe that energy enters from the top and is advected downwards to the shear layer. The map of the dissipation term for this flow looks quite like the production term. We see that most of the dissipation we are able to capture happens at the center of the cylinder but, as we said before, we miss all the dissipation that happens near the lower impeller where there is the shear layer. At first sight, maps (b)  (d) and (f) are just symmetric of maps (a)  (c) and (e) with respect to  $z\rightarrow -z$. However, by looking closely, tiny differences are observed, especially near the walls and impellers. These differences explain the asymmetry between the two bifurcated states shown on Fig. \ref{Dissi_log(Re)}. \medbreak

\subsection{Influence  of the forcing asymmetry $\theta$}

We now investigate the case where the impellers are not exactly counter-rotating, so that the parameter $\theta$ varies from $-0.5$ to $0.5$. In the (-) case, the geometry observed at $\vert \theta\vert > 0$ is always bifurcated (see Fig. \ref{epsilonReet theta}(b)), with the shear layer located at the top or bottom impeller. In the (+) case, the transition is more gradual: the shear layer is increasingly shifted upwards (resp. downwards) as $\theta$ goes from $0$ to $1$ (resp. $-1$), allowing for finer tests of the accuracy of our PIV-estimates as a function of the flow geometry.\

\begin{figure}[!h]
\centering
\includegraphics[width=0.49\textwidth]{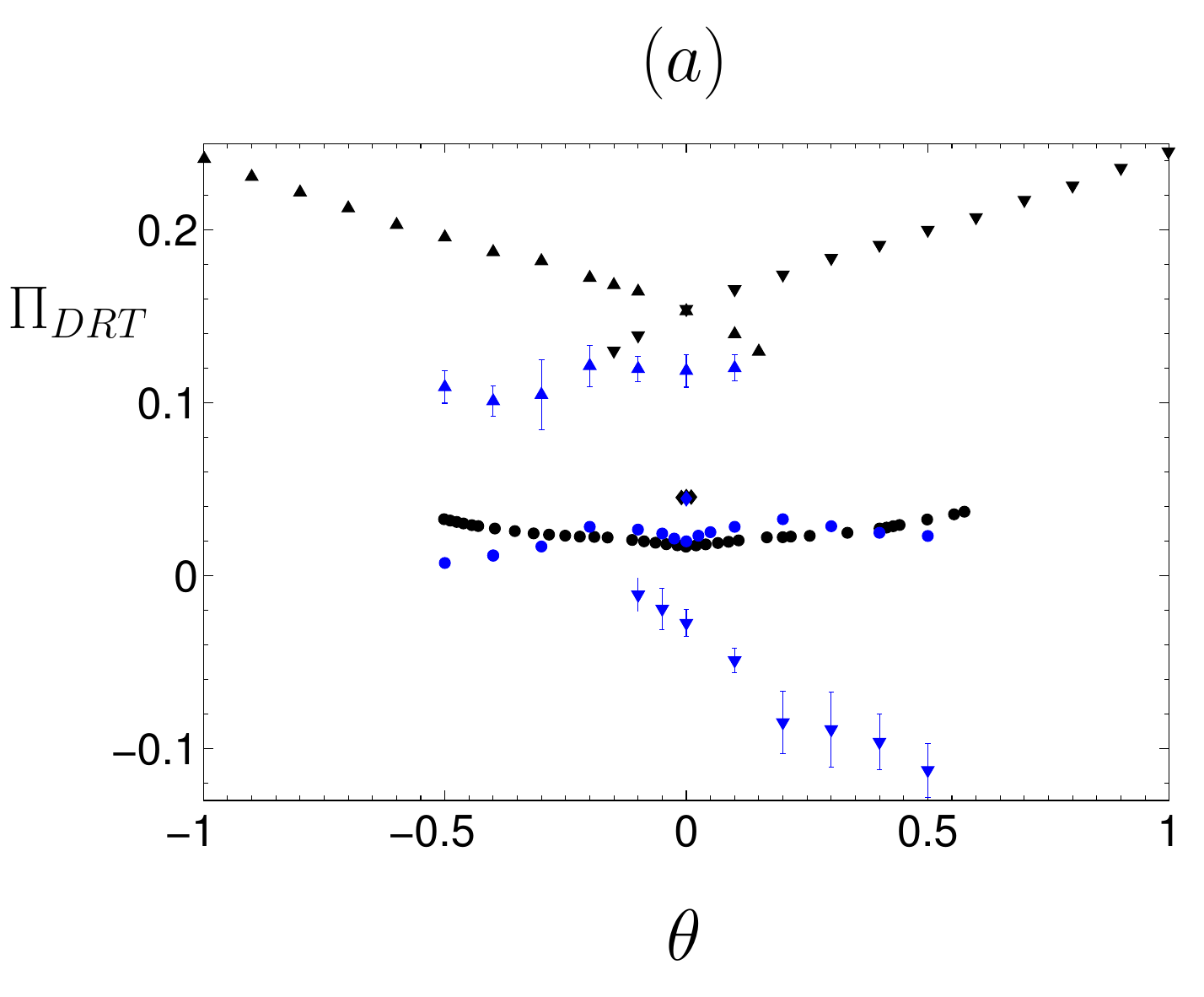}
\includegraphics[width=0.49\textwidth]{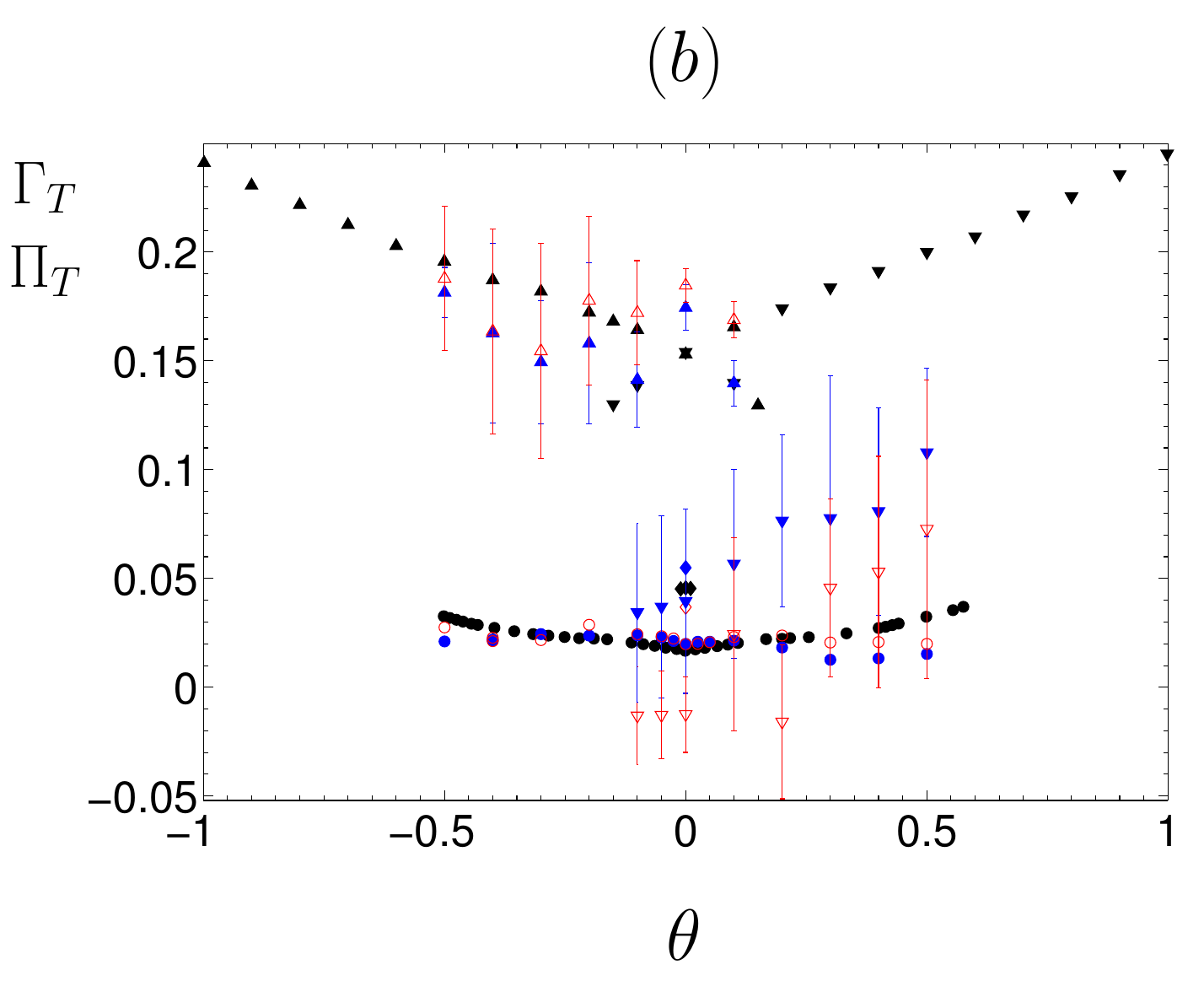}
\caption{Comparison between direct measurements of energy injection obtained through torque measurements (black symbols) and PIV based estimates at various $\theta$, for TP87 impellers for the four different mean state geometries of Fig. 2: disks (+). Rhombi: (-) symmetric. Up triangles: (-) shear layer sent downwards. Down triangle: (-) shear layer sent upwards. a) energy dissipation using DR method (blue symbols). b) energy injection $\Gamma_T$ (red symbols) and dissipation $\Pi_T$ (blue symbols) using the LES-PIV method. The estimates are computed based on 1 to 2 realizations of the same experiment where at least 600 instantaneous velocity snapshots have been taken for each of them. The symbols represent the mean of our computations while the error bars represent the standard deviation.}
\label{Dissi(theta)}
\end{figure}

On Fig. \ref{Dissi(theta)} are plotted our estimates for the total injected (red symbols) and dissipated power (blue symbols) using both the LES-PIV method (Fig. \ref{Dissi(theta)}(b)) and the DR formula (Fig. \ref{Dissi(theta)}(a)). For each $\theta$, our computations are based on sets of at least 600 instantaneous velocity snapshots. Because of the symmetry $z\rightarrow -z$, we expect all estimates to be symmetric with respect to $\theta\rightarrow -\theta$, provided the statistics are well converged and that the shear layer is sufficiently resolved. We see that this is indeed the case for (+) forcing condition, where the shear layer always lies in between the two impellers. However, we observe that both methods become inaccurate when $\vert \theta\vert$ is too high i.e $\vert \theta\vert > 0.3$. For the (-) forcing condition, the estimates give very good agreement for the symmetric state. However, when the phase transition occurs, one observes the same asymmetry as in the $\theta=0$ case between the two possible states: when the shear layer is sent downwards we get a good agreement between measurements and PIV estimates with the LES method. However, the DR method systematically underestimates the dissipation. In the case where with the shear layer is sent upwards, the estimates are really bad with both methods.

\begin{figure}[!h]
\begin{minipage}{0.99\textwidth}
\includegraphics[width=0.43\textwidth]{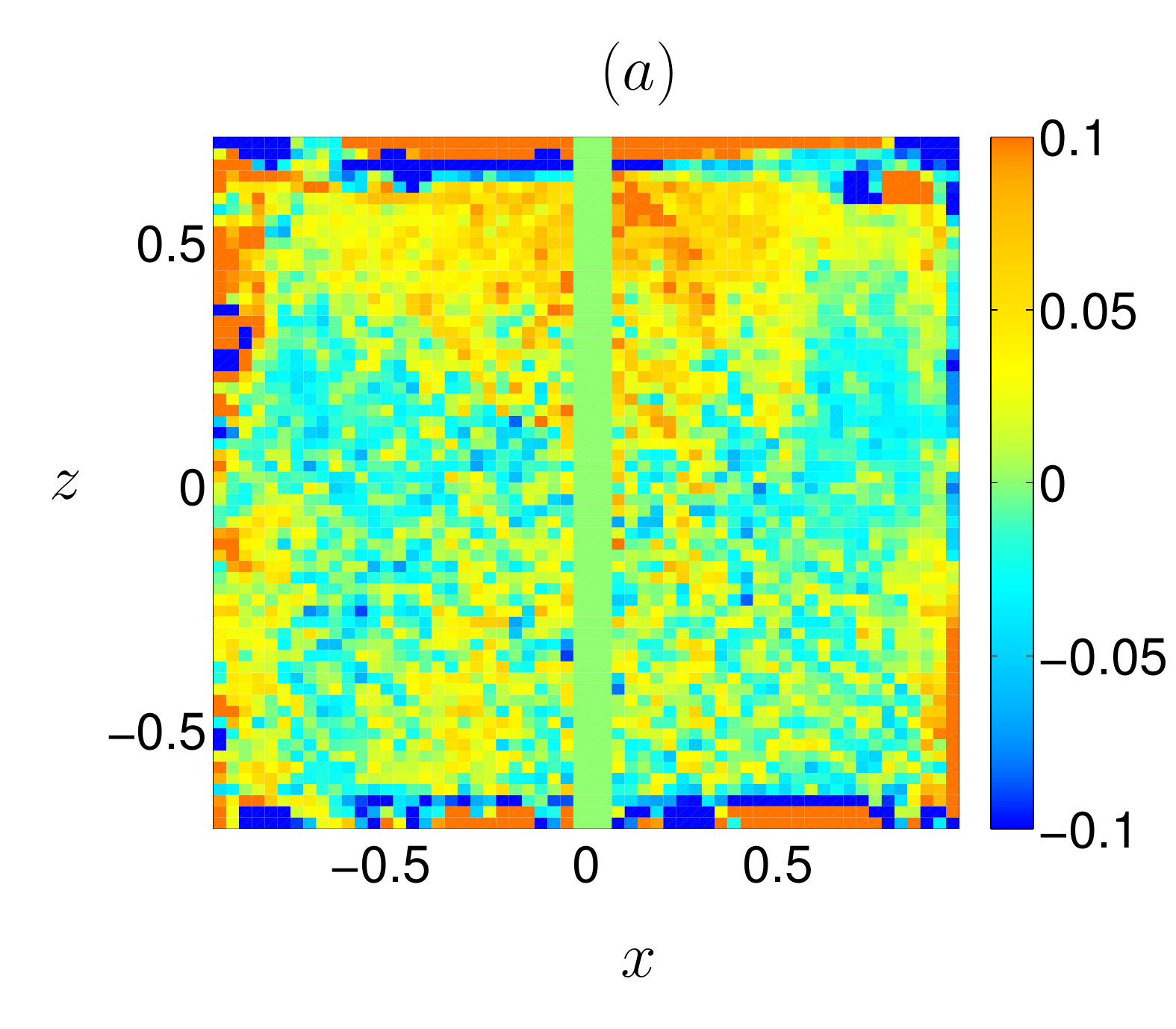}
\includegraphics[width=0.43\textwidth]{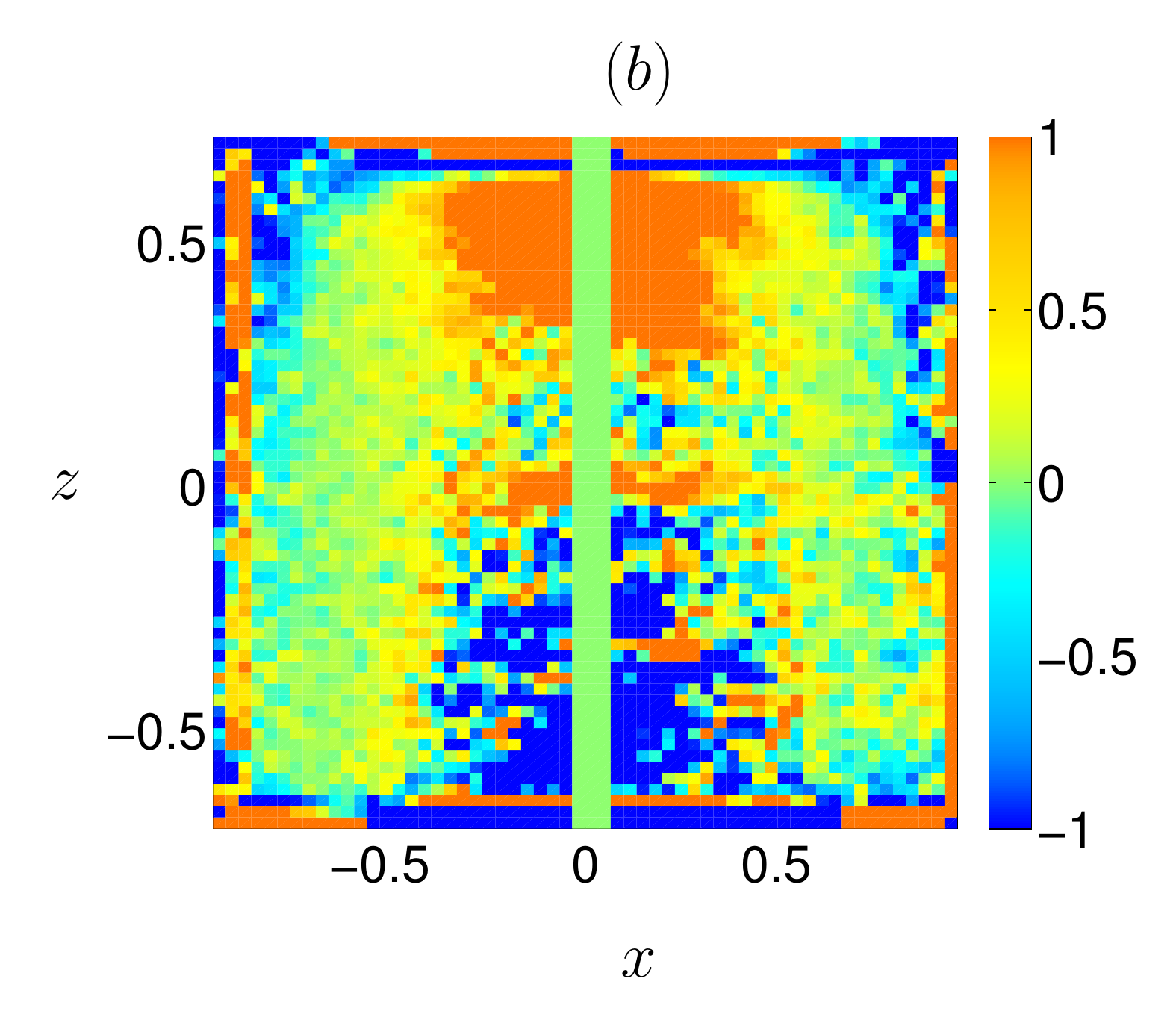}
\includegraphics[width=0.43\textwidth]{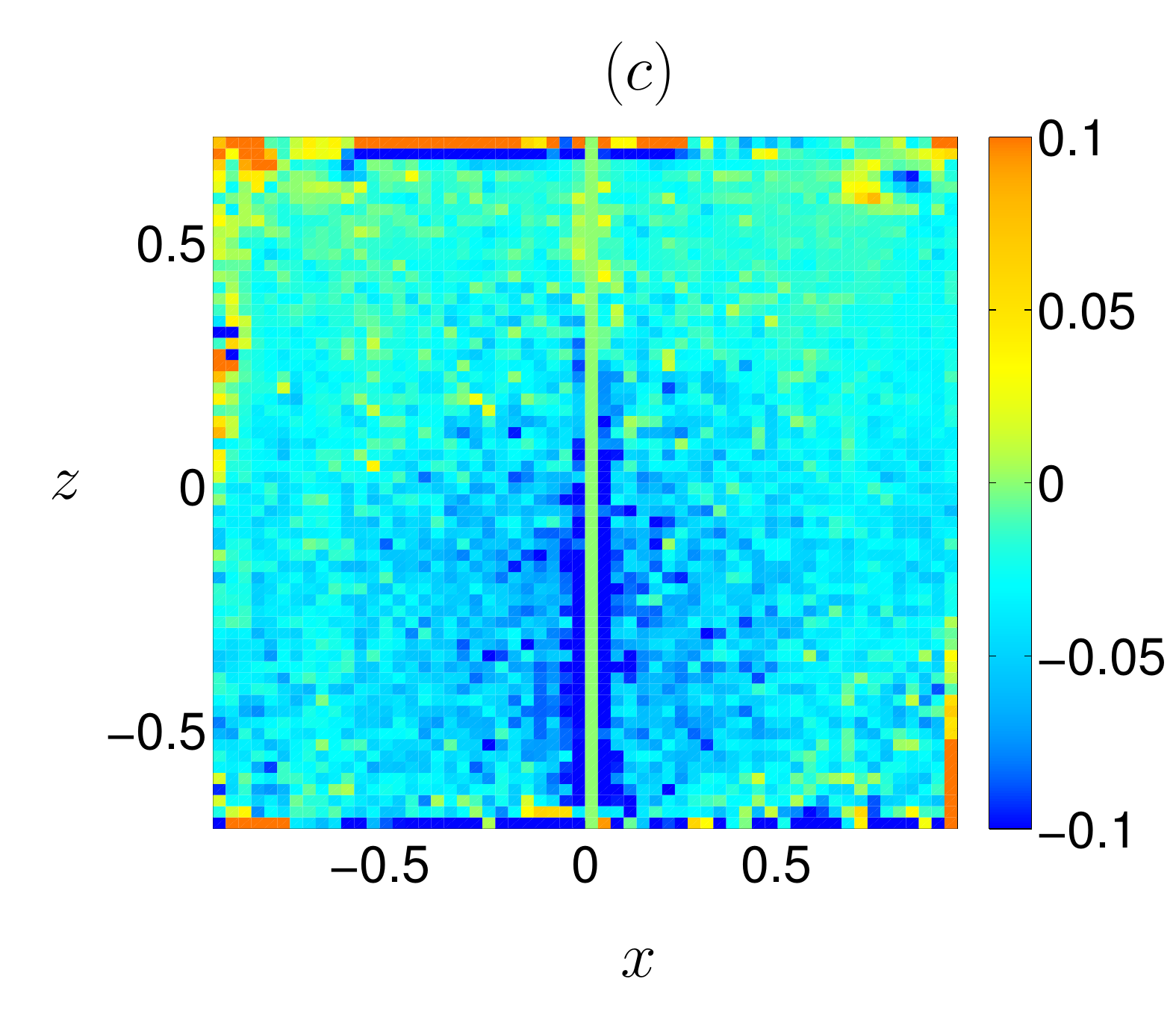}
\includegraphics[width=0.43\textwidth]{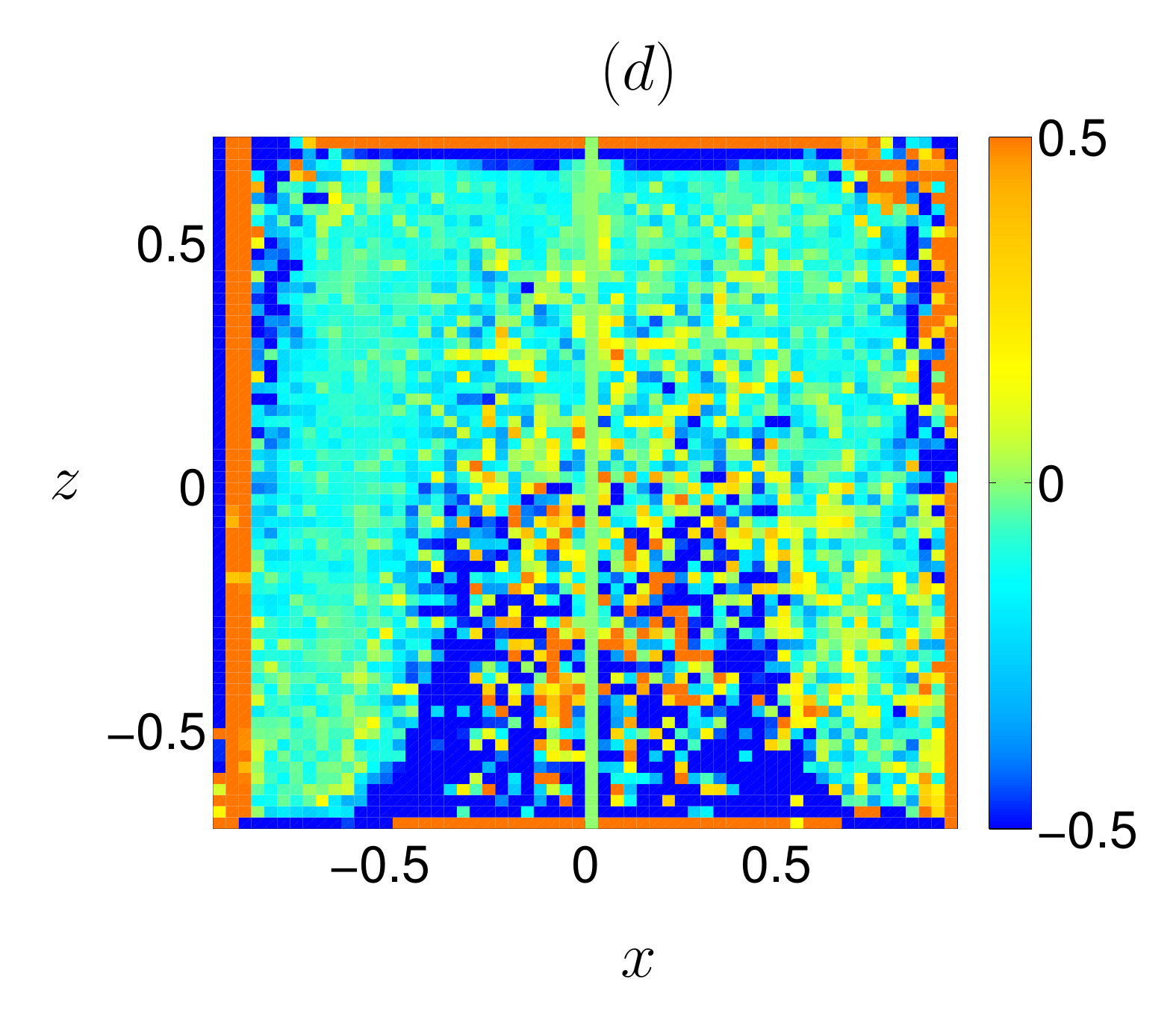}
\includegraphics[width=0.43\textwidth]{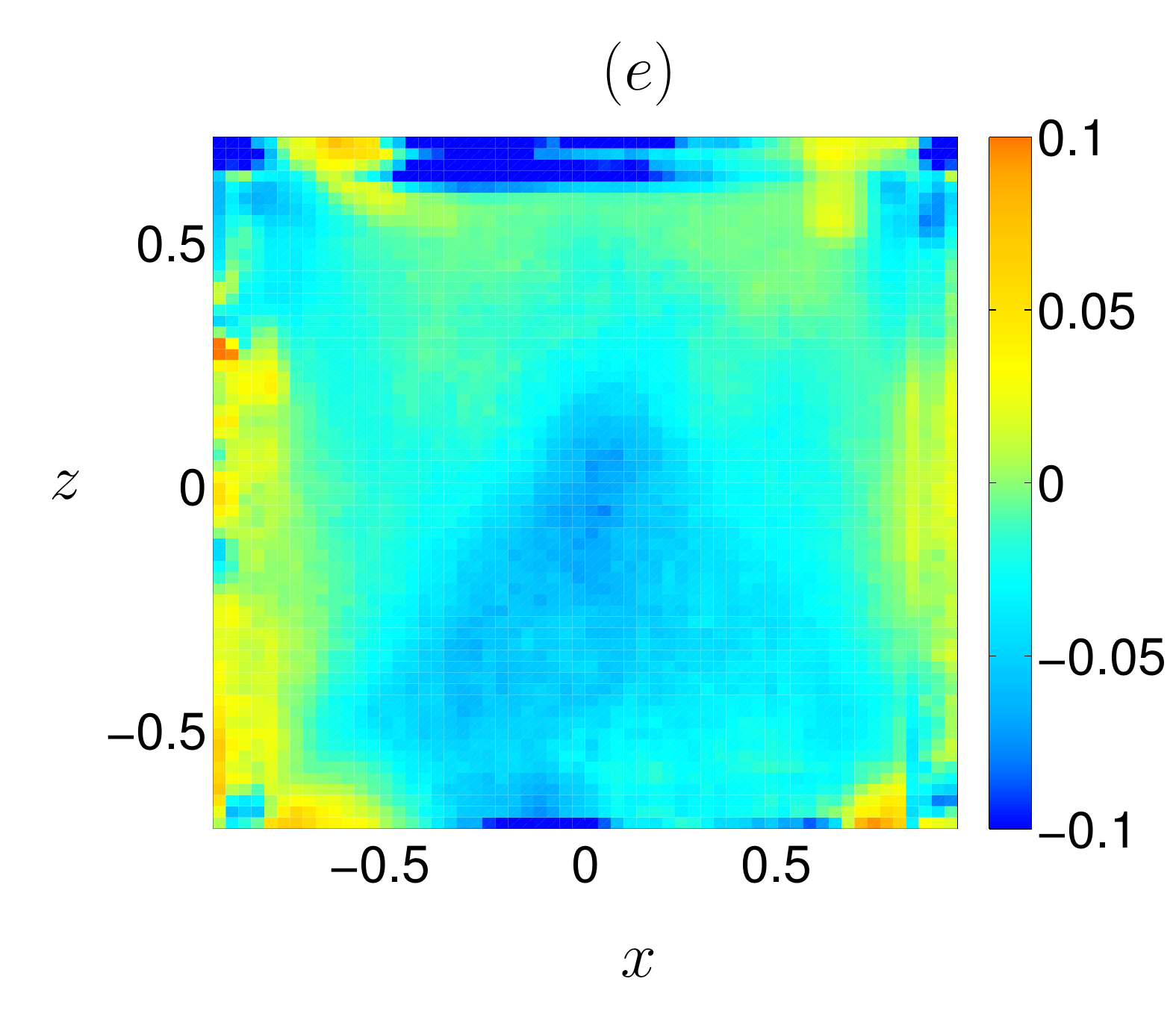}
\includegraphics[width=0.43\textwidth]{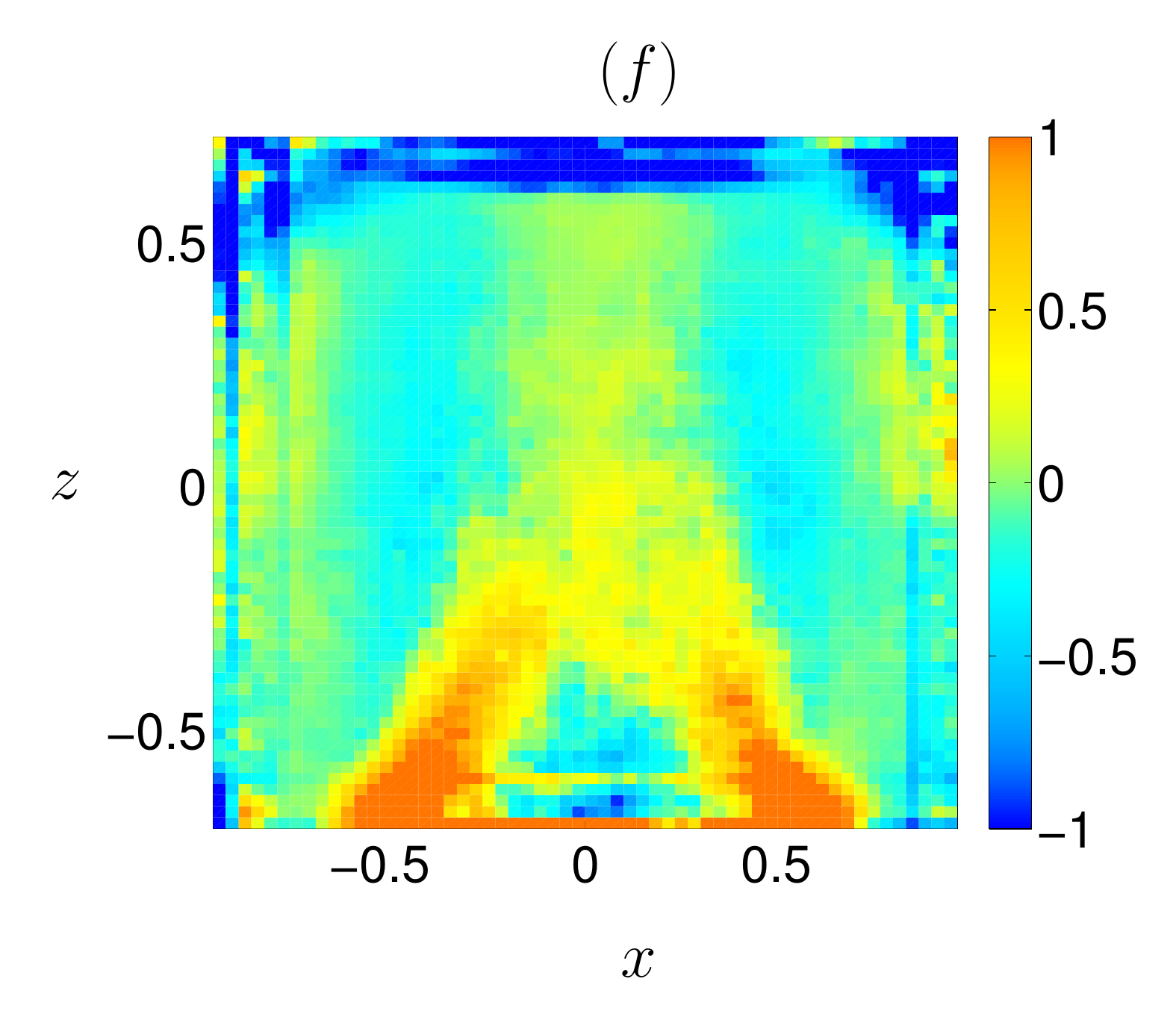}
\end{minipage}
\caption{Maps of the energy production rate and energy transfer rate at $\theta=-0.1$ for the two rotation directions: Left: (+). Right: (-). Top line represents the estimated injected energy, using LES-PIV method. Middle line represents the estimated dissipated energy using the LES-PIV method. Bottom line represents the estimated dissipated energy using the DR method. Areas where energy accumulates are represented in red, and those where energy leaves are represented in blue.}
\label{Mapvartheta}
\end{figure}

Corresponding maps of the injected and dissipated power are provided in Fig. \ref{Mapvartheta} at $\theta=-0.1$ in the (+) and (-) sense. They obey qualitatively the same behavior as in the symmetric case, so that the energy cycle description is qualitatively the same, with the shear layer location being moved as $\theta$ varies.

\section{Conclusion}

In this paper, we investigate the energy cycle of a turbulent von Kármán flow. In this kind of set-up, it is possible to control the mean energy flux inside the flow in order to get a statistically stationary regime. It is also possible to monitor the torque and frequency of the impellers. In our case, we use Particle Image Velocimetry (PIV) measurements to study local and global energy transfers inside the flow. This kind of measurements gives us access to the effective velocity field on a grid that does not resolve the dissipative scale. Therefore, we use a Large Eddy Simulation (LES) approximation to model the influence of unresolved scales. This procedure involves a free parameter which has to be calibrated for our set-up. This is achieved by imposing angular momentum balance at the smallest resolved scale \cite{marie2004}. After deriving an energy balance equation at a fixed scale $\ell$, we proceed to estimate four quantities from our PIV measurements: the local and global mean power injected by the impellers and the local and global mean dissipated power. This computation is performed for various Reynolds numbers and for various flow topologies. These PIV-estimates are then compared with direct injected power estimates provided by torque measurements at the impellers. The agreement between PIV-estimates and direct measurements depends on the flow topology. In symmetric situations, we capture up to $90\%$ of the actual energy dissipation. However, our results become increasingly inaccurate as the shear layer responsible for most of the dissipation approaches one of the impellers, and cannot be resolved by our PIV set-up. At the same time, we explore a new method for PIV-estimates of the energy dissipation, based on the  work of Duchon and Robert \cite{duchonrobert} that generalizes the Kármán-Howarth equation to nonisotropic, non homogeneous flows. This method provides parameter-free estimates of the energy dissipation as long as the smallest resolved scale lies in the inertial range and the shear layer is resolved by the PIV set-up. With this method, we obtain a very good agreement between PIV estimates and direct measurements, and we are able to capture up to $98\%$ of the actual dissipated power in symmetric situations. However, this method also gives increasingly inaccurate results as the mixing layer approaches one of the impellers. These results are used to evidence a well-defined stationary energy cycle within the flow in which the energy is injected by the top and bottom impellers towards the shear layer. There, turbulent fluctuations dissipate energy, and the flow is then pumped towards the impellers, closing the energy cycle.

\newpage

\bibliography{biblio}

\begin{thebibliography}{10}

\bibitem{tokgoz2012}
S.~{Tokgoz}, G.~E. {Elsinga}, R.~{Delfos}, and J.~{Westerweel}.
\newblock {Spatial resolution and dissipation rate estimation in
  Taylor--Couette flow for tomographic PIV}.
\newblock {\em Experiments in Fluids}, 53:561--583, September 2012.

\bibitem{sheng2000}
J.~Sheng, H.~Meng, and R.~O. Fox.
\newblock A large eddy piv method for turbulence dissipation rate estimation.
\newblock {\em Chemical Engineering Science}, 55:4423--4434, 2000.

\bibitem{saa2000}
P.~{Saarenrinne} and M.~{Piirto}.
\newblock {Turbulent kinetic energy dissipation rate estimation from PIV
  velocity vector fields}.
\newblock {\em Experiments in Fluids}, 29:300--307, December 2000.

\bibitem{tanaka2007}
T.~{Tanaka} and J.~K. {Eaton}.
\newblock {A correction method for measuring turbulence kinetic energy
  dissipation rate by PIV}.
\newblock {\em Experiments in Fluids}, 42:893--902, June 2007.

\bibitem{delafosse2011}
A.~{Delafosse}, M.~L. {Collignon}, M.~{Crine}, and D.~{Toye}.
\newblock Estimation of the turbulent kinetic energy dissipation rate from
  2d-piv measurements in a vessel stirred by an axial mixel ttp impeller.
\newblock {\em Chemical Engineering Science}, 66:1728--1737, 2011.

\bibitem{baldihann}
S.~{Baldi} and M~{Yianneskis}.
\newblock {On the Direct Measurement of Turbulence Energy Dissipation in
  Stirred Vessels with PIV}.
\newblock {\em Industrial and Engineering Chemistry Research},
  42(26):7006--7016, novembre 2003.

\bibitem{cortet2009}
P.-P. {Cortet}, P.~{Diribarne}, R.~{Monchaux}, A.~{Chiffaudel}, F.~{Daviaud},
  and B.~{Dubrulle}.
\newblock {Normalized kinetic energy as a hydrodynamical global quantity for
  inhomogeneous anisotropic turbulence}.
\newblock {\em Physics of Fluids}, 21(2):025104, February 2009.

\bibitem{roussetrsi}
B.~{Rousset}, P.~{Bonnay}, P.~{Diribarne}, A.~{Girard}, J.~M. {Poncet},
  E.~{Herbert}, J.~{Salort}, C.~{Baudet}, B.~{Castaing}, L.~{Chevillard},
  F.~{Daviaud}, B.~{Dubrulle}, Y.~{Gagne}, M.~{Gibert}, B.~{H{\'e}bral},
  T.~{Lehner}, P.-E. {Roche}, B.~{Saint-Michel}, and M.~{Bon Mardion}.
\newblock {Superfluid high REynolds von K{\'a}rm{\'a}n experiment}.
\newblock {\em Review of Scientific Instruments}, 85(10):103908, October 2014.

\bibitem{duchonrobert}
J.~{Duchon} and R.~{Robert}.
\newblock {Inertial energy dissipation for weak solutions of incompressible
  Euler and Navier-Stokes equations}.
\newblock {\em Nonlinearity}, 13:249--255, January 2000.

\bibitem{germano92}
M.~{Germano}.
\newblock {Turbulence - The filtering approach}.
\newblock {\em Journal of Fluid Mechanics}, 238:325--336, May 1992.

\bibitem{lesieurbook}
M.~Lesieur.
\newblock {\em Turbulence in Fluids}.
\newblock Kluwer Academic Publishers, 1990.

\bibitem{eyink2005}
G.~L. {Eyink}.
\newblock {Multi-scale gradient expansion of the turbulent stress tensor}.
\newblock {\em Journal of Fluid Mechanics}, 549:159--190, 2006.

\bibitem{clark79}
R.~A. {Clark}, J.~H. {Ferziger}, and W.~C. {Reynolds}.
\newblock {Evaluation of subgrid-scale models using an accurately simulated
  turbulent flow}.
\newblock {\em Journal of Fluid Mechanics}, 91:1--16, March 1979.

\bibitem{onsager49}
L.~Onsager.
\newblock Statistical hydrodynamics.
\newblock {\em Nuovo Cimento (Suppl.)}, 6:279--287, 1949.

\bibitem{mariephd}
L.~Mari{\'e}.
\newblock {\em Transport de moment cin{\'e}tique et de champ magn{\'e}tique par
  un {\'e}coulement tourbillonnaire turbulent: influence de la rotation}.
\newblock PhD thesis, Universit{\'e} Paris 7, 2003.

\bibitem{raveletphd}
F.~Ravelet.
\newblock {\em Bifurcations globales hydrodynamiques et
  magn{\'e}tohydrodynamiques dans un {\'e}coulement de von K{\'a}rm{\'a}n
  turbulent}.
\newblock PhD thesis, Ecole doctorale de l'Ecole Polytechnique, 2005.

\bibitem{monchauxphd}
R.~Monchaux.
\newblock {\em M{\'e}canique statistique et effet dynamo dans un {\'e}coulement
  de von K{\'a}rm{\'a}n turbulent}.
\newblock PhD thesis, Universit{\'e} Paris 7, 2007.

\bibitem{bricephd}
B.~Saint-Michel.
\newblock {\em L'{\'e}coulement de von K{\'a}rm{\'a}n comme paradigme de la
  physique statistique hors {\'e}quilibre}.
\newblock PhD thesis, Universit{\'e} Paris 6, 2013.

\bibitem{cortet2010}
P.-P. {Cortet}, A.~{Chiffaudel}, F.~{Daviaud}, and B.~{Dubrulle}.
\newblock {Experimental Evidence of a Phase Transition in a Closed Turbulent
  Flow}.
\newblock {\em Physical Review Letters}, 105(21):214501, November 2010.

\bibitem{brice2014}
B.~{Saint-Michel}, F.~{Daviaud}, and B.~{Dubrulle}.
\newblock {A zero-mode mechanism for spontaneous symmetry breaking in a
  turbulent von K{\'a}rm{\'a}n flow}.
\newblock {\em New Journal of Physics}, 16(1):013055, January 2014.

\bibitem{brice2014shrek}
B.~{Saint-Michel}, E.~{Herbert}, J.~{Salort}, C.~{Baudet}, M.~{Bon Mardion},
  P.~{Bonnay}, M.~{Bourgoin}, B.~{Castaing}, L.~{Chevillard}, F.~{Daviaud},
  P.~{Diribarne}, B.~{Dubrulle}, Y.~{Gagne}, M.~{Gibert}, A.~{Girard},
  B.~{H{\'e}bral}, T.~{Lehner}, and B.~{Rousset}.
\newblock {Probing quantum and classical turbulence analogy in von
  K{\'a}rm{\'a}n liquid helium, nitrogen, and water experiments}.
\newblock {\em Physics of Fluids}, 26(12):125109, December 2014.

\bibitem{ravelet2005}
F.~{Ravelet}, A.~{Chiffaudel}, F.~{Daviaud}, and J.~{L{\'e}orat}.
\newblock {Toward an experimental von K{\'a}rm{\'a}n dynamo: Numerical studies
  for an optimized design}.
\newblock {\em Physics of Fluids}, 17(11):117104, November 2005.

\bibitem{ravelet2008}
F.~{Ravelet}, A.~{Chiffaudel}, and F.~{Daviaud}.
\newblock {Supercritical transition to turbulence in an inertially driven von
  K?rm?n closed flow}.
\newblock {\em Journal of Fluid Mechanics}, 601:339--364, 2008.

\bibitem{marie2004}
L.~{Mari{\'e}} and F.~{Daviaud}.
\newblock {Experimental measurement of the scale-by-scale momentum transport
  budget in a turbulent shear flow}.
\newblock {\em Physics of Fluids}, 16:457--461, February 2004.

\bibitem{leonard74}
A.~{Leonard}.
\newblock {Energy cascade in large-eddy simulations of turbulent fluid flows}.
\newblock In F.N. Frenkiel and R.E. Munn, editors, {\em Turbulent Diffusion in
  Environmental Pollution}, volume~18 of {\em Advances in Geophysics}, pages
  237--248. Academic Press, New York, 1974.

\bibitem{simon2015}
S.~{Thalabard}, B.~{Saint-Michel}, E.~{Herbert}, F.~{Daviaud}, and
  B.~{Dubrulle}.
\newblock {A statistical mechanics framework for the large-scale structure of
  turbulent von K{\'a}rm{\'a}n flows}.
\newblock {\em New Journal of Physics}, 17(6):063006, June 2015.

\end{thebibliography}
\bibliographystyle{unsrt}			%unsrt permet de faire en sorte que l'ordre de la bibliographie 
							%soit l'ordre d'apparition des références dans le texte

\end{document}